\documentclass[a4paper,12pt]{article}
\usepackage{amsmath,amssymb,mathrsfs,graphicx}
\usepackage[top=1.in, bottom=1.in, left=0.75in, right=0.75in]{geometry}
\usepackage{tikz}
\usepackage{outline}
\usetikzlibrary{shapes}
\usepackage[english]{babel}
\usetikzlibrary{backgrounds}
\usepackage{hyperref}
\usepackage{caption}
\usepackage{subcaption}
\usepackage{ulem}
\usepackage{authblk}
\usepackage{xcolor}

\newcommand{\ub}{\overline{u}}

\numberwithin{equation}{section}

\begin{document}

\title{Soliton-mean field interaction in Korteweg-de Vries dispersive hydrodynamics}
\author[1]{Mark J. Ablowitz}
\author[2]{Justin T. Cole}
\author[3]{Gennady A. El}
\author[1]{Mark A. Hoefer}
\author[4]{Xu-dan Luo}
\affil[1]{Department of Applied Mathematics, University of Colorado, Boulder}
\affil[2]{Department of  Mathematics, University of Colorado, Colorado Springs}
\affil[3]{Department of Mathematics, Physics, and Electrical Engineering, Northumbria University}
\affil[4]{Chinese Academy of Sciences}

\maketitle

\begin{abstract}
  The propagation of localized solitons in the presence of large-scale
  waves is a fundamental problem, both physically and mathematically,
  with applications in fluid dynamics, nonlinear optics and condensed
  matter physics.  Here, the evolution of a soliton as it interacts
  with a rarefaction wave or a dispersive shock wave, examples of
  slowly varying and rapidly oscillating dispersive mean fields, for
  the Korteweg-de Vries equation is studied.  Step boundary conditions
  give rise to either a rarefaction wave (step up) or a dispersive
  shock wave (step down).  When a soliton interacts with one of these
  mean fields, it can either transmit through (tunnel) or become
  embedded (trapped) inside, depending on its initial amplitude and
  position.  A comprehensive review of three separate analytical
  approaches is undertaken to describe these interactions.  First, a
  basic soliton perturbation theory is introduced that is found to
  capture the solution dynamics for soliton-rarefaction wave
  interaction in the small dispersion limit. Next, multiphase Whitham
  modulation theory and its finite-gap description are used to
  describe soliton-rarefaction wave and soliton-dispersive shock wave
  interactions.  Lastly, a spectral description and an exact solution
  of the initial value problem is obtained through the Inverse
  Scattering Transform. For transmitted solitons, far-field
  asymptotics reveal the soliton phase shift through either type of
  wave mentioned above.  In the trapped case, there is no proper
  eigenvalue in the spectral description, implying that the evolution
  does not involve a proper soliton solution. These approaches are
  consistent, agree with direct numerical simulation, and accurately
  describe different aspects of solitary wave-mean field interaction.
\end{abstract}

\tableofcontents

\section{Introduction}

The interaction of small-scale dispersive waves with large-scale mean
fields is a fundamental process in nonlinear wave systems with a
number of applications.  Traditionally, this multiscale problem
involves mean fields that are either externally prescribed, such as a
current, or that are induced by a finite amplitude wavetrain.  A
different class of nonlinear wave-mean field interactions has recently
been identified in which a localized soliton or, more generally, a
solitary wave, and the dynamic mean field evolve according to the same
evolutionary equation \cite{Hoefer2}.  A suitable equation to describe
soliton-mean field interaction is the Korteweg-de Vries (KdV) equation
\begin{equation}
  \label{kdv}
  u_t + 6 u u_x + \varepsilon^2 u_{xxx} = 0 \; ,
\end{equation}
where $t > 0$ is the temporal variable, $x \in \mathbb{R}$ is the
spatial variable, and $u$ is proportional to the wave amplitude.
Equation \eqref{kdv} is presented in non-dimensional, scaled form with
the dispersion parameter $\varepsilon > 0$ measuring the relative
strength of dispersion and nonlinearity.  The KdV equation \eqref{kdv}
admits soliton solutions whose width is proportional to $\varepsilon$.

In the small dispersion regime where $\varepsilon \ll 1$, the soliton
width $\mathcal{O}(\varepsilon)$ is small relative to $\mathcal{O}(1)$
mean field spatial variation.  In this context, the mean field can be
slowly varying itself or can exhibit expanding, rapid, dispersive
oscillations such as for a rarefaction wave (RW) or a dispersive shock
wave (DSW), respectively.  The term mean field applies to both the RW
and DSW in the small dispersion regime because, in the RW, dispersion
is negligible and the DSW is locally described by rapid oscillations
with wavelength $\mathcal{O}(\varepsilon)$ whose parameters, e.g., its
period-mean, change slowly, $\mathcal{O}(1)$, in comparison
\cite{El2016}.

\begin{figure}
  \centering
  \includegraphics[scale=0.75]{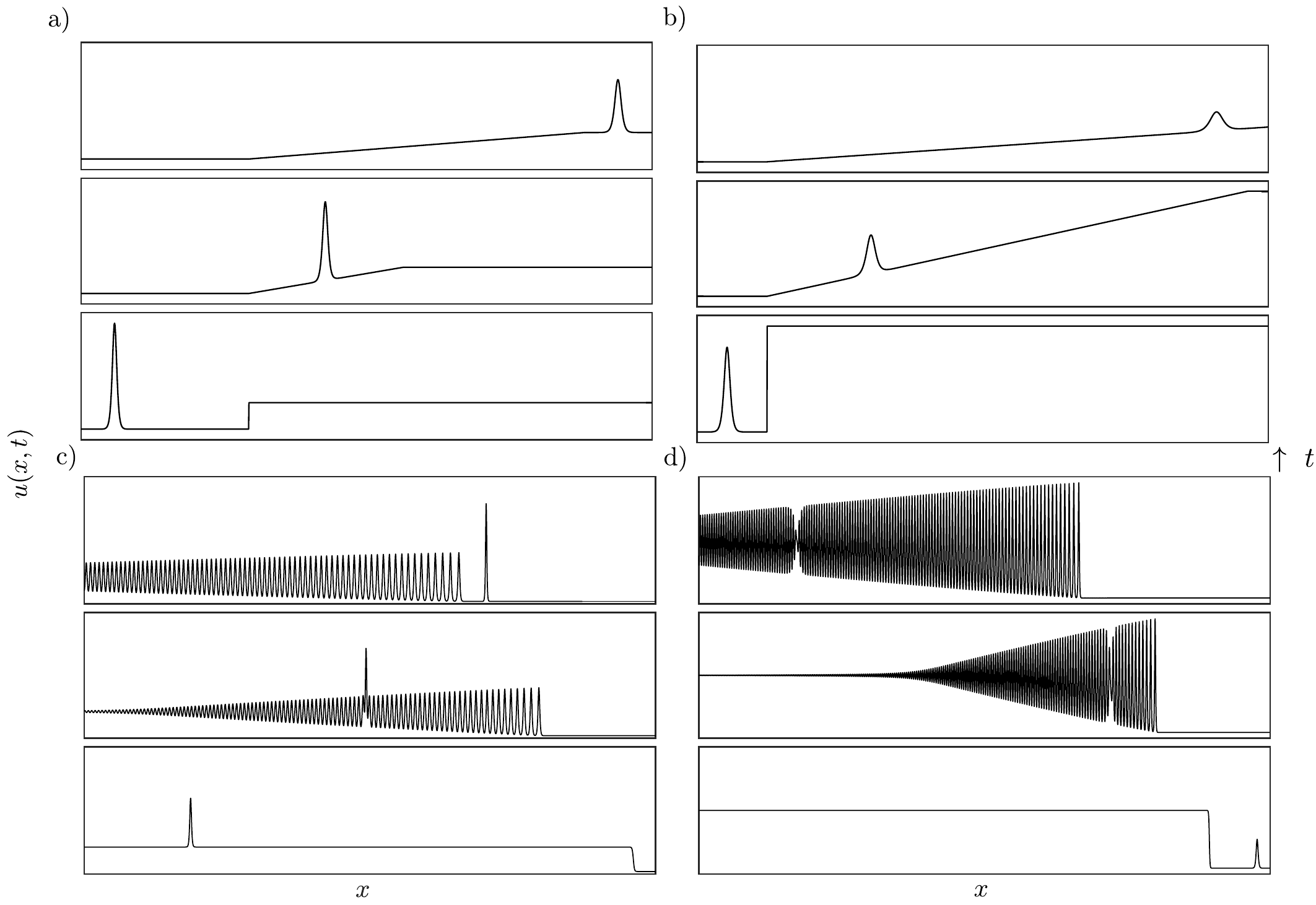}
  \caption{Soliton-mean field interaction scenarios.  a)
    Soliton-RW tunneling. b) Soliton-RW trapping. c) Soliton-DSW
    tunneling. d) Soliton-DSW trapping.}
  \label{fig:interaction_scenarios}
\end{figure}

In \cite{Hoefer2}, Whitham modulation theory
\cite{whitham_linear_1974}---an approximate method for studying
modulated nonlinear wavetrains---was applied to a fluid dynamics
experiment in which the free interface between an interior, buoyant,
viscous fluid and an exterior, much more viscous fluid were found to
exhibit solitary wave, RW, and DSW interaction dynamics
\cite{Hoefer1}.  In both cases of solitary wave-mean field interaction
considered, two possibilities emerge.  Either the solitary wave
incident upon the mean field transmits or tunnels through the mean
field to then propagate freely on the other side with an altered
amplitude and speed.  Or, the incident solitary wave remains embedded
or trapped within the interior of the mean field.  Sketches of these
scenarios are depicted in Figure \ref{fig:interaction_scenarios}.
Soliton transmission and trapping can also be interpreted as a form of
soliton steering by the mean field.

Applications of the soliton-mean field problem range from geophysical
fluid dynamics to photonic/matter waves and material science.
Wherever the fundamental processes of dispersive hydrodynamics
\cite{biondini_dispersive_2016}---multiscale, nonlinear, dispersive
waves---arise, solitons and large-scale mean fields can occur.  We
highlight some recent examples of environments in which solitons and
DSWs have been observed and studied. If solitons and DSWs are studied
separately, their interaction, and the interaction of solitons with
other mean fields, are additional dynamical processes that are
important to understand.  In geophysical fluid dynamics, applications
include gravity water waves and tsunamis
\cite{Trillo2016,Berchet2018,Chassagne2019,Bruhl2022} as well as
internal ocean waves \cite{Carr2019,Li2018,Harris2017,Jackson2004}
where DSWs are also termed undular bores.  There has been significant
interest in the propagation of large scale mean fields in spatial and
fiber nonlinear optics
\cite{Wetzel2016,Nuno2019,Bienaime2021,Bendahmane2022}, where
fundamental DSW properties have been favorably compared with Whitham
modulation theory.  Superfluids are another medium in which
nonlinearity due to particle interactions and dispersion resulting
from quantum matter wave interference lead to solitons and DSWs
\cite{Mossman2018,Buelna2019,DiCarli2020,Mossman2020}.  Solitons and
dispersive shock waves have also been studied in stressed solids and
magnetic materials \cite{Qiu2022,Hooper2021,Janantha2017}.

The motivation for this review is the rapid and varied mathematical
developments on this problem that have occurred within only a few
years.  This justifies a presentation of several mathematical
approaches to the problem of soliton-mean field interaction for the
KdV equation, which is arguably the simplest and most fundamental
nonlinear dispersive wave model.  First, we highlight the recent
results obtained using Whitham modulation theory.  The analysis of
soliton-mean field interaction for a general class of unidirectional
evolutionary equations that was initiated in \cite{Hoefer2} has since
been extended to the bidirectional case for the defocusing nonlinear
Schr\"odinger (NLS) equation \cite{sprenger_hydrodynamic_2017} that
includes head-on and overtaking interactions.  The modified KdV
equation has been used to study the interactions of new types of
solitons and mean fields that come about because of nonconvex flux
\cite{vandersande}.  An extension to the two-dimensional, oblique
interaction of solitons and mean fields was obtained in the context of
the Kadomtsev-Petviashvili (KP) equation \cite{Ryskamp2021}.  An
analogous problem involving linear wave-mean field interaction
exhibits similar behavior to soliton-mean field interaction, which was
studied for the KdV equation in \cite{congy,kamchatnov}.  While
(m)KdV, NLS, and KP are all integrable equations, modulation theory
can be applied to nonintegrable equations and has been successfully
used to analyze soliton-mean field interaction in the conduit equation
\cite{Hoefer2}, a model of viscous core-annular fluids, and the
Benjamin-Bona-Mahoney equation in \cite{Gavrilyuk2021}, both nonlocal
equations.  The inclusion of external non-uniformities via a
Hamiltonian-based modulation approach to soliton-mean field
interaction for the non-integrable Gross-Pitaevskii equation was
obtained in \cite{Ivanov2022}.

While modulation theory has proven to be an effective method to
analytically describe soliton-mean field interaction in a wide class
of model equations, it is a formal approach in the sense that its
results are not rigorously proven.  A parallel set of rigorous
mathematical developments for integrable equations has been achieved
using the Inverse Scattering Transform (IST) \cite{Ablowitz_book}.
The exact solutions for soliton-RW interaction in the KdV equation and
soliton-DSW interaction in the focusing NLS equation were obtained in
\cite{ALC} and \cite{biondini_2018,biondini_2019}, respectively.  In
these cases, small dispersion asymptotics provide strong justification
for the modulation theory results.  Another IST-related approach that
leverages the integrability of the KdV equation is the Darboux
transformation, which was used in \cite{Mucalica2022} to obtain a
nonlinear superposition of a soliton and rarefaction wave at $t = 0$
for the transmission case.

The problem of a soliton interacting with a nonlinear wavetrain that
asymptotes to a cnoidal wave in the mKdV equation was studied in
\cite{Girotti2022}.  This soliton-mean field problem is equivalent to
a test soliton propagating through a soliton condensate, a special
kind of soliton gas \cite{El2021}, linking soliton-mean interaction to
another rapidly growing field of nonlinear wave research.  In fact,
soliton-cnoidal wave interaction in the KdV equation was studied some
time ago \cite{Kuznetsov1975} in which exact solutions corresponding
to $N$ soliton ``dislocations'' to a cnoidal wave were obtained.  For
the $N=1$ case, we recognize these solutions as breathers, bright or
dark, exhibiting two time scales associated with their propagation and
background oscillations.  As we will see, breathers play an important
role in soliton-DSW interaction.

This manuscript provides a comprehensive description of solitary
waves/solitons as they interact with either RWs or DSWs, the simplest
class of mean fields, in the KdV equation.  We review and compare both
previously obtained \cite{Hoefer2,ALC} and new results for
soliton-mean field interaction using modulation theory and IST.  In
addition, we develop another analytical approach to the problem using
soliton perturbation theory.  Here we focus on the KdV equation as it
is integrable and can be solved exactly by the IST
\cite{Ablowitz_book}.  The exact solution provides proof of the
effectiveness of our approximate methods and makes precise through
integrals of motion the origin of soliton-mean field trapping and
tunneling phenomena.  All of these results are further elucidated by
comparison with direct numerical simulation.  Finally, the KdV
equation is the canonical model of nonlinear wave trains in weakly
dispersive media.  Indeed, in a small amplitude regime, it is possible
to recover the KdV equation \cite{whitehead} from the conduit equation
modeling the interfacial fluid dynamics of solitary wave-mean field
interaction experiments \cite{Hoefer1,Hoefer2}.

\subsection{Initial Value Problem}
\label{sec:init-value-probl}

In order to provide a heuristic sketch of the soliton-mean field
interaction's multiscale structure, we begin with the soliton solution
to the KdV equation \eqref{kdv}
\begin{equation} 
  \label{soliton_const_back}
  u_s(x,t) = B +  A_0 ~{\rm sech}^2\left( \sqrt{\frac{A_0}{2
        \varepsilon^2}}  \left[ x - (2 A_0 + 6B) t - x_0 \right]
  \right) ~ , 
\end{equation}
where the parameter $B \in \mathbb{R}$ corresponds to the background,
constant mean field, $A_0 > 0$ is the soliton amplitude and
$x_0 \in \mathbb{R}$ is the soliton's initial position.  The soliton's
width is proportional to $\varepsilon$ so, in the small dispersion
regime, the soliton is a rapidly decaying, rapidly varying, finite
amplitude disturbance.  The simplest class of slowly varying mean
fields are those in which $\varepsilon \to 0$ and $u \to \overline{u}$
in eq.~\eqref{kdv} corresponding to the Hopf or inviscid Burgers'
equation
\begin{equation}
  \label{eq:hopf}
  \overline{u}_t + \overline{u}~\overline{u}_x = 0 ~,
\end{equation}
where we identify $\overline{u}(x,t)$ as the slowly varying mean
field.  The solution to the Hopf equation
$\overline{u} = B_0(x-\overline{u}t)$ for smooth initial data
$\overline{u}(x,0) = B_0(x)$ corresponds to a slowly varying mean
field until the point of gradient catastrophe
$t = t_b = -\max_x 1/B_0'(x)$.  For $t > t_b$, the dispersive term in
\eqref{kdv} becomes important and a DSW is formed.  The leading order
behavior for the soliton-mean field problem can be described by the
initial value problem $u(x,0) = u_s(x,0)$ for eq.~\eqref{kdv} in the
small dispersion regime $\varepsilon \ll 1$ for which the initial mean
field in \eqref{soliton_const_back} is slowly varying $B \to B_0(x)$
relative to the rapid $\mathcal{O}(\varepsilon^{-1})$ variation of the
soliton.  Due to scale separation, the leading order evolution of the
mean field $\overline{u}(x,t)$ is independent of the soliton,
including when $t > t_b$.  In contrast, the soliton is significantly
influenced by the evolving mean field, which changes the soliton's
amplitude and speed during the course of interaction.  Although this
presentation pre-supposes an initial, slowly varying mean field
$B_0(x)$, a suitable limit extends this heuristic description to step
initial conditions $B_0(x) \to \pm c^2 H(x)$ for the mean field
provided the soliton is well-separated from the step, i.e.,
$|x_0| \gg \varepsilon$.  The Heaviside step function is defined as
\begin{equation}
\label{Heaviside}
H(x) =
\begin{cases}
0 & x < 0 \\
1 & x > 0 
\end{cases} .
\end{equation}
This is the canonical problem of a soliton interacting with either a
RW or DSW mean field.  It consists of three inherent length scales:
the separation between the soliton and the step ($|x_0|$), the soliton
width ($\varepsilon/\sqrt{A_0}$), and the width of the RW or DSW
($L \sim t$) that emerges during the course of evolution of the
initial step.  The soliton-mean field problem requires scale
separation, implying
\begin{equation}
  \label{eq:scale-separation}
  |x_0| \gg \varepsilon/\sqrt{A_0}, \quad L \sim t \gg
  \varepsilon/\sqrt{A_0}.
\end{equation}
All of the analysis that follows assumes the multiscale structure in \eqref{eq:scale-separation}.

We now precisely state the initial value problems under consideration
in this review.  The KdV equation \eqref{kdv} is subject to the
boundary conditions (BCs) $u \rightarrow u_{\pm}$ as
$x \rightarrow \pm \infty$, where $u_+ \not= u_-.$ By utilizing the
Galilean invariance of the KdV equation, the left boundary condition
can always be set to zero without loss of generality: $u_- = 0$.  At
the right boundary, we consider two cases: $u_+ = c^2 > 0 $ (step up)
and $u_+ = - c^2 < 0$ (step down) for $c > 0$.  For $t>0$, these
boundary conditions lead to a RW and DSW, respectively
\cite{Leach2008,baldwin,Egorova2013}. Using the scaling symmetry
$u' = u/c^2$, $x' = c x$, $t' = c^3 t$, which leaves the KdV equation
\eqref{kdv} unchanged in primed coordinates, we can always set $c = 1$
without loss of generality.  The initial conditions for the step up
and step down cases are of the form
\begin{equation}
  \label{general_IC}
  u(x,0) = \pm c^2 H(x) + v(x,0;x_0) \; , 
\end{equation}
respectively, for step height $c^2$ and a localized solitary mode
$v(x,t;x_0)$ initially centered at the point $x_0 \in \mathbb{R}$ with
initial amplitude $A_0 > 0$.  We deliberately refer to $v(x,t;x_0)$ as
a solitary mode because its precise form will be determined during the
course of our analysis.  If $|x_0| \gg \varepsilon$, it can initially
be approximated as $v(x,0;x_0) \sim u_s(x,0)$ from
eq.~\eqref{soliton_const_back} in which the constant mean field
$B \to \pm c^2 H(x)$.

Throughout this work, we shall use the term {\it soliton} somewhat
loosely to describe a nonlinear localized mode that travels with
velocity directly proportional to its amplitude. The term {\it proper
  soliton} is reserved for modes corresponding to eigenvalues of the
associated Schr\"odinger operator scattering problem.  Overall, we
find that proper solitons only occur for sufficiently large amplitude
initial data. When the initial soliton amplitude is small enough, we
find soliton-like modes called {\it pseudo solitons}, which do {\it
  not} correspond to proper eigenvalues of the scattering problem, yet
can propagate similar to solitons \cite{ALC}.  It turns out that
trapping is always associated with a pseudo soliton and a proper
soliton always tunnels through the RW or DSW.

\begin{figure}
  \centering
  \includegraphics[scale=0.55]{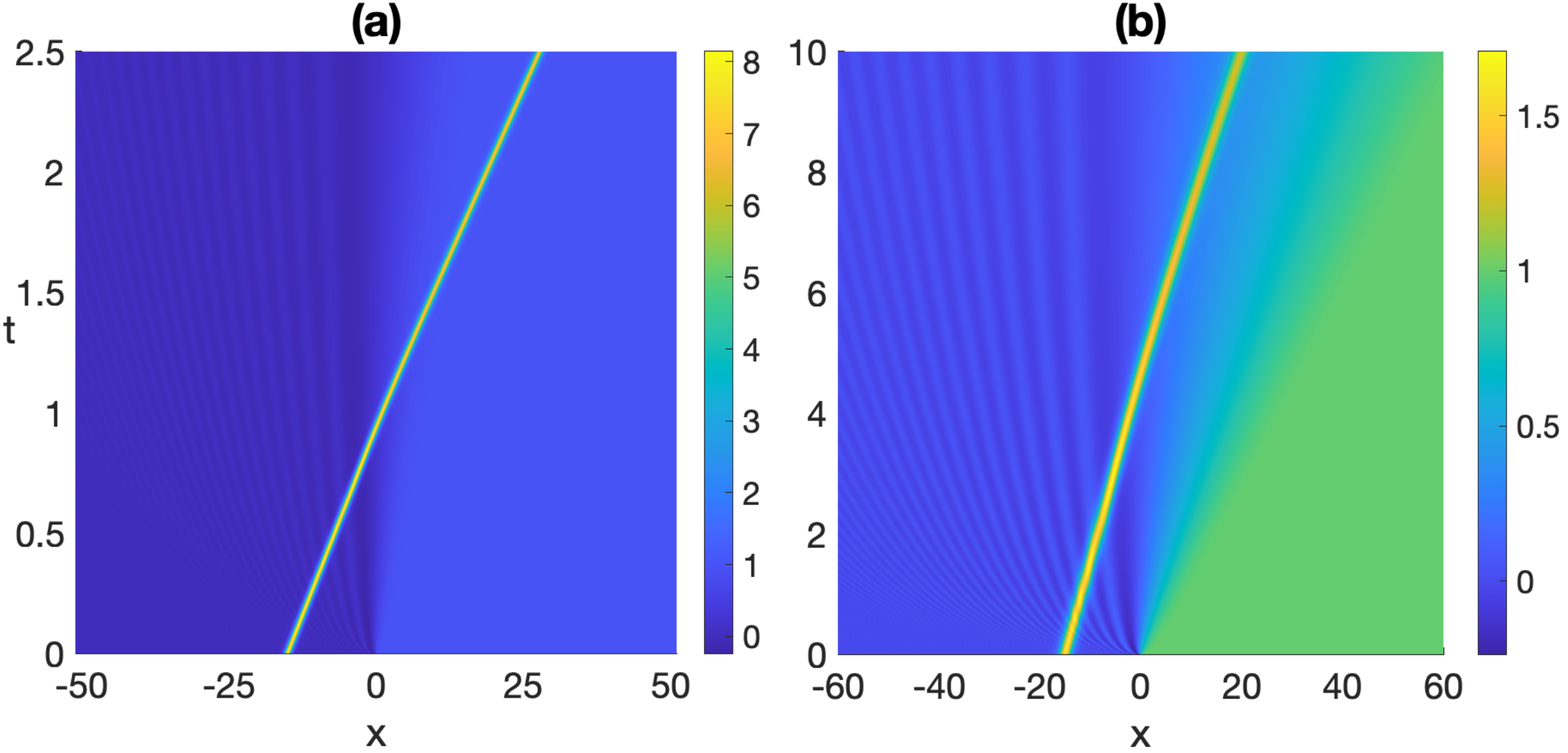}
  \caption{(a) Transmission and (b) trapping of a soliton-RW
    interaction. Initial data is given in (\ref{general_IC_soliton})
    with $c = 1, \varepsilon = 1$ $ x_0 = -15$ and initial amplitude:
    (a) $A_0 = 8$ and (b) $A_0 = 1.62$.} 
  \label{KdV_RW_tunnel_trap}
\end{figure}

The first question we address is whether a soliton will or will not
become trapped.  Broadly speaking, we initially place a soliton mode
to one side of the jump (\ref{general_IC}) and examine whether the
soliton completely propagates through the resulting RW or DSW mean
field in finite time. If the soliton can not travel fast enough to
escape the RW or DSW, we call this a {\it trapped} or an embedded
soliton.  Examples are shown in the right panels of Figure
\ref{fig:interaction_scenarios}. When a soliton completely passes
through one of the step-induced mean fields, this is called a {\it
  transmitted} or tunneling soliton.  Examples are shown in the left
panels of Figure \ref{fig:interaction_scenarios}. In general,
transmitted modes correspond to large amplitude initial states.
Whether or not a soliton becomes trapped depends on its initial data,
e.g.~its initial position and amplitude.

A typical case of soliton trapping and tunneling with a RW is shown in
Fig.~\ref{KdV_RW_tunnel_trap}. The initial condition is
Eq.~(\ref{general_IC}) with step up BCs. A relatively large amplitude
solitary wave placed to the left of the jump will pass through the
ramp region and reach the upper plateau region. On the other hand, a
small amplitude soliton will never exit the ramp region of the RW in
finite time. As shown below, the precise condition for a soliton
tunneling through a RW is that the soliton amplitude be at least twice
the step height, i.e. $A_0 > 2 c^2$.

\begin{figure}
  \centering
  \includegraphics[scale=0.55]{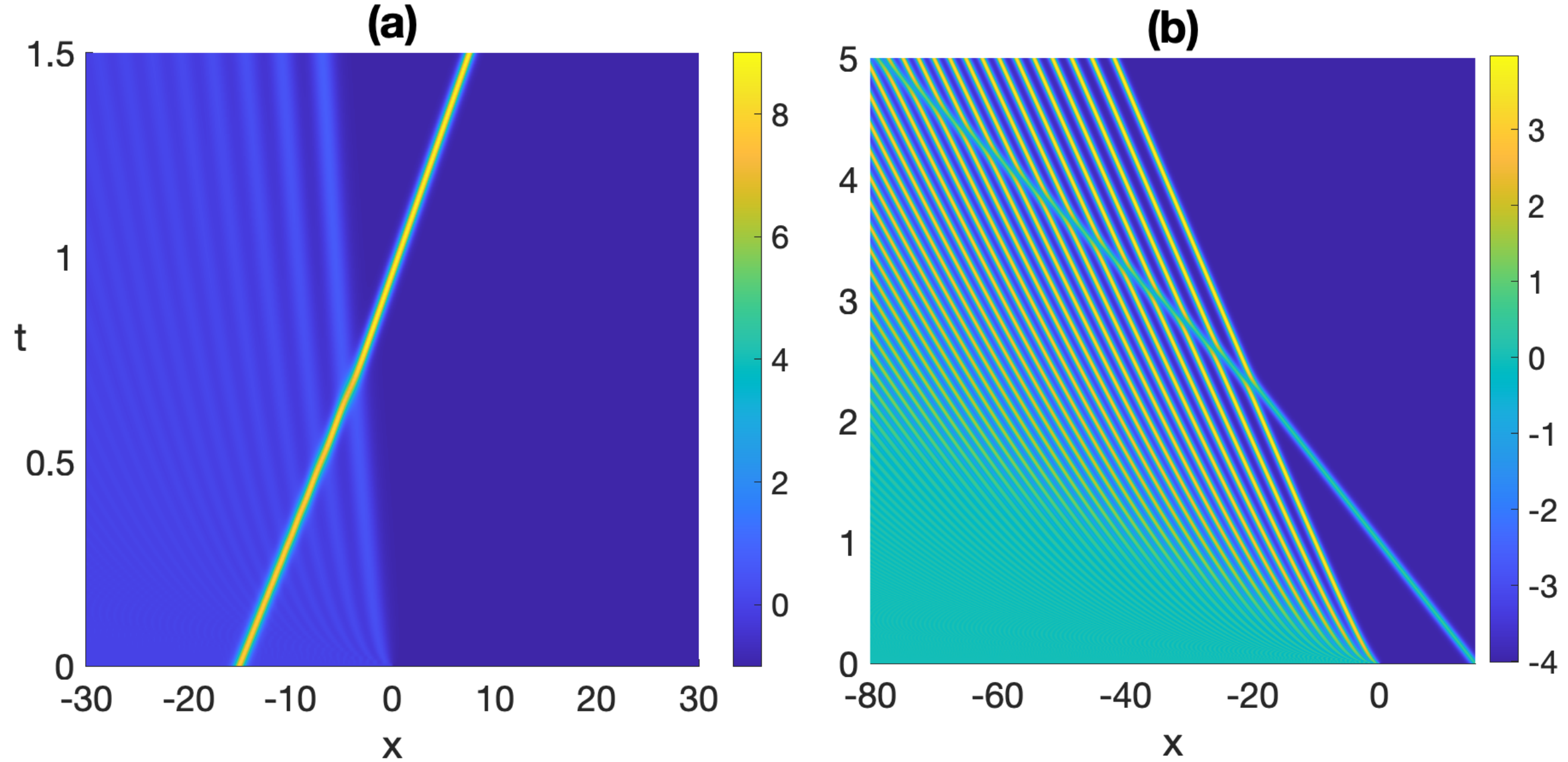}
  \caption{(a) Transmission and (b) trapping of a soliton-DSW. Initial
    data is given in (\ref{general_IC_soliton}) with parameters: (a)
    $ x_0 = -15, A_0 = 8, c = 1, \varepsilon = 1$ and (b)
    $ x_0 = 15, A_0 = 4.5, c = 2, \varepsilon = 1$.}
  \label{KdV_DSW_tunnel_trap}
\end{figure}

Next, typical soliton interactions with a DSW are presented in
Fig.~\ref{KdV_DSW_tunnel_trap}. Here, the initial state is
Eq.~(\ref{general_IC}) with step down BCs. Any soliton initially
placed to the left of the jump will tunnel through the DSW. On the
other hand, a soliton with small amplitude placed to the right of the
jump may become trapped in the DSW region. As shown below, a soliton
in the latter case will become trapped if its initial amplitude is
strictly less than twice the step height, i.e.  $0< A_0 < 2c^2$.

We note that the oscillations in Figure \ref{KdV_RW_tunnel_trap} for
soliton-RW interaction are a result of the sharp, step-like initial
transition that has been approximated by a tanh function for accurate
numerical simulations.  These oscillations give rise to higher order
effects on the soliton-mean field interaction problem hence are not
considered in our asymptotic analysis.  The oscillation amplitude
decays with increasing $t$ and its largest value is inversely
proportional to the initial step width.  In contrast, the large
amplitude DSW oscillations in Fig.~\ref{KdV_DSW_tunnel_trap} persist
as $t$ increases.

The remainder of this work is divided into three parts where the
soliton-mean field interaction is treated using different techniques:
Sec.~\ref{sol_pert_th}, soliton perturbation theory;
Sec.~\ref{sec:solit-disp-hydr}, Whitham modulation theory;
Sec.~\ref{IST_section}, inverse scattering transform.  Within each
section, a comparison between analytical predictions and direct
numerics is given.  The perturbative approach is found to accurately
describe the soliton dynamics in the small dispersive and relatively
large amplitude limits.  The modulation theory approach is shown to
provide an accurate description of the solution dynamics when a single
phase (soliton-RW) or two-phase (soliton-DSW) ansatz is taken, both
interpreted within the context of periodic spectral theory.  The IST
method is found to yield an exact formula for the solutions as well as
a spectral description of the soliton-RW and soliton-DSW
interactions. It is here that the notions of proper and pseudo
solitons arise.  In the case of transmitted solitons, a direct
connection between the asymptotic (soliton perturbation and modulation
theories) and exact (IST) approach is found in the small dispersion
limit.  We conclude in Sec.~\ref{conclude}.

\section{Soliton Perturbation Theory}
\label{sol_pert_th} 

In this section, we present a simple perturbative approach to
approximating the dynamics of a soliton as it passes through a RW or
DSW.  The general perturbation theory for slowly varying solitary
waves was originally introduced in \cite{Grimshaw1979} via
multiple-scale expansions, providing a detailed description of
solitary wave behavior under weak perturbations/slowly varying
coefficients. Here we apply a simpler, formal approach that directly
yields the leading order terms describing the soliton's variations due
to its interaction with the slowly varying mean field. This setting
yields approximate results for the soliton-RW interaction that compare
well with other, more sophisticated approaches described in the
subsequent sections. The interaction with the rapidly oscillating mean
field in a DSW is more complicated, but the direct perturbative
approach provides some useful insights in this case as well.

The method assumes: (a) the solution can be expressed as the linear
combination of a soliton plus a step-induced wave; (b) the soliton
maintains a hyperbolic secant profile (albeit with slowly varying
parameters), and (c) the RW or DSW portion of the solution does {\it
  not} depend on the soliton. The above assumptions are quite
intuitive and, as we show later, this perturbation theory gives good
agreement with the exact solution in the weak dispersion and large
amplitude limits, where the soliton profile is narrow relative to the
variation in the mean field. Below, we give a general framework that
holds for either step up or step down boundary conditions. In the case
of step up boundary conditions, we obtain an explicit analytical
approximation of the soliton dynamics, whereas for step down, we derive
a set of governing differential equations that are solved numerically.

\subsection{General Integral Asymptotics Formulation}
\label{gen_form_asym_sec}

To begin, we express the solution of Eq.~(\ref{kdv}) as the sum
\begin{equation}
  \label{general_perturb_soln}
  u(x,t) = w(x,t) + v(x,t;w(x,t)) \; ,
\end{equation}
where $w(x,t)$ is an approximation of either a RW or DSW and $v(x,t)$
is a solitary mode ansatz with boundary conditions that decay to zero
as $|x| \rightarrow \infty$. Notice that the RW/DSW is assumed to be
independent of the soliton, but not vice versa. Substituting
(\ref{general_perturb_soln}) into (\ref{kdv})
{yields 
\begin{equation}
\label{vw_eqn}
v_t + 6 (wv)_x + 6 v v_x + \varepsilon^2 v_{xxx} =  - F[w(x,t)] ~ ,
\end{equation}
where 
\begin{equation}
\label{RHS_asym}
F[w] \equiv w_t + 3  (w^2)_x + \varepsilon^2 w_{xxx} ~ .
\end{equation}
Note that {$F[w]$ is  zero  
if} $w$ is a solution of (\ref{kdv}).} 

Next, we derive two integral relations {from Eq.~(\ref{vw_eqn})}.
Multiplying (\ref{vw_eqn}) by $v$ and integrating {over $\mathbb{R}$ we obtain
\begin{equation}
\label{energy_evolve}
 \frac{d}{dt} \int_{-\infty}^{\infty} v^2 dx = 6 \int_{- \infty}^{\infty} w (v^2)_x  ~dx  -  \int_{- \infty}^{\infty} 2v F[w] ~dx  ~ ,
\end{equation}
}utilizing the decaying BCs 
of the soliton.
This equation describes how the total momentum of
the soliton changes with time. 
The second integral relation we derive is the time evolution of the soliton center of mass (first moment), given by
\begin{align}
\label{moment_1_norm}
\frac{d}{dt} \left( \frac{\int_{-\infty}^{\infty} x v^2 dx}{\int_{-\infty}^{\infty} v^2 dx} \right) & = \frac{ \frac{d}{dt}  \int_{- \infty}^{\infty} xv^2 dx }{\int_{- \infty}^{\infty} v^2 dx} -  \frac{ (\int_{- \infty}^{\infty} xv^2 dx) \frac{d}{dt} \int_{- \infty}^{\infty} v^2 dx }{(\int_{-\infty}^{\infty} v^2 dx)^2} ~ .
\end{align}
The first term on the right-hand side of the equation above may be simplified by noting 
\begin{align}
\frac{d}{dt}  \int_{- \infty}^{\infty} xv^2 dx  
  \label{moment_1}
  & =   \int_{- \infty}^{\infty}( 12 w v (x v)_x  + 4 v^3 - 3  \varepsilon^2 v_x^2 + x v F[w])  ~dx  ~ ,
\end{align}
where (\ref{vw_eqn}) has been applied.
Substituting Eqs.~(\ref{energy_evolve}) and (\ref{moment_1}) into (\ref{moment_1_norm}) gives 
\begin{align}
\label{locate_eqn}
\frac{d}{dt} \left( \frac{\int_{- \infty}^{\infty} x v^2 dx}{\int_{- \infty}^{\infty} v^2 dx} \right)  =  & \frac{\int_{- \infty}^{\infty} (12 w v (x v)_x  +4 v^3 - 3  \varepsilon^2 v_x^2 + x v F[w])~  dx}{\int_{- \infty}^{\infty} v^2 ~dx} \\ \nonumber
& - \frac{[\int_{- \infty}^{\infty} x v^2 dx] [6 \int_{- \infty}^{\infty} w (v^2)_x  - \int_{- \infty}^{\infty} 2v F[w]  ~dx] }{[\int_{- \infty}^{\infty} v^2 ~dx]^2} ~ .
\end{align}
Equation (\ref{locate_eqn}) can be further simplified by assuming that $v^2(x,t)$ is even-symmetric about the point $x = z(t),$ i.e. $v(x,t) = v(x-z(t)) = v(-(x- z(t)))$, hence Eq.~(\ref{locate_eqn}) becomes 
\begin{align}
\label{rarefaction_first_moment2}
\frac{dz}{dt} = &  \frac{\int_{- \infty}^{\infty}  12 w v (x v)_x  dx}{\int_{- \infty}^{\infty} v^2~ dx}   + \frac{ \int_{- \infty}^{\infty} 4 v^3 - 3  \varepsilon^2 v_x^2   dx}{\int_{- \infty}^{\infty} v^2 ~dx} + \frac{ \int_{- \infty}^{\infty}x v F[w] ~dx}{\int_{- \infty}^{\infty} v^2 ~dx}  \\ \nonumber
& - z(t) \frac{ \left(  \int_{- \infty}^{\infty}6 w (v^2)_x  - 2v F[w]  ~dx \right)}{\int _{- \infty}^{\infty} v^2~ dx} ~ .
\end{align}
 Motivated by soliton solutions on a constant background (\ref{soliton_const_back}), 
we look for soliton modes with the secant hyperbolic form
\begin{equation} 
\label{soliton_define}
v(x,t) = 2 \kappa^2(t) {\rm sech}^2\left( \frac{\kappa(t)}{\varepsilon} \left[ x - z(t) \right] \right) ~ ,
\end{equation}
{whose parameters $\kappa, z$ depend on time. Notice that Eq.~(\ref{soliton_define})  solves (\ref{kdv}) exactly when $w = 0$,} 
{$\kappa(t) = \kappa_0$ and $z(t) = 4 \kappa_0^2 t + x_0$ for $\kappa_0 > 0, x_0 \in \mathbb{R}$.}
Substituting the soliton ansatz (\ref{soliton_define}) into Eqs.~(\ref{energy_evolve}) and (\ref{rarefaction_first_moment2}), we obtain a coupled system of equations that determine $\kappa(t)$ and $z(t)$, amplitude and position, respectively. First, we consider the case when $w$ is a rarefaction wave (step up BC); later, we investigate {the DSW case} 
(step down BC).

\subsection{Soliton Interaction with Rarefaction Wave}
\label{asym_theory_sec}

First, we study a soliton-RW interaction  with initial condition
\begin{equation}
\label{rare_IC}
u(x,0) = c^2 H(x) + v(x,0; x_0) ~ ,
\end{equation}
for a soliton located to the left or right of the origin at time $t = 0$. For $t >0$, a RW forms and connects the left (zero) and right ($+c^2$) boundary conditions. The RW that develops  is approximated by \\
\begin{equation}
\label{define_rare_wave}
w(x,t) = 
\begin{cases}
\vspace{0.1 in}
0 \; , & x \le 0  \\ \vspace{0.15 in}
\frac{x}{6t} \; , & 0 < x \le 6c^2 t   \\ 
c^2 \; , & 6c^2 t  < x 
\end{cases} ~ ,
\end{equation}
which is a continuous, global solution of the KdV equation with
neglected dispersive term --- the approximation is valid away from the
points $x = 0, 6c^2t$, namely $F[w] = 0$ in Eq.~(\ref{RHS_asym}) away
from these points. Thus we will neglect the small,
$\mathcal{O}(\varepsilon^2)$ terms in $\int v F[w]dx$.  Near the edges
of the RW \eqref{define_rare_wave} the KdV dispersive term must be
taken into account and the weak discontinuities at $x=0, 6c^2t$ are
smoothed. Such higher order regularisation can be achieved, for
example, by matched asymptotic methods \cite{Leach2008}. Small
amplitude oscillations at the left edge of the RW decay according to
the typical linear dispersive, long-time estimate
$\mathcal{O}(t^{-1/2})$ and the right edge decays exponentially to
$c^2$.  The linear middle region $0 < x \le 6c^2 t$ of the RW
\eqref{define_rare_wave} is referred to as the {\it ramp} region
below.  Having an explicit and simple approximation formula for the RW
allows us to derive a complete characterization of the soliton-RW
interaction.

\subsubsection{Soliton-Rarefaction Wave Dynamics}
\label{asymptotic_sec}

 The case of the soliton with  amplitude $a_0 \equiv 2 \kappa_0^2$,  
 placed initially to the right of the step ($x_0 > 0$) is rather uninteresting {since it moves away from the ramp; so} we do not consider it in much detail. Through Galilean invariance, the approximate solution is found to be
\begin{equation}
\label{RW_soln_x0_pos}
u(x,t) = w(x,t) + 2 \kappa_0^2~ \text{sech}^2\left( \frac{\kappa_0}{\varepsilon} \left[ x - (  4 \kappa_0^2 + 6c^2) t - x_0 \right]  \right)  ~ ,
\end{equation}
where $\kappa_0 , x_0 > 0$ and the RW is described by (\ref{define_rare_wave}). There is no trapping whatsoever since the soliton is moving faster than the rarefaction ramp.

Now consider a soliton initially placed to the left of the jump
($x_0 < 0$). To obtain a description of the slowly varying soliton
(\ref{soliton_define}) on the ramp portion of the rarefaction wave, we
employ the integral relations (\ref{energy_evolve}) and
{(\ref{rarefaction_first_moment2})}. To begin, rewrite
(\ref{energy_evolve}) as
\begin{equation}
\label{energy_RW1}
 \frac{d}{dt} \int_{- \infty}^{\infty} v^2 dx =  - 6 \int_{- \infty}^{\infty} w_x v^2  dx ~ ,
\end{equation}
and (\ref{rarefaction_first_moment2}) as
\begin{equation}
\label{rarefaction_first_moment}
\frac{dz}{dt} =   \frac{- \int_{-\infty}^{\infty}  12x v (w v)_x   dx}{\int_{- \infty}^{\infty} v^2 dx}   + \frac{ \int_{- \infty}^{\infty} (4 v^3 - 3  \varepsilon^2 v_x^2 ) dx}{\int_{- \infty}^{\infty} v^2 dx} + z(t) \frac{  6\int_{- \infty}^{\infty} w (v^2)_x dx }{\int_{- \infty}^{\infty} v^2 dx} ~ ,
\end{equation}
neglecting $\mathcal{O}(\varepsilon^2)$ terms as explained above.
The soliton in (\ref{soliton_define}) propagates with constant
positive velocity until it encounters the bottom of the rarefaction
ramp at the origin. The time at which the soliton peak reaches the origin is $t = T_1$ where 
\begin{equation}\label{T1}
T_1 = - \frac{x_0}{4\kappa_0^2} > 0 \, .
\end{equation}
 Substituting the 
RW solution \eqref{define_rare_wave} into (\ref{energy_RW1}) yields
\begin{equation}
\label{energy}
 \frac{d}{dt} \int_{-\infty}^\infty v^2 dx  = - \frac{1}{t} \int_0^{6c^2 t} v^2 dx ~ . 
\end{equation}
To get a tractable solution, we extend the domain of integration for
the second integral to $\mathbb{R}.$ Errors in this approximation
occur at the exponentially small {tail} portion of the soliton,
especially when the soliton maximum is near the edges of the ramp
region.
Then the solution of \eqref{energy} is approximately
\begin{equation}
\label{energy_evolve_eqn}
\int_{-\infty}^\infty v^2(x,t) dx =   \frac{T_1}{ t } \int_{-\infty}^\infty v^2(x,T_1) dx ~ .
\end{equation}
Below, we find this approximation improves as
$\varepsilon \rightarrow 0$ or $\kappa \rightarrow \infty$ for fixed
$x_0$ and final time $t$, which corresponds to a narrow soliton width
in (\ref{soliton_define}) and is consistent with the asymptotic
ordering \eqref{eq:scale-separation}.  Substituting the soliton ansatz
(\ref{soliton_define}) into Eq.~(\ref{energy_evolve_eqn}) and
rearranging yields a simple formula
\begin{equation}\label{kappa_define}
\kappa(t) = \kappa_0 \left( \frac{T_1}{t} \right)^{1/3} ~ ,
\end{equation}
where $\kappa_0  > 0$ defines the incoming soliton amplitude $a_0=2 \kappa_0^2$. 
This means that the soliton amplitude on the rarefaction ramp decreases {as $t$ increases:}
\begin{equation}\label{amp_sol_rar}
a(t) =  2 \kappa^2(t) = 2 \kappa_0^2 \left( \frac{T_1}{t} \right)^{2/3} ,
\end{equation}
where $t \ge T_1 > 0$. Next, the soliton position while on the linear ramp is computed from (\ref{rarefaction_first_moment}). Again, using the  soliton ansatz over the linear portion of the RW gives 
\begin{equation}
\label{asym_RW_vel}
\frac{dz}{dt} =  4\kappa^2(t) + \frac{z(t)}{t} \; ,
\end{equation}
whose solution for $z(T_1) = 0$ and $\kappa(t)$ given by
(\ref{kappa_define}) is
\begin{equation}
\label{pos_define_rare_ramp}
z(t) = 6 \kappa_0^2 t \left[ 1 - \left( \frac{T_1}{t} \right)^{2/3} \right]    .
\end{equation}
Equation (\ref{asym_RW_vel}) tells us that the soliton speed is proportional to the amplitude of the soliton {plus the RW solution value at the soliton peak.} 
Note that  
this equation is not valid for $t < T_1.$

The soliton dynamics are now broken down into three regions: ({\bf I}) soliton traveling on zero background, left of the ramp, where $w(x,t) = 0$; ({\bf II}) soliton propagating on the linear ramp, where $w(x,t) = \frac{x}{6t} $; ({\bf III}) soliton propagating {to} the right of the ramp, where $w(x,t) = c^2.$ To match these three regions, continuity of $\kappa(t)$ and $z(t)$ 
is assumed. \\

\noindent
{\bf Region I: $0 \le t < T_1, ~ z(t) < 0 $}

For step-up initial condition (\ref{rare_IC}), the soliton is initially placed to the left of the origin at $x = x_0 < 0 $. The soliton given in (\ref{soliton_define}) travels with constant velocity $4 \kappa_0^2$ a distance $-x_0$ until it reaches the bottom {of} the ramp at $t=T_1$. 
In this time interval the global solution is approximately
\begin{align}
u(x,t) = ~& w(x,t)  +  2 \kappa_0^2 ~{\rm sech}^2\left( \frac{\kappa_0}{\varepsilon} \left[ x -z(t) \right] \right) ~ , \\ \nonumber
 & z(t) = 4 \kappa_0^2 t + x_0 ~ ,
\end{align}
where $w(x,t)$ is given by \eqref{define_rare_wave}. \\

\noindent
{\bf Region II: $ T_1 \le t < T_2, ~ 0 < z(t) < 6c^2 t $}

The soliton enters the ramp region  at time $t = T_1$; the bottom of the ramp is located at the origin. The precise moment the soliton reaches the top of the ramp, if it does, is 
$t = T_2$, to be determined. The results in Eqs.~(\ref{kappa_define}) and (\ref{pos_define_rare_ramp}) are used to describe the soliton in this region. The approximate global solution in this time interval is 
\begin{align}
\label{full_rare_soln_on_ramp}
 u(x,t) = &~ w(x,t) + 2 \kappa_0^2 \left( \frac{T_1}{t} \right)^{2/3} {\rm sech}^2\left( \frac{\kappa_0}{\varepsilon  } \left( \frac{T_1}{t} \right)^{1/3} \left[ x - z(t) \right]\right) \; , \\ \label{z_define}
& z(t)  =  6 \kappa_0^2 t \left[ 1  - \left( \frac{T_1}{t} \right)^{2/3} \right]   ~ .
\end{align}
The soliton tunnels through the RW if it reaches the point $x =  6c^2 t  $ i.e. the top of the ramp. Otherwise, the soliton becomes trapped within the RW. For soliton tunneling to occur there must exist a finite time $t = T_2> T_1$ such that $z(T_2) =  6c^2 T_2$. Using (\ref{z_define}), we find this condition is met at
\begin{equation}
\label{top_time}
T_2 =  T_1 \left( \frac{\kappa_0^2}{\kappa_0^2 - c^2} \right)^{3/2} ~ .
\end{equation}
Notice this formula admits real and positive values for $T_2$ {only}
when the tunneling condition $\kappa_0 > c$ is satisfied.  The
tunneling condition says that only solitons with initial amplitude
$2c^2$ (twice as large as the step height) will make it to the top of
the RW ramp. Otherwise, when $\kappa_0 \le c$ this perturbation theory
predicts that the soliton will never reach the top in finite time and
instead it becomes trapped on the ramp and the approximate solution
continues to be described by (\ref{full_rare_soln_on_ramp}) and
(\ref{z_define}) for all $t > T_1$.  In this case, the soliton
amplitude decays algebraically $\propto t^{-2/3}$.

In the case of tunneling, the soliton peak reaches the top of the ramp at 
\begin{equation}
\label{phase_shift}
z(T_2) = \frac{6c^2 \kappa_0^3 T_1}{( \kappa_0^2 - c^2)^{3/2}} = - \frac{3 c^2 \kappa_0 x_0}{2 (\kappa_0^2 - c^2)^{3/2}}  ~ .
\end{equation}
If the soliton reaches the top of the ramp (from  $2\kappa^2(T_2)= 2 \kappa_0^2(T_1/T_2)^{2/3}$) we see that it has an amplitude parameter of $\kappa(T_2) = (\kappa_0^2 - c^2)^{1/2}$ which means
\begin{equation}
\label{eigenvalue_relation}
\kappa^2(T_2) =  \kappa_+^2 \equiv \kappa_0^2 - c^2 > 0  ,
\end{equation}
or, equivalently, $a(T_2)=2\kappa^2(T_2)>0$, i.e. the existence of $T_2>0$ is equivalent to the condition that the amplitude $a(T_2)$ of the soliton exiting the ramp is positive.
As we shall see,  relation \eqref{eigenvalue_relation} agrees {\it exactly} with the Whitham and IST results in 
Secs.~\ref{sec:solit-rw-inter} and \ref{sec_IST_left_scatter}, respectively.\\

\noindent
{\bf Region III: $T_2 < t , ~ 6c^2 t < z(t)$}

Only the tunneling soliton, with $\kappa_0 > c$, reaches this region at the top of the rarefaction ramp.
At this point, the soliton is traveling on a constant background and will now travel with constant velocity and amplitude. 
The approximate solution here is
\begin{align}
\label{full_soltion_perturb_soln_RW}
 &u(x,t) = w(x,t) +  2 \kappa_+^2 ~{\rm sech}^2\left( \frac{\kappa_+}{\varepsilon } \left[ x - z(t) \right]\right) ~ , \\
& z(t)  =  \left( 6c^2 + 4 \kappa_+^2 \right) \left( t - T_2 \right) + z(T_2) = (6c^2 + 4 \kappa_+^2) t + x_s^+  ~ ,
\end{align}
where $\kappa_+ > 0$ is defined in (\ref{eigenvalue_relation}), and
$z(T_2)$ in (\ref{phase_shift}). The term
$x_s^+ = z(T_2) - \left( 6c^2 + 4 \kappa_0^2 \right) T_2$ can be
expressed as
\begin{equation}
\label{rare_phase_shift}
x_s^+  =\frac{\kappa_0 x_0}{\kappa_+} =  \frac{x_0}{\sqrt{1 - \left(\frac{c}{\kappa_0}\right)^2 }}  ~ ,
\end{equation}
which is less than than $x_0$ {since $x_0<0$}.  The total phase shift  $\Delta  \equiv x_s^+  - x_0$ is a negative quantity indicating delay or retardation due to the rarefaction ramp.


\subsubsection{Comparison with Numerics}

In this section, the soliton perturbation theory results are compared
with numerical simulations of the KdV equation (\ref{kdv}). We also
present here a comparison with relevant IST and Whitham theory
predictions to have an early idea on how the three main analysis
methods in this review compare.

\begin{figure}
  \centering
  \includegraphics[scale=0.6]{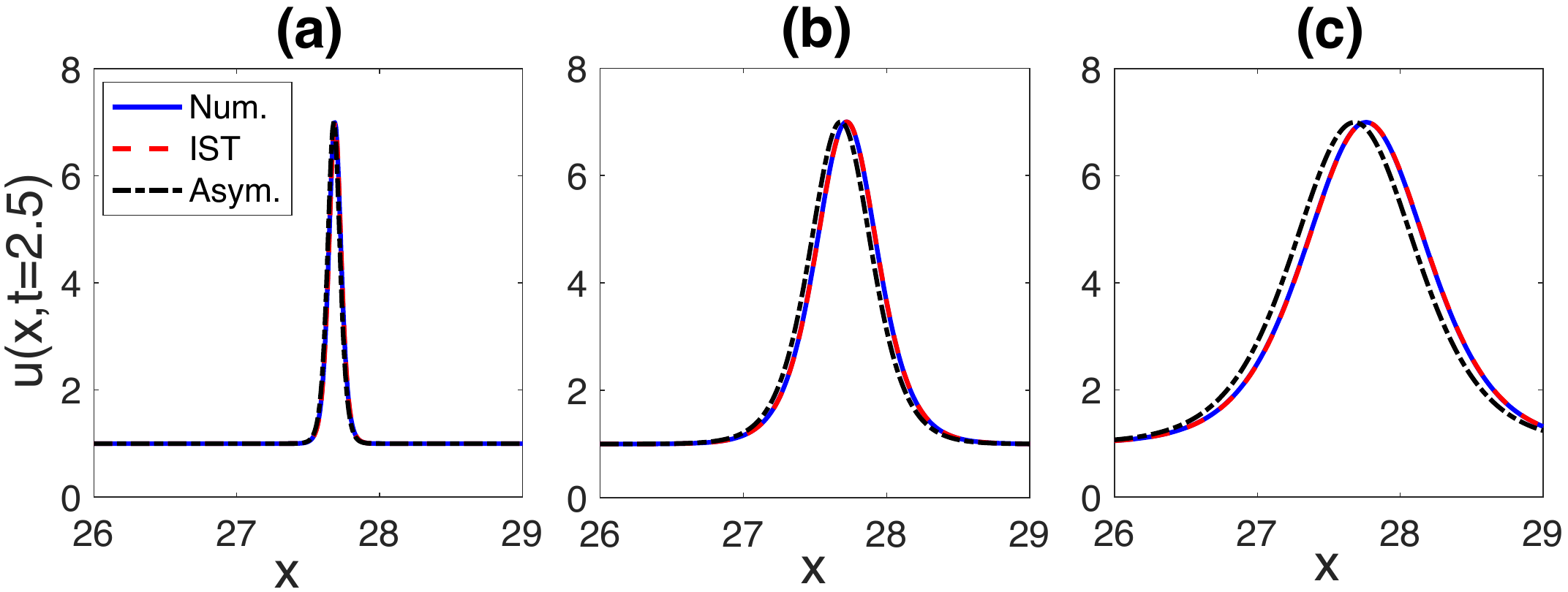}
  \caption{Soliton (post transmission) portion of the solution for (a)
    $\varepsilon = 1/10$, (b) $\varepsilon = 1/2$, and (c)
    $\varepsilon = 1$. Shown are numerical, IST, and asymptotic
    (soliton perturbation and Whitham theories) results. Initial data
    is given in (\ref{general_IC_soliton}) with parameters:
    $ x_0 = -15, \kappa_0 = 2, c = 1$.}
  \label{soliton_transmit_profiles}
\end{figure}

Consider the initial condition in \eqref{rare_IC} with step up BCs,
$\kappa_0 >c$, and $x_0 < 0.$ The depiction of a typical soliton
transmission through a RW is shown in
Figs.~\ref{fig:interaction_scenarios}(a) and
\ref{KdV_RW_tunnel_trap}(a).  The soliton travels with velocity
$4 \kappa_0^2$ until it reaches the bottom of the ramp. As the soliton
travels up the ramp, its amplitude decreases continuously from
$2 \kappa_0^2$ to $2 \kappa_+^2 = 2( \kappa_0^2 - c^2)$. As this
occurs, the velocity of the soliton increases from $4 \kappa_0^2$ to
$4\kappa_+^2 + 6c^2 = 4 \kappa_0^2 + 2 c^2$. To the right of the ramp,
the solution is described by (\ref{full_soltion_perturb_soln_RW}).
Note that the integral asymptotic results in this section are
identical to the results from Whitham modulation theory presented in
Sec.~\ref{sec:solit-rw-inter}.  Several transmitted soliton profiles
are shown in Fig.~\ref{soliton_transmit_profiles} for different values
of $\varepsilon$. The IST (see Sec.~\ref{IST_section}) and numerical
results are found to be indistinguishable, while the asymptotic
approximations improve as $\varepsilon$ decreases. Note that the
soliton perturbation, Whitham, and the small dispersion IST approaches
give the same asymptotic description of the soliton.

\begin{figure} [h]
\centering
\includegraphics[scale=0.5]{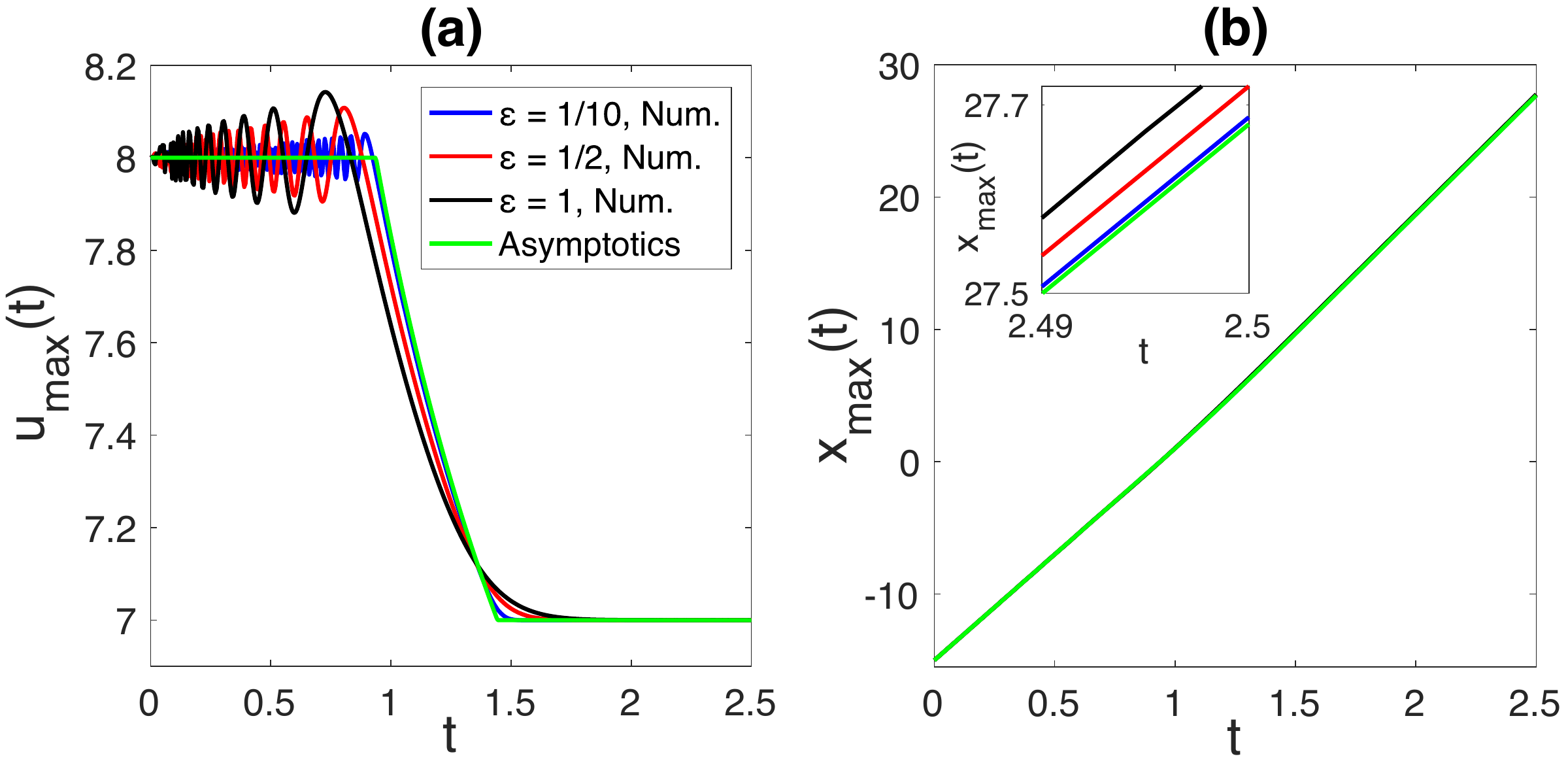}
\caption{Comparison of (a) amplitude and (b) position corresponding to
  a transmitted soliton through RW at different values of
  $\varepsilon$. Initial data is given in (\ref{general_IC_soliton})
  with parameters: $x_0 = -15, \kappa_0 = 2, c = 1$.}
\label{max_soliton_evolve_tunnel}
\end{figure}

While the soliton is traveling up the ramp, the asymptotic
approximation in Eqs.~(\ref{full_rare_soln_on_ramp}) and
(\ref{z_define}), or equivalently (\ref{eq:33}) and (\ref{eq:34}),
analytically describes the soliton motion. Define
$u_{\max}(t) \equiv \max_{x} u(x,t)$, which corresponds to the soliton
peak, located at the point $x_{\max}(t) $. The amplitude and position
of the soliton peak are numerically computed and displayed in
Fig.~\ref{max_soliton_evolve_tunnel} for fixed $\kappa_0$. Note that
the amplitude shown in Fig.~\ref{max_soliton_evolve_tunnel}(a)
consists of the soliton plus the RW. Initially, the amplitude
oscillates due to the small dispersive undulations at the bottom of
the ramp (see Fig.~\ref{KdV_RW_tunnel_trap}, $x< 0$). Next, the
amplitude monotonically decreases until the soliton reaches the top of
the ramp, at which point the amplitude is
$u_{\max}(t) = c^2 + 2 (\kappa_+)^2 = 2 \kappa_0^2 - c^2.$ What is
apparent from Fig.~\ref{max_soliton_evolve_tunnel}(a) is that the
solution behavior is approaching the asymptotic predictions as
$\varepsilon \rightarrow 0$. Furthermore, from
Fig.~\ref{max_soliton_evolve_tunnel}(b) it is seen {that even though}
the difference between soliton position for different values of
$\varepsilon$ is rather small, it too is approaching the asymptotic
prediction in the small dispersion limit.  It is striking how accurate
the asymptotic results are, even when $\varepsilon = \mathcal{O}(1)$.
This is because the asymptotic ordering in \eqref{eq:scale-separation}
is well-maintained when $|x_0| \gg \varepsilon/\kappa_0$.

\begin{figure} [h]
  \centering
  \includegraphics[scale=0.5]{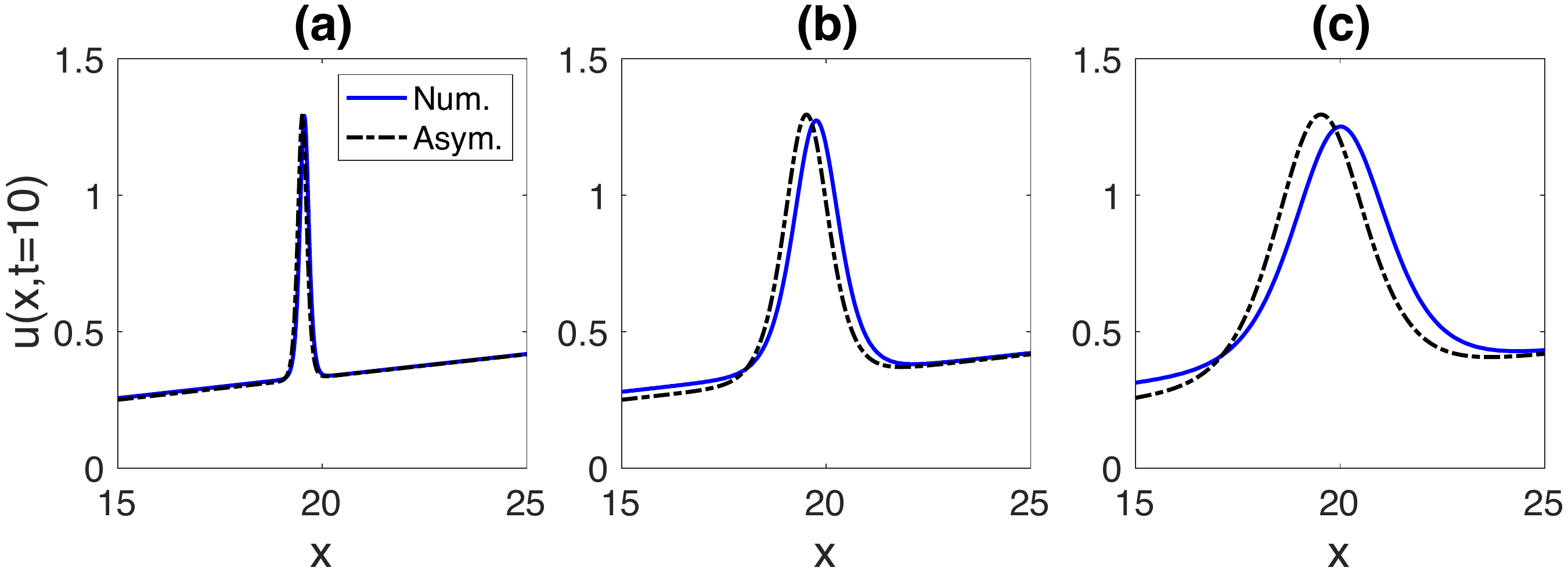}
  \caption{Trapped soliton portion of the solution for (a)
    $\varepsilon = 1/10$, (b) $\varepsilon = 1/2$, and (c)
    $\varepsilon = 1$.  Shown are numerical and asymptotic
    results. Initial data is given in (\ref{general_IC_soliton}) with
    parameters: $ x_0 = -15, \kappa_0 = 0.9, c = 1$.}
  \label{soliton_trap_profiles}
\end{figure}

\begin{figure} [h]
\centering
\includegraphics[scale=0.5]{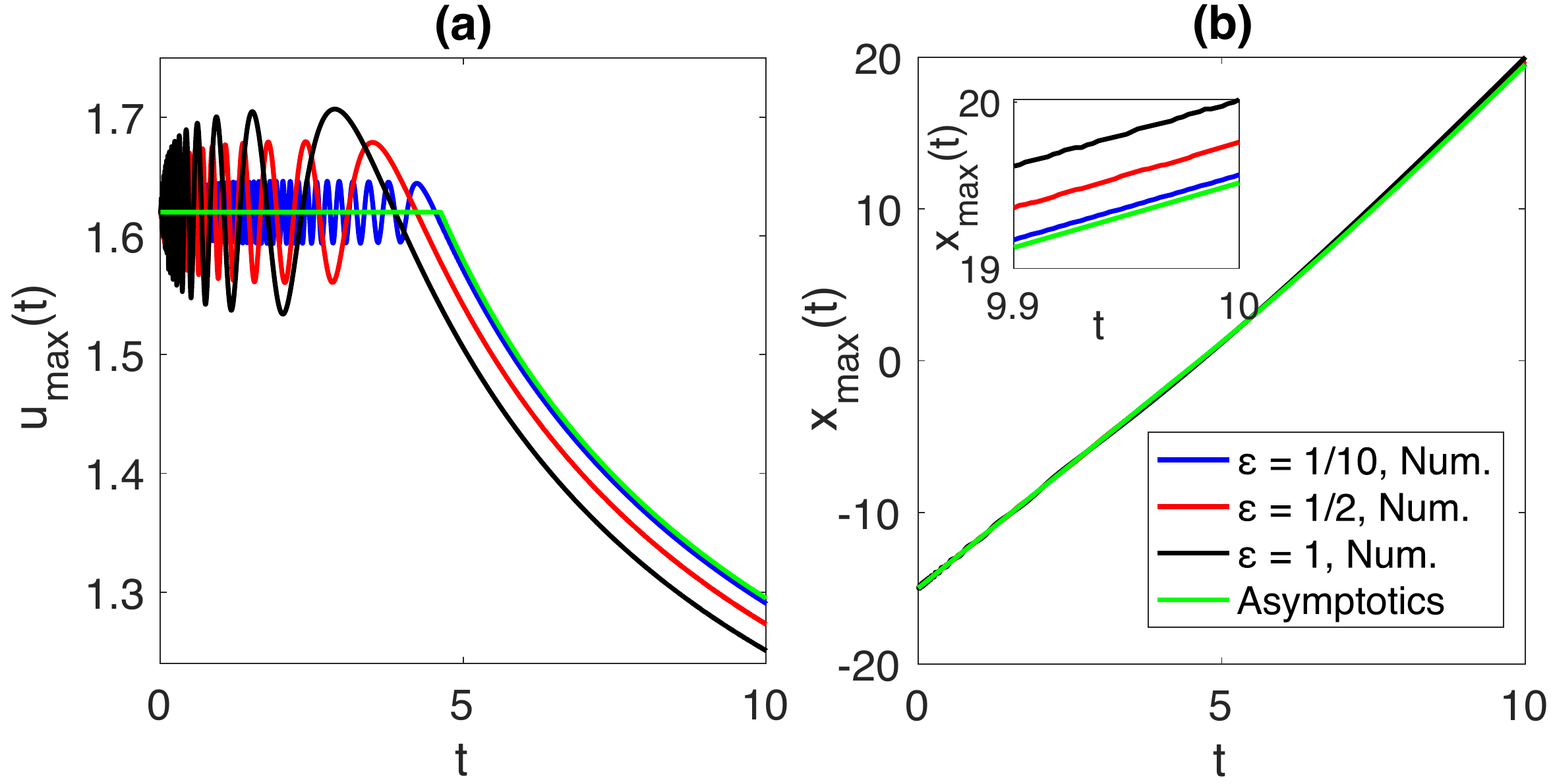}
\caption{Comparison of (a) amplitude and (b) position of a trapped soliton on top of a RW at different values of $\varepsilon$. Initial data is given in (\ref{general_IC_soliton}) with parameters:  $ x_0 = -15, \kappa_0 = 0.9, c = 1$.}
\label{max_soliton_evolve_trap}
\end{figure}

In the case of soliton trapping, soliton perturbation theory continues
to describe the soliton evolution.  Indeed, comparing the asymptotic
predictions with the numerical results in
Fig.~\ref{soliton_trap_profiles} shows excellent agreement as
$\varepsilon \rightarrow 0.$ Recall, the asymptotics predict in the
$t \rightarrow \infty$ limit that the soliton amplitude
$2 \kappa(t)^2$ decreases like $t^{-2/3}$ and the velocity approaches
$6 \kappa_0^2 $, which is slower than the top of the ramp that moves
with velocity $6 c^2$. Comparison between the asymptotic approximation
and the direct numerics is shown in
Fig.~\ref{max_soliton_evolve_trap}. The amplitude and position of the
solution are found to approach the asymptotic, small dispersion
limit. Even when $\varepsilon$ is not so small, the numerically
computed soliton amplitude and position exhibit good agreement with
that of the asymptotic prediction because the asymptotic ordering
\eqref{eq:scale-separation} is maintained.

\subsection{Soliton Interaction with a Dispersive Shock Wave}
\label{sec:sol_int_DSW_cons}

\begin{figure}
  \centering
  \includegraphics[scale=0.5]{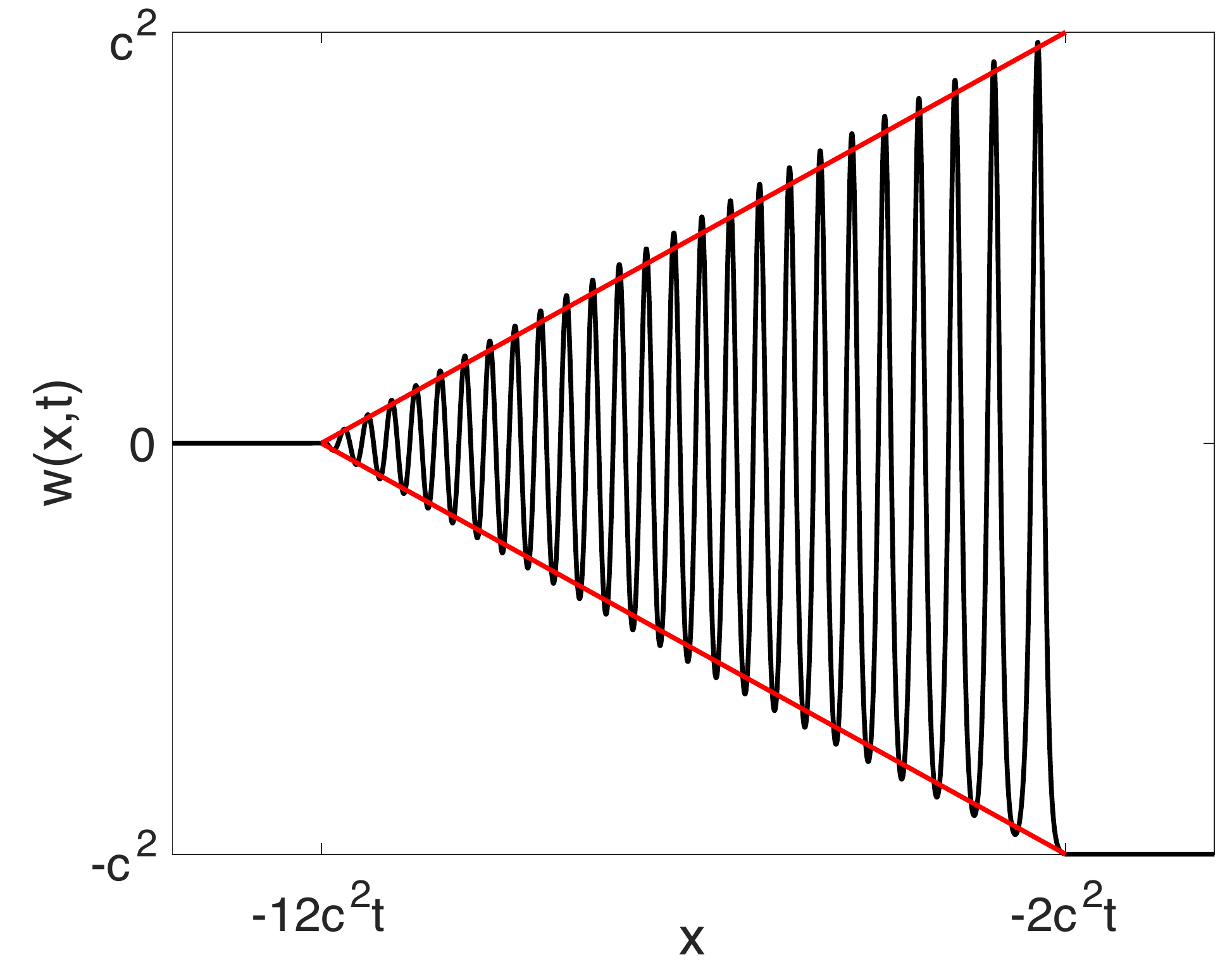}
  \caption{Dispersive mean field $w(x,t)$---a DSW profile obtained from
    the modulation solution \eqref{eq:17}, \eqref{eq:36}. Included are
    linear approximations of the DSW envelope.}
  \label{fig:dsw_sketch}
\end{figure}

Let us now consider how a soliton interacts with a DSW.  The initial
conditions considered are 
\begin{equation}
\label{dsw_IC}
u(x,0) = - c^2 H(x) + v(x,0; x_0) ~ ,
\end{equation}
where the position of the soliton in (\ref{soliton_define}) is taken
to the left ($x_0 < 0$) or right ($x_ 0 > 0$) of the step down. For
$t >0,$ a DSW forms and connects the left (zero) and right ($-c^2$)
boundary conditions. The problem of dispersive regularization of a
compressive initial step was first studied by Gurevich and Pitaevskii
(GP) in \cite{Gurevich} where an asymptotic DSW solution was
constructed using Whitham modulation theory
\cite{whitham_linear_1974}.  Remarkably, it was shown that the DSW
modulation is described by a simple {\it rarefaction wave solution} of
the Whitham equations.  Later, the KdV step problem was studied using
rigorous IST, Riemann-Hilbert methods \cite{Egorova2013}, enabling a
detailed description of the arising oscillations, which all include
the slowly modulated DSW region as the salient, persistent-in-time
feature. Here, we take advantage of the results of GP theory combined
with the above soliton perturbation approach to describe soliton-DSW
interaction.


According to the modulation theory solution \cite{Gurevich} (see
Section~\ref{sec:solit-non-oscill} for details) the evolution of the
step down initial condition can be split into three regions:
\begin{equation}
  \label{DSW_define}
  w(x,t) =
  \begin{cases}
    0 \; , & ~~~~~ x < - 12 c^2 t  \\
    w_D(x,t) \; ,  & -12 c^2 t \le x  < -2c^2 t  \\
    -c^2  \; , & ~~~~ -2c^2t  \le x
  \end{cases} ~ ,
\end{equation}
where the {interval} $ (- 12 c^2 t, -2 c^2 t)$ is the DSW region with
the mean field $w_{D}(x,t)$.

The DSW region is described by a modulated elliptic solution (see
\eqref{eq:17}, \eqref{eq:36}), which is shown in
Fig.~\ref{fig:dsw_sketch}.  Despite a rather complicated form of the
analytical solution, both the upper and lower DSW envelopes can be
reasonably well approximated by the linear functions
$\pm (x + 12 c^2 t)/(10 t)$ for $t>0$.  Using this simple heuristic
approximation, the soliton-DSW interaction can be viewed as the
interaction of a soliton with a descending ramp---the lower DSW
envelope---so that we can apply the same soliton perturbation approach
as in Section~\ref{asym_theory_sec}.  As we shall see, this simple
decreasing ramp approximation of the DSW mean field gives
quantitatively correct predictions for the post-interaction soliton
amplitude in the appropriate asymptotic regime as verified by 
comparison with direct numerical simulations.


As was already noted, an advantage of the soliton perturbation
approach is that it does not rely on integrability and hence can be
applied to various dispersive systems for which an explicit solitary
wave solution is known. This is particularly pertinent to the
soliton-DSW interaction as the two other approaches to this problem
presented later (2-phase Whitham theory and IST) essentially require
the complete integrability of the KdV equation. Indeed, for many
non-integrable systems, the edge speeds and the lead solitary wave
amplitude necessary for the triangular DSW envelope approximation are
available via the DSW fitting method \cite{el_resolution_2005,El2016}.
\begin{figure} [h]
  \centering
  \includegraphics[scale=0.5]{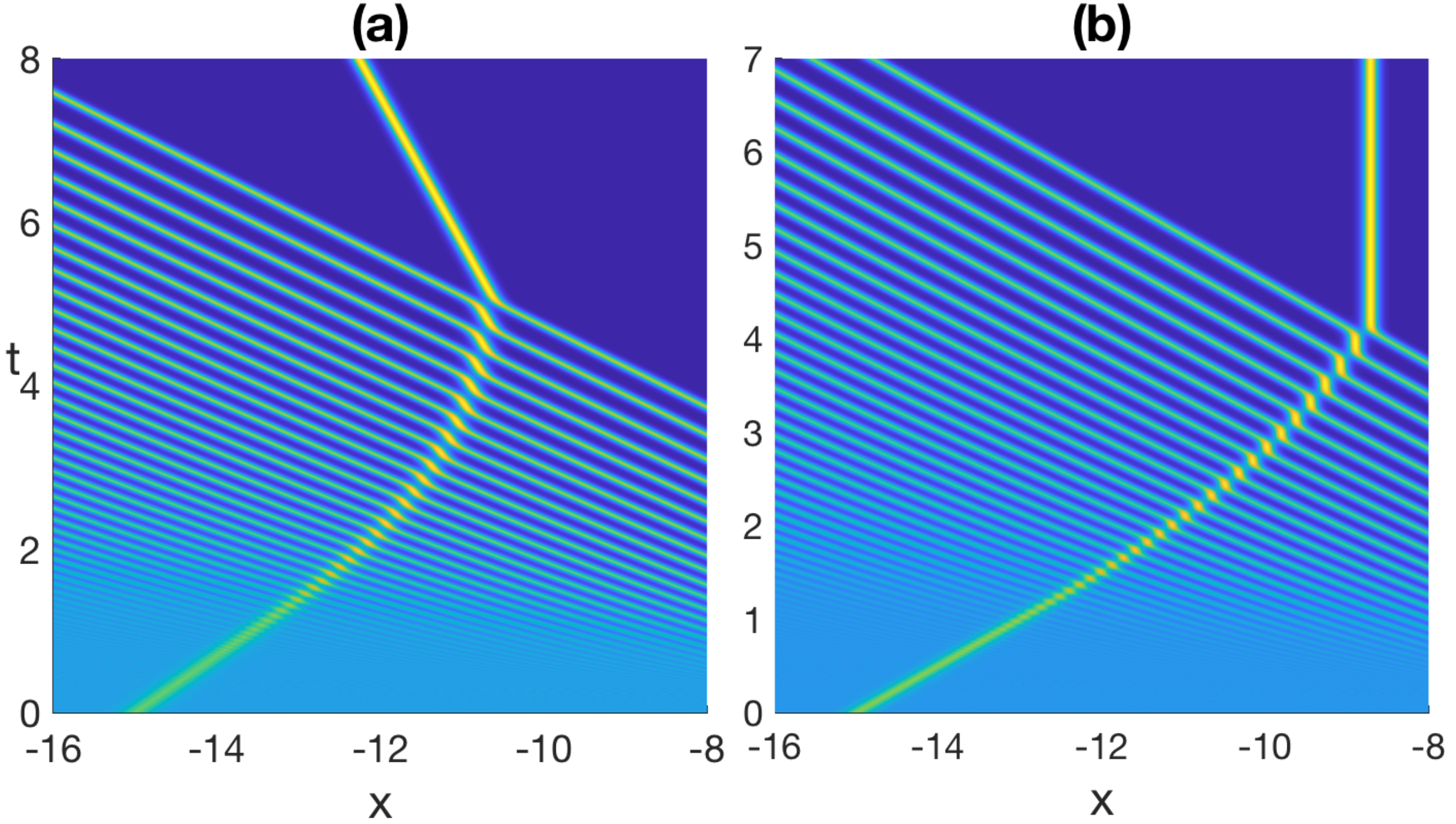}
  \caption{{Direct numerical simulations for (a) $\kappa_0 = 3c/5$
      (negative exit velocity predicted) and (b)
      $\kappa_0 = c/\sqrt{2}$ (zero exit velocity predicted).  The
      initial condition is (\ref{general_IC_soliton}) with parameters:
      $x_0 = -15, \varepsilon = 1/10, c = 1$.}}
  \label{DSW_l2r_small_amp_change_direc}
\end{figure}

\subsubsection{Soliton-DSW Dynamics}

Motivated by the discussion in the previous Section and in analogy with the rarefaction wave case,  we will  
model the soliton-DSW interaction with a solitary wave ansatz \eqref{soliton_define} on a descending ramp described by the linear approximation of the lower DSW envelope. Specifically, we shall consider
\begin{equation}
\label{DSW_define_ramp}
w_D(x,t) = - \frac{ (x + 12 c^2 t) }{10 t}  , ~~ t > 0,
\end{equation}
as the mean field, active in the DSW region, in the calculation of the
integral relations (\ref{energy_evolve}) and
(\ref{rarefaction_first_moment2}). The motivation lies in the
observation that a soliton tunnels through a DSW, see
Fig.~\ref{DSW_l2r_small_amp_change_direc}. There are two alternating
types of soliton evolution here: propagation down a ramp-like region
(approximated by \eqref{DSW_define_ramp}), interspersed by nearly
instantaneous phase shifts forward due to nonlinear interaction with
individual DSW oscillations. The soliton perturbation approach used
here only accounts for the average, slow dynamics of the soliton-ramp
propagation as in the previous case of the interaction with the
rarefaction ramp.  Note, however, a key difference is that
(\ref{DSW_define}), unlike the rarefaction wave
\eqref{define_rare_wave}, is {\it not} an approximate solution of the
KdV equation (\ref{kdv}).

Consider solitons of the form (\ref{soliton_define}) in relations
(\ref{energy_evolve}) and (\ref{rarefaction_first_moment2}).  If we
assume a narrow soliton width, i.e. $\kappa(t) / \varepsilon \gg 1$,
then the the soliton-DSW interaction is local, i.e. the solitary wave
can be effectively considered as a particle that has nontrivial
interaction with only one region of $w$ in Eq.~(\ref{DSW_define}) at a
time. Hence, the integrals in these equations are evaluated over
$\mathbb{R}$ for a narrow soliton located in one of the three
regions. In the DSW region, these relations yield the linear coupled
system
\begin{align}
\label{explicit_sys_kap}
\frac{d \kappa }{d t} &= - \frac{  2 z  + 9 c^2 t}{25 \kappa t^2 } + \frac{ \kappa}{5 t} , \\  
\label{explicit_sys_z}
\frac{d z}{d t} & = - \frac{3 ( z + 12 c^2 t)}{5 t} + 4 \kappa^2  ,
\end{align}
in the small dispersion limit. Note: to obtain
Eq.~(\ref{explicit_sys_z}), we additionally assume that the soliton
has a large amplitude ($\kappa \gg 1$) so that the integrals in
(\ref{rarefaction_first_moment2}) involving $F[w]$, which are
$O(\kappa^{-2})$, are neglected in comparison with the other
integrals. Also note: to see the linearity, multiply
(\ref{explicit_sys_kap}) by $\kappa$. This approach is simple and
yields explicit dynamical equations. As we shall see, it correctly
predicts the amplitude via $\kappa(t)$, even for moderate amplitudes,
despite the formal large amplitude assumption used in the derivation
of \eqref{explicit_sys_kap}, \eqref{explicit_sys_z}.  However, as
highlighted below, it fails to account for the nonlinear phase shift
between the soliton and the local oscillations in the DSW. A
correction to account for the phase shifts is necessary to accurately
describe the soliton position within the DSW.  This would require a
significant modification of the linear ramp approximation of the DSW
mean field, which would compromise the simplicity of the direct
soliton perturbation approach. Instead, in
Sec.~\ref{sec:solit-oscill-mean}, 2-phase Whitham modulation theory,
which is based on the integrable theory of KdV, will be used to
correctly predict the soliton position within the DSW region. An
alternative approach to handle soliton propagation through a DSW
involves soliton gas theory \cite{El2021} and yields, in the end, the
same description \cite{Congy2022} as that described in
Sec.~\ref{sec:solit-oscill-mean}.


Within the soliton perturbation approach, the soliton-DSW interaction
spatial domain can be split into three parts: ({\bf I}) to the left of
the DSW region, where $w = 0$; ({\bf II}) within the DSW region, where
$w = w_D$; and ({\bf III}) to the right of the DSW region, where
$w = -c^2$. The soliton to be considered is {initially} placed either
to the left or right of the jump. When placed to the left of the step,
the soliton {\it always} tunnels through the DSW. If the soliton is
initially placed to the right of the step, then two scenarios are
possible. If the soliton's amplitude is sufficiently small, then it
will become trapped inside the DSW. Conversely, sufficiently large
amplitude solitons never enter the DSW region when initially placed to
the right of the jump. The dynamics within each resion are described
in more detail below.

\medskip
\noindent
{\bf Region I: $  z(t) < -12 c^2 t$}
\medskip

A soliton with amplitude $2 \kappa_0^2$ is initially centered at
$x = x_0 < 0$ and travels with constant positive velocity
$4 \kappa^2_0$ until it's peak reaches the left edge of the DSW. This
edge of the DSW moves with constant negative velocity $-12 c^2 $, so
the time at which the soliton peak meets the DSW is
$4 \kappa_0^2 t + x_0 = -12 c^2 t$, or
\begin{equation}
  \label{time_T1}
  T_1 = - \frac{x_0}{4 \kappa_0^2 + 12 c^2 }  > 0 ~ .
\end{equation}
When the soliton is in Region I, the solution is approximated by
\begin{align}
  u(x,t) = & w(x,t) + 2 \kappa_0^2 ~ {\rm sech}^2 \left(
             \frac{\kappa_0}{\varepsilon} \left[ x - z(t) \right]
             \right) \; , \\ \nonumber 
           & z(t) = 4 \kappa_0^2 t + x_0  ~ ,
\end{align}
where $w(x,t)$ is given in Eq.~(\ref{DSW_define}). We point out that
this region occurs only when a soliton is initially placed to the left
of the jump. A soliton initially placed to the right of the jump will
never be able to tunnel all the way backward through the DSW.

\medskip

\noindent
{\bf Region II: $  -12 c^2 t < z(t) < - 2c^2 t$}

\medskip

In this region, the soliton is interacting with the DSW. Depending on
the initial position and amplitude, the soliton can enter the DSW
region from either direction.  The IST results from
Sec.~\ref{IST_section} tell us that a soliton that starts to the left
of a jump ($x_0 < 0$) moves in the positive direction, hence it will
tunnel through the DSW, which moves to the left in the chosen
reference frame specified by the initial jump $-c^2H(x)$. On the other
hand, for $x_0 >0$, sufficiently small amplitude solitons with
$\kappa_0 < c$ will eventually be trapped within the DSW. If on the
other hand, $\kappa_0 \ge c$, then the soliton's speed is larger than
the DSW's right edge speed, hence it will never interact the DSW
region.

For both cases, the soliton perturbation theory model in
(\ref{explicit_sys_kap})--(\ref{explicit_sys_z}) provides useful
predictions for the dynamics of soliton-DSW interaction.  In the case
of large, transmitting ($x_0 <0$) solitons, the local, soliton-DSW
oscillatory phase shift is small, so the dynamics are effectively
described by the dynamical system \eqref{explicit_sys_kap},
\eqref{explicit_sys_z}.  If, on the other hand, the soliton amplitude
is suitably small, then the soliton interacts with the DSW's
oscillatory region for an extended period of time.  As such, it
experiences many phase shifts forward that can appreciably accumulate.
This model does not incorporate the effect of these phase shifts.  A
more sophisticated theory is needed (see
Sec.~\ref{sec:solit-oscill-mean}). Regardless, for the initial soliton
amplitudes examined, soliton perturbation theory provides a good,
quantitative prediction for the amplitude of the transmitted soliton;
see Fig.~\ref{DSW_tunnel_c1_x0n15b}.


In the case of solitons initialized to the right of the jump
($x_0 > 0$), the interesting case is that of trapping. The asymptotic
model predicts a localized solitary wave that gradually loses all its
amplitude. While numerics appear to suggest a loss of amplitude, the
form of the solitary wave no longer resembles the hyperbolic secant
function in (\ref{soliton_define}). Rather, the solution appears to
take the form of a DSW modulated by a depression or dark envelope type
structure (see Fig.~\ref{fig:interaction_scenarios}(d)).  This is
different from the perturbed soliton form assumed in
(\ref{general_perturb_soln}) and is not captured by the simple soliton
perturbation theory developed here.

\medskip
\noindent
{\bf Region III: $ - 2 c^2 t < z(t) $}
\medskip

When the soliton propagates to the right of the DSW region, it is
effectively on the constant background $-c^2$.  When a soliton
transmits through the DSW and $x_0 < 0$, the approximate solution is
\begin{align}
  \label{DSW_asym_soln_x0_neg}
  x_0 < 0: ~~~~~~~
  &u(x,t) =  w(x,t) + 2 (\kappa_-)^2 ~ {\rm sech}^2 \left(
    \frac{\kappa_-}{\varepsilon} \left[ x - z(t) \right] \right) \; ,
  \\ \nonumber &~~~~~~ z(t) = (4 (\kappa_-)^2 - 6 c^2)  t + x_s^-   ~
                 , 
\end{align}
where $\kappa_-$ is the soliton amplitude parameter upon exiting the
DSW region. In this soliton perturbation theory, we do not have an
analytical formula for $\kappa_-$; instead, we must numerically
compute it by integrating \eqref{explicit_sys_kap},
\eqref{explicit_sys_z}. One can see from
Fig.~\ref{DSW_tunnel_c1_x0n15b} that the model here agrees with direct
numerical simulations of soliton-DSW tunneling.  We note that the
exact, analytical result $\kappa_- = \sqrt{\kappa_0^2 + c^2}$ is
available through the IST and Whitham theory approaches described
later. The phase shift $\Delta x_- = x_s^- - x_0$ is not captured by
the simple soliton perturbation approach employed here; we will
compute it using Whitham theory in Section~\ref{sec:solit-dsw-transm}
and using IST asymptotics.

If, on the other hand, the soliton starts to the right of the
jump, $x_0 > 0$, two outcomes are possible: $\kappa_0 \ge c$ and the
soliton never reaches the DSW; or $\kappa_0 < c$ and the soliton will
eventually embed itself within the DSW. The first case is trivial as
the approximate solution is just a superposition of a well-separated
soliton and a DSW.  The second case is not tractable using the soliton
perturbation approach presented here.

Here, the initial soliton amplitude is expressed in terms of
$\kappa_0$ in order to be consistent with the IST convention
below. The solution in this region is described by
\begin{align}
\label{DSW_asym_soln_x0_pos}
x_0 > 0: ~~~~~~~
&u(x,t) =  w(x,t) + 2 \kappa_0^2 ~ {\rm sech}^2 \left( \frac{\kappa_0}{\varepsilon} \left[ x - z(t) \right] \right) \; , \\ \nonumber
& ~~~~~~ z(t) = (4 \kappa_0^2 - 6 c^2)  t + x_0  .
\end{align}
for short times. This approximation holds for all $t > 0$ when
$\kappa_0 > c$ since the soliton velocity will always be larger than
the rightmost DSW edge velocity, i.e.  $4 \kappa_0^2 - 6c^2 > -
2c^2$. In the trapping case, when
$- 2 c^2 t = (4 \kappa_0^2 - 6 c^2 ) t + x_0$, or equivalently at time
\begin{equation}
\label{trapping_time_DSW}
T_D = - \frac{x_0}{ 4( \kappa_0^2 -  c^2)} ~ ,
\end{equation}
the soliton reaches the DSW's rightmost edge and embeds itself inside
the DSW region (Region II).

\subsubsection{Comparison with Numerics}
\label{DSW_asym_numer_compare}

\begin{figure} [h]
\centering
\includegraphics[scale=0.6]{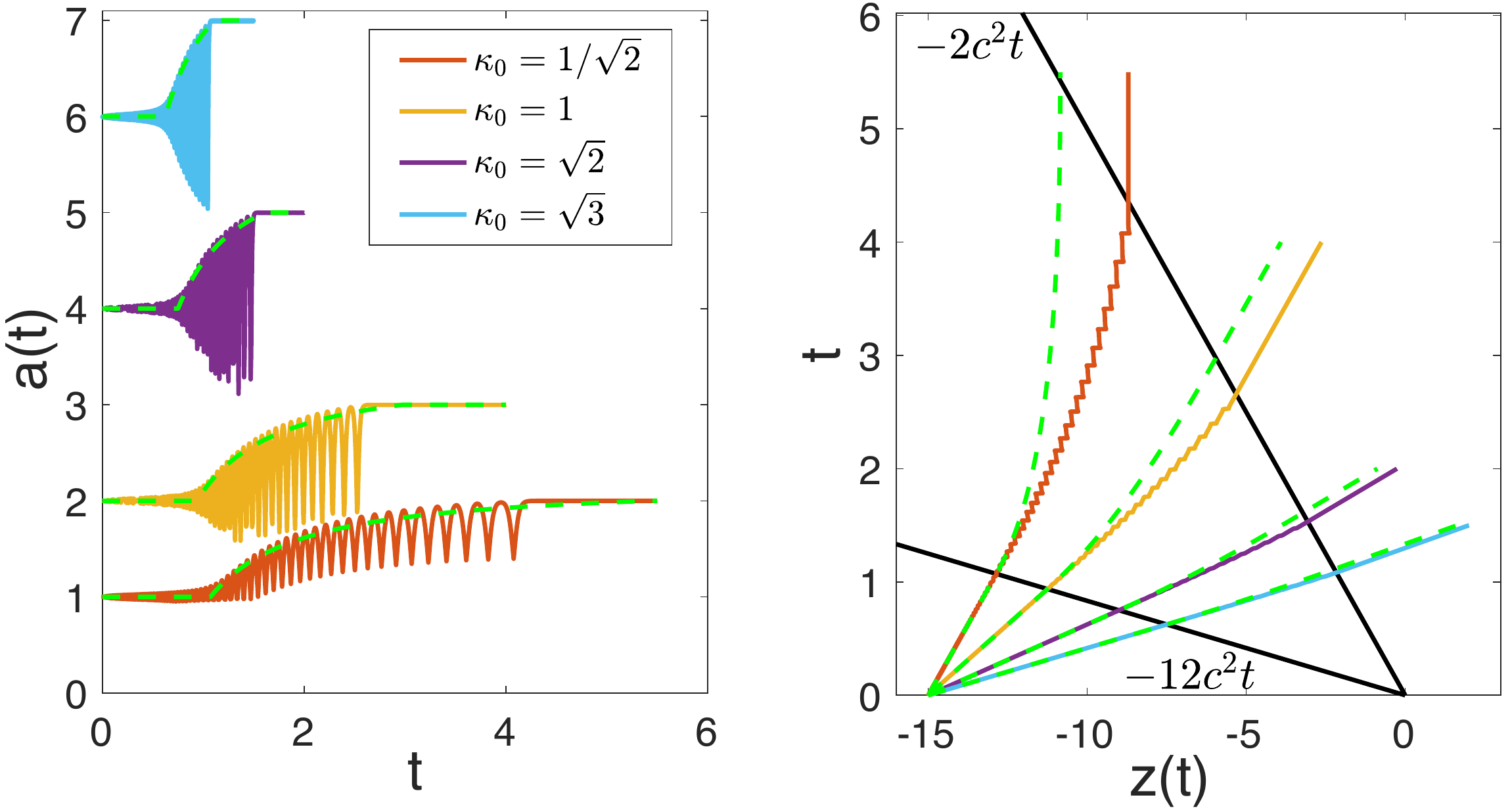}
\caption{Comparison of soliton amplitude $a(t)$ and position $z(t)$
  for direct numerical simulations (solid curves) and asymptotic
  predictions of soliton perturbation theory (dashed curves). The
  initial condition used is (\ref{general_IC_soliton}) with
  parameters: $x_0 = -15, \varepsilon = 1/10, c = 1$.}
\label{DSW_tunnel_c1_x0n15b}
\end{figure}

A summary of typical results produced by this model is shown in
Fig.~\ref{DSW_tunnel_c1_x0n15b}. The predicted soliton amplitude,
$a(t) = 2\kappa(t)^2 + w(z(t),t)$, and position, $z(t)$, are shown as
functions of time and are compared with direct numerical
simulations. The full solution can be reconstructed from the ansatz
(\ref{general_perturb_soln}). Note that the model predicts the
``effective'', average soliton amplitude shown in
Fig.~\ref{DSW_tunnel_c1_x0n15b}, left panel, with dashed line, whereas
the local, instantaneous amplitude, shown with the solid line, is
oscillating due to the interaction of the tunneling soliton with
individual oscillations within the DSW. Such oscillations were
recently described analytically for the modified KdV equation using
rigorous asymptotic theory in \cite{Girotti2022}.

As the soliton passes through the DSW region, its average amplitude
increases, while its velocity decreases. Depending on the incoming
amplitude, a transmitted soliton can have either negative
($\kappa_0 < c/\sqrt{2}$), zero ($\kappa_0 = c/\sqrt{2}$), or positive
($\kappa_0 > c/\sqrt{2}$) exit velocity.

Interestingly, regardless of the initial amplitude, the system
(\ref{explicit_sys_kap})--(\ref{explicit_sys_z}) does an excellent job
of predicting the transmitted soliton amplitude. The errors in the
exit amplitudes are less than $O(10^{-6})$ for all cases
checked. Moreover, for a large incoming soliton, this system is found
to describe its position $z(t)$ well (see
Fig.~\ref{DSW_tunnel_c1_x0n15b}, right panel).  For smaller values of
$\kappa$, the soliton remains inside the DSW region for too long.  As
stated earlier, nonlinear phase shift interactions between the soliton
and the two successive minima of the DSW are required. This effect is
not captured by the soliton perturbation model. However, the total
phase shift is predicted by modulation theory presented in
Section~\ref{sec:solit-oscill-mean}.
 
Finally, the trapped soliton case ($x_0 > 0$) is also out of reach of
the soliton perturbation method as the hyperbolic secant ansatz
(\ref{soliton_define}) does not accurately describe the spatial
profile of the trapped soliton.  A more sophisticated analysis is
needed. Such analysis based on the spectral finite-gap modulation
theory of the soliton-DSW interaction is presented in
Section~\ref{sec:solit-oscill-mean}.

\section{Whitham Modulation Theory}
\label{sec:solit-disp-hydr}

In this section, we study the problem of soliton-RW and soliton-DSW
interaction using multiphase Whitham modulation theory developed for
the KdV equation in \cite{FFM}.  Whitham theory, originally introduced
in \cite{whitham65, whitham_linear_1974} for the single-phase case,
has been successfully applied to the asymptotic description of
dispersive shock waves in the zero dispersion limit
\cite{Gurevich,El2016}.  Multiphase Whitham theory has been utilized
to describe wavebreaking in the Whitham equations \cite{Grava2002},
DSW-DSW interaction \cite{Baldwin2013}, and admits a thermodynamic
limit describing a gas of solitons \cite{El2003}.  Here, we utilize
certain degenerate limits of the 1- and 2-phase Whitham modulation
equations to asymptotically describe the interaction of a soliton with
a RW and a DSW, respectively.

The KdV equation \eqref{kdv} admits a family of quasi-periodic or
multiphase solutions in the form \cite{Nov74, Lax75, ItsMat75,  Date} 
\begin{subequations}
  \label{eq:83}
  \begin{equation}
    \label{eq:14}
    u(x,t) =
    F_N(\theta_1/\varepsilon,\theta_2/\varepsilon,\ldots,\theta_N/\varepsilon) .
  \end{equation}
  The integer $N \in \{0,1,2, \cdots\}$ corresponds to the number of
  nontrivial, independent variables (called phases)
  $\theta_j = k_j x - \omega_j t + \theta_{0j}$, $j = 1, \ldots, N$
  required to describe the solution.  The case $N=0$ corresponds to a
  constant solution.  The $N$-phase solution is normalized so that
  $F_N$ is $2\pi$-periodic in each phase, i.e., rapidly varying for
  $0< \varepsilon \ll 1$.  For example, the case $N = 1$ corresponds to
  the well-known KdV cnoidal traveling wave solution.  The
  $j^{\rm th}$ phase's wavenumber $k_j$, frequency $\omega_j$, and
  phase shift $\theta_{0j}$ at the origin for $j = 1, \ldots, N$ and
  the mean
  $\int_0^{2\pi} \cdots \int_0^{2\pi}
  F_N(\theta_1/\varepsilon,\ldots,\theta_N/\varepsilon)\,\mathrm{d\theta_1}\cdots
  \mathrm{d\theta_N}/(2\pi)^N$ completely determine the $N$-phase
  solution.  The representation of the $N$-phase solution in the form
  \eqref{eq:14} requires use of multidimensional theta functions
  \cite{ItsMat75}.  Although it obscures the dependence of the
  solution on the independent phases $\{\theta_j\}_{j=1}^N$, an
  alternative representation that provides practical advantages is the
  so-called trace formula \cite{Dubr75}
  \begin{equation}
    \label{eq:8}
    u(x,t) = \Lambda - 2 \sum_{j=1}^N \mu_j(x,t), \quad \Lambda \equiv
    \sum_{k=1}^{2N+1} \lambda_k,
  \end{equation}
\end{subequations}
where the constants $\lambda_k$, $k = 1, \ldots, 2N+1$ bound the
functions $\mu_j(x,t) \in [\lambda_{2j-1},\lambda_{2j}]$, $j = 1,2,
\ldots, N$, which satisfy the Dubrovin system \cite{Dubr75}
\begin{equation}
  \label{eq:9}
  \begin{split}
    \varepsilon \partial_x \mu_j &= \frac{2 \sigma_j R(\mu_j)}{\prod_{i
        \ne j}
      (\mu_i - \mu_j)}, \\
    \varepsilon \partial_t \mu_j &= \left ( u + 2 \mu_j \right ) \frac{ 4
      \sigma_j R(\mu_j)}{\prod_{i \ne j}(\mu_i - \mu_j)} , \quad
    R(\lambda) = \sqrt{\prod_{k=1}^{2N+1}(\lambda - \lambda_k)} .
  \end{split}
\end{equation}
The coefficient $\sigma_j = \pm 1$ changes sign so that $\mu_j(x,t)$
oscillates between its maximum $\lambda_{2j}$ and minimum
$\lambda_{2j-1}$. In the form \eqref{eq:8}, the $2N+1$ constants
$\{\lambda_k\}_{k=1}^{2N+1}$ and the $N$ initial conditions
$\{\mu_j(0,0)\}_{j=1}^N$ completely determine the $N$-phase solution.
These $3N+1$ constants are in one-to-one correspondence with the
wavenumbers, frequencies, mean and phase shifts associated with the
solution in the form \eqref{eq:14} \cite{FFM}. 

The above results on multiphase KdV solutions were obtained in the
framework called finite-gap spectral theory.  Within this theory, the
parameters $\lambda_k$ in \eqref{eq:8}, \eqref{eq:9} are the band
edges of the Schr\"odinger operator
\begin{equation}
  \label{eq:26}
  \mathcal{L} = \varepsilon^2 \partial_{xx} + u(x,t)
\end{equation}
with the potential \eqref{eq:83}. Note that $\mathcal{L} + \lambda$ is
also the scattering operator that will be used in (\ref{E:lax 1}). The
real spectral parameter $\lambda \in \mathcal{S}_N$ corresponding to
$L(\mathbb{R})^\infty$ eigenfunctions of (\ref{eq:26}) with $u = F_N$
lies in the union of bands
\begin{equation}
  \label{eq:27}
  \mathcal{S}_N = (-\infty,\lambda_1] \cup [\lambda_2,\lambda_3] \cup
  \cdots \cup [\lambda_{2N},\lambda_{2N+1}] .
\end{equation}
Thus, the band edges $\{\lambda_k\}_{k=1}^{2N+1}$ parametrize the
$N$-phase solution $u$ in \eqref{eq:8} (up to the $N$ initial
conditions $\mu_j(0,0)$, $j = 1, \ldots, N$). Finite-gap theory can be
regarded as a periodic analogue of IST on the real line that will be
used in the next section for the exact description of the soliton-mean
field interaction. In what follows, we take advantage of some results
from KdV finite-gap theory and apply them to the approximate,
modulation description of the soliton-mean interaction.

The $N$-phase KdV-Whitham equations \cite{FFM}
\begin{equation}
  \label{eq:13}
  \frac{\partial \lambda_k}{\partial t} + v_k(\boldsymbol{\lambda})
  \frac{\partial \lambda_k}{\partial x} = 0, \quad k = 1, \ldots,
  2N+1, 
\end{equation}
asymptotically describe modulations of $N$-phase solutions
\eqref{eq:83} to the KdV equation (\ref{kdv}) in the limit
$\varepsilon \to 0$ \cite{LaxLevermore,Venakides,Deift1997,El2001}.  The
characteristic velocities are
\begin{equation}
  \label{eq:29}
  v_k = -6 \Lambda + 12 \lambda_k + 12 \frac{\det
    \mathrm{Q}(\lambda_k)}{\det \mathrm{P}(\lambda_k)},
\end{equation}
where the $N\times N$ matrices $\mathrm{P}(\lambda_k)$ and
$\mathrm{Q}(\lambda_k)$ are defined component-wise by elliptic
integrals, in the $N=1$ case, or hyperelliptic integrals otherwise
\begin{equation}
  \label{eq:35}
  \mathrm{P}(\lambda_k)_{ij} = \int_{\lambda_{2i-1}}^{\lambda_{2i}}
  \frac{(\lambda_k-\mu) 
    \mu^{j-1}}{R(\mu)} \,\mathrm{d}\mu, \quad 
  \mathrm{Q}(\lambda_k)_{ij} = \int_{\lambda_{2i-1}}^{\lambda_{2i}}
  \frac{(\lambda_k-\mu) 
    \mu^{j-1+\delta_{jN}}}{R(\mu)} \,\mathrm{d}\mu .
\end{equation}
Note that $v_k$ depends on the $2N+1$ modulation variables
$\lambda_k$, $k = 1, \ldots, 2N+1$.  We highlight a very special
property of the KdV-Whitham modulation equations \eqref{eq:13}, which
is their diagonal structure. The remarkable fact that the Riemann
invariants of the KdV-Whitham system $\{\lambda_k\}_{k=1}^{2N+1}$ are
precisely the band edges of the finite-gap Schr\"odinger operator
\eqref{eq:26} was discovered in \cite{FFM}.  The modulation variables
and velocities are ordered
\begin{equation}
  \label{eq:28}
  \lambda_1 \le \lambda_2 \le \cdots \le \lambda_{2N+1}, \quad v_1 \le
  v_2 \le \cdots \le v_{2N+1} 
\end{equation}
and the modulation equations \eqref{eq:13} are strictly hyperbolic and
genuinely nonlinear \cite{Levermore}.  The notion of strict
hyperbolicity is defined to be $v_1 < v_2 < \cdots < v_{2N+1}$ if and
only if $\lambda_1 < \lambda_2 < \cdots < \lambda_{2N+1}$.  It has
been shown that for $N$-phase modulations where
$\lambda_i \to \lambda_{j}$ with $j = i \pm 1$, then $v_i \to v_{j}$
and the remaining modulation variables and velocities correspond to
the $N-1$-phase modulation equations \cite{Dubr97, El2001, Grava2002}.
Thus, the modulations of the $N-1$-phase solution described by the
$2N-1$ variables
$\{\lambda_k\}_{k=1}^{2N+1} \setminus \{\lambda_{ij} \equiv \lambda_i
= \lambda_j\}$ completely decouple from and evolve independently of
the degenerate phase described by the single, coalesced modulation
variable $\lambda_{ij}$.  In this section, we identify $\lambda_{ij}$
with a soliton interacting with a RW and a DSW in the degeneration of
$1$- and $2$-phase modulations, respectively.  We determine the
respective characteristic velocities $v_{ij} \equiv v_i=v_j$ and use
the constant Riemann invariant $\lambda_{ij}$ to completely describe
the transmission and trapping of a soliton by a RW and a DSW.

We note that, due to the ordering \eqref{eq:28}, all limits of the
form $\lambda_i \to \lambda_j$ are to be understood as one-sided
limits:  $\lambda_i \to \lambda_{i+1}^-$ or $\lambda_i \to
\lambda_{i-1}^+$.

While the Whitham velocities $v_k$ in \eqref{eq:29} are generally
expressed in terms of complete hyperelliptic integrals, they can be
simplified in certain important cases.  For the soliton-mean field
interaction problem, we will make particular use of the Whitham
equations with $N \in \{0,1,2\}$.

Recall that the term \textit{mean field} is used to describe both RWs
and oscillatory DSWs whose mean vary on a much slower spatial scale
than the soliton width.  As we will see, this is reflected in the
structure of the modulation equations that decouple the descriptions
of the RW and DSW from soliton propagation.  On the other hand, the
soliton dynamics will be significantly altered by the presence of the
mean field.

\subsection{Mean Fields:  0-Phase Modulations}
\label{sec:non-oscillatory-mean}

The $N = 0$ Whitham equation is
\begin{subequations}
  \label{eq:86}
  \begin{equation}
    \label{eq:15}
    \frac{\partial \lambda_1}{\partial t} + v_1 \frac{\partial
      \lambda_1}{\partial x} = 0, \quad v_1(\lambda_1) = 6 \lambda_1,
  \end{equation}
  which is obtained by simply sending $\varepsilon \to 0$ in the KdV
  equation (\ref{kdv}) and identifying $u \to \lambda_1$.  Thus, the
  Hopf equation \eqref{eq:15} describes slowly varying $|u_x/u| \sim
  |u_t/u| = {\cal O}(1)$, dispersionless dynamics.  We identify these
  dynamics with mean fields and write the dependent variable in the
  suggestive form $\overline{u} = \lambda_1$ so that the 0-phase
  equation (\ref{eq:15}) becomes
  \begin{equation}
    \label{eq:41}
    \overline{u}_t + 6 \overline{u} \,  \overline{u}_x = 0,
  \end{equation}
\end{subequations}
and describes the modulations of the constant ($0$-phase) solution \eqref{eq:14} $u=\ub$ of the KdV equation.

\begin{figure}
  \centering
  \includegraphics[scale=0.6]{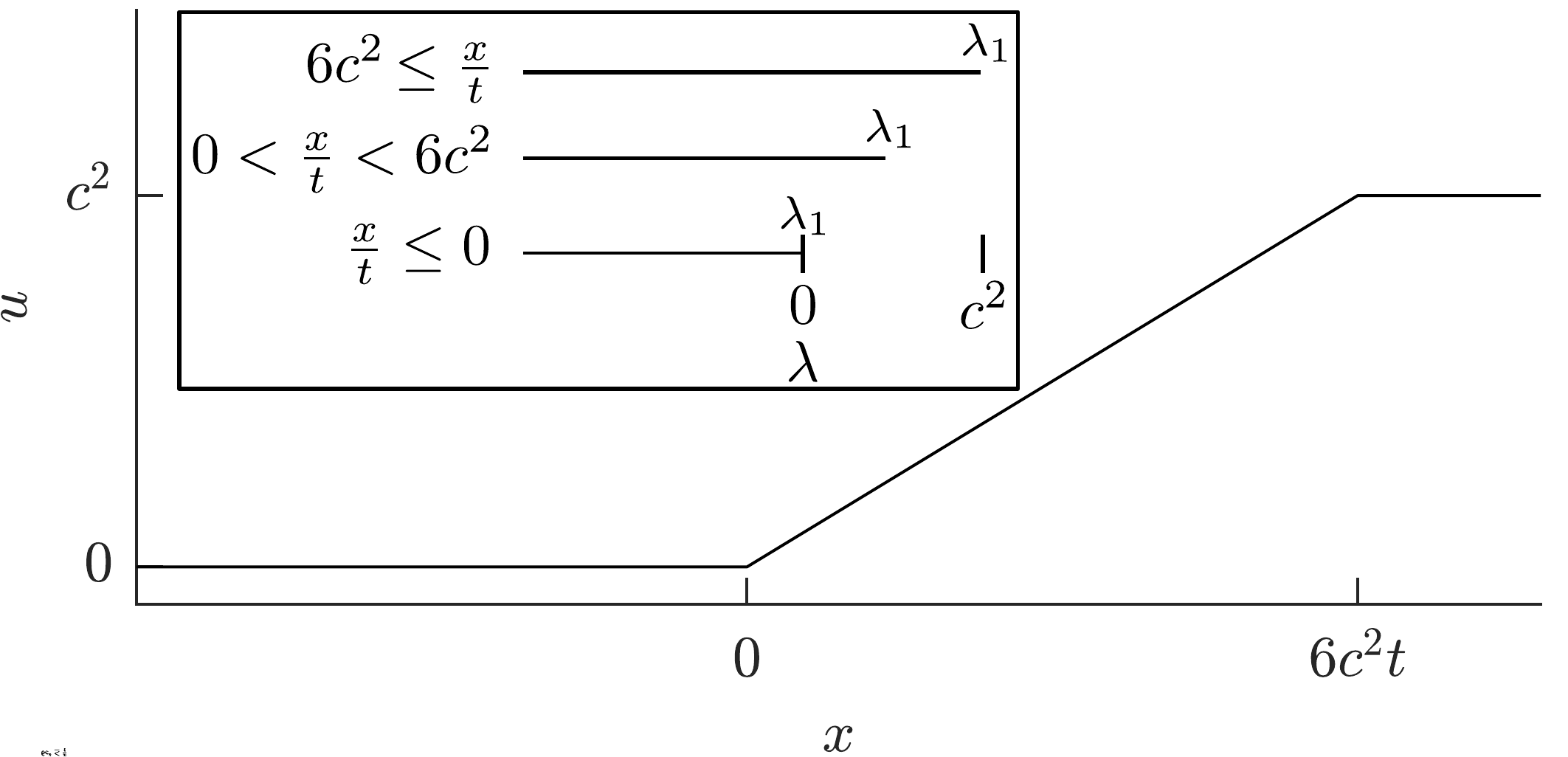}
  \caption{Self-similar RW solution.  The inset represents the
    associated variation of the 0-phase spectrum $\mathcal{S}_0$ in
    the self-similar variable $x/t$.}
  \label{fig:RW_spectrum}
\end{figure}

The solutions of \eqref{eq:86} are simple waves and include the
centered rarefaction wave (\ref{define_rare_wave}), which we reproduce
here for convenience using the notation $\ub(x,t)$ for the mean field,
\begin{equation}
  \label{eq:37}
  \frac{x}{t} = 6\overline{u}(x,t), \quad 0 \le \overline{u}(x,t) \le c^2 ,
  \quad \overline{u}(x,t) =
  \begin{cases}
    0 & x < 0 \\ c^2 & x > 6c^2t
  \end{cases}
  .
\end{equation}
In this case, the spectrum of $\mathcal{L}$ \eqref{eq:26} consists of
the single semi-infinite band $\mathcal{S}_0 = (-\infty,\lambda_1]$
whose band edge is the slowly varying mean $\overline{u}$ depicted in
Fig.~\ref{fig:RW_spectrum}.

Any initial condition for the 0-phase modulation equation
\eqref{eq:86} with a decreasing part exhibits gradient catastrophe in
finite time.  This singularity is regularized by adding another phase
to the modulated solution \eqref{eq:14}.  Thus we are led to a
1-phase, i.e., periodic, modulated wave and the $N=1$ Whitham
equations.

\subsection{Soliton-Mean Field Interaction:  1-Phase Modulations}
\label{sec:solit-non-oscill}

The $N = 1$ Whitham equations describe slow modulations of the
periodic cnoidal wave solution, which is expressed in terms of the
modulation variables
$\boldsymbol{\lambda} = (\lambda_1,\lambda_2,\lambda_3)$ as
\begin{equation}
  \label{eq:17}
  u(\Theta/\varepsilon,x,t) = \lambda_1 - \lambda_2 + \lambda_3 + 2(\lambda_2 -
  \lambda_1) \mathrm{cn}^2 \left ( \frac{K(m)}{\varepsilon \pi} \Theta,
    m \right ) , \quad m = \frac{\lambda_2 - \lambda_1}{\lambda_3 -
      \lambda_1} . 
\end{equation}
The wavenumber and frequency are related to the modulation variables
according to
\begin{equation}
  \label{eq:18}
  \Theta_x = k = \frac{\pi \sqrt{\lambda_3 - \lambda_1}}{K(m)}, \quad
  \Theta_t = - \omega, \quad V \equiv \frac{\omega}{k} = 2 (\lambda_1
  + \lambda_2 + \lambda_3) . 
\end{equation}
The Whitham velocities can be expressed in terms of complete elliptic
integrals
\begin{equation}
  \label{eq:19}
  \begin{split}
    v_1 &= V - 4 (\lambda_2 - \lambda_1 ) \frac{K(m)}{K(m) - E(m)}, \\
    v_2 &= V - 4 (\lambda_2 - \lambda_1 ) \frac{(1-m) K(m)}{E(m) -
      (1-m)K(m)}, \\
    v_3 &= V + 4 (\lambda_3 - \lambda_2 ) \frac{K(m)}{E(m)} .
  \end{split}
\end{equation}
These expressions were originally obtained in the foundational Whitham
paper \cite{whitham65}. The mean and amplitude of the cnoidal solution
(\ref{eq:17}) are
\begin{equation}
  \label{eq:20}
  \overline{u} \equiv \frac{1}{2\pi} \int_0^{2\pi} u(\Theta/\varepsilon,x,t)\,
  \mathrm{d}(\Theta/\varepsilon) = \lambda_1 + \lambda_2 - \lambda_3 + 2
  (\lambda_3 - \lambda_1) \frac{E(m)}{K(m)} ,
\end{equation}
and
\begin{equation}
  \label{eq:21}
  a = 2(\lambda_2 - \lambda_1),
\end{equation}
respectively.

\begin{figure}
  \centering
  \includegraphics[scale=0.6]{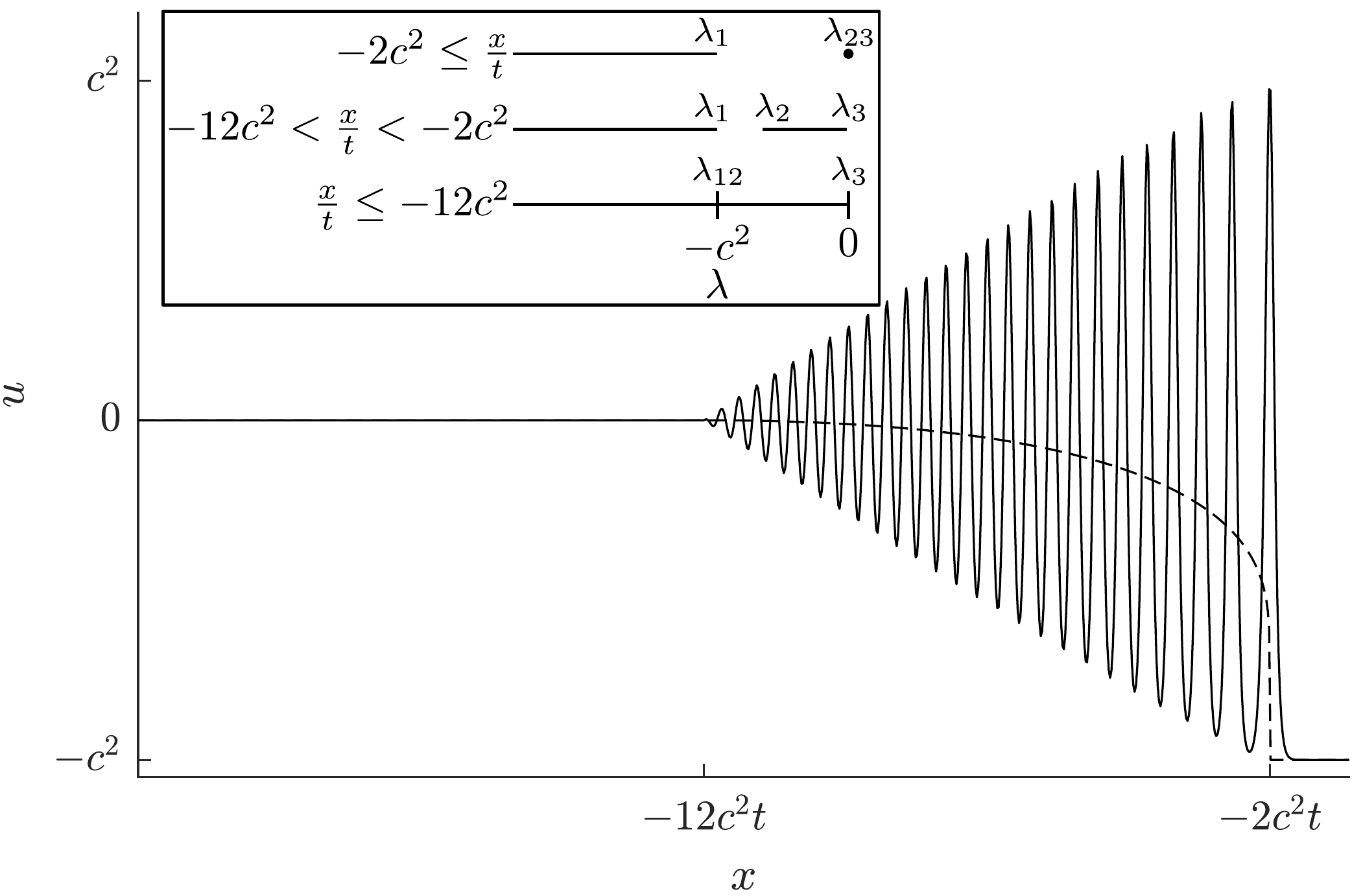}
  \caption{Reconstruction of the KdV DSW (solid) from the self-similar
    GP modulation solution and its mean $\overline{u}$ (dashed).  The
    inset represents the associated variation of the 1-phase spectrum
    $\mathcal{S}_1$.}
  \label{fig:DSW_spectrum_evolution}
\end{figure}

The 1-phase Whitham equations serve two purposes for us.  First, the
centered rarefaction solution with constant $\lambda_1 = -c^2$,
$\lambda_3 = 0$, and $\lambda_2$ implicitly defined by
\begin{equation}
  \label{eq:36}
  \frac{x}{t} = v_2(-c^2,\lambda_2(x,t),0), \quad
  -c^2 \le \lambda_2(x,t) \le 0 , \quad \lambda_2(x,t) =
  \begin{cases}
    -c^2 & x < v_{12}(-c^2,-c^2,0) t \\ 0 & x > v_{23}(-c^2,0,0) t
  \end{cases},
\end{equation}
is the celebrated Gurevich-Pitaevskii modulation solution for a DSW
\cite{Gurevich}.  
The DSW, reconstructed from the modulation solution \eqref{eq:36} in
Fig.~\ref{fig:DSW_spectrum_evolution}, is an example of a dispersive
mean field exhibiting rapid oscillations.  The spatiotemporal evolution
of the DSW's 1-phase spectrum $\mathcal{S}_1$ according to the GP
solution \eqref{eq:36} is shown in the inset of
Fig.~\ref{fig:DSW_spectrum_evolution}.

The second use of the 1-phase Whitham equations is to describe the
interaction of a soliton and a mean field.  For both purposes, we need
to understand the limiting behavior of the 1-phase equations and the
associated spectrum $\mathcal{S}_1$ when $\lambda_2 \to \lambda_3 =
\lambda_{23}$.  For the DSW, we also need to consider the limit
$\lambda_2 \to \lambda_1 = \lambda_{12}$, which we consider first.

\subsubsection{Harmonic Limit:  Merged 1-Phase Spectrum}
\label{sec:harm-limit:-merg}

When $\lambda_2 \to \lambda_1$, the cnoidal wave \eqref{eq:17} is a
vanishing harmonic wave
\begin{equation}
  \label{eq:38}
  u(\Theta/\varepsilon,x,t) \sim \lambda_3 + (\lambda_2 - \lambda_1)
  \cos \left ( \Theta/\varepsilon \right ), \quad \lambda_2 \to \lambda_1,
\end{equation}
with wavenumber
\begin{equation}
  \label{eq:39}
  k = 2\sqrt{\lambda_3 - \lambda_{12}} .
\end{equation}
Passing to the limit in the Whitham velocities (\ref{eq:19}), we have
$v_1 \to v_2 = v_{12}$ where
\begin{equation}
  \label{eq:40}
  v_{12} = 6(\lambda_3 - 2 \lambda_1), \quad
  \lim_{\lambda_2\to\lambda_1} v_3 = 6 \lambda_3 .
\end{equation}
The limit $\lambda_2 \to \lambda_1$ results in two modulation
equations describing a linear wave on a slowly varying mean
(\ref{eq:38}). To see this, we substitute $K = k/\varepsilon$ and
$\Omega = \omega/\varepsilon$ from (\ref{eq:18}) and \eqref{eq:39},
$\overline{u} \to \lambda_3$ from (\ref{eq:20}) and recognize one
characteristic velocity as the group velocity
$v_{12} = \Omega'(K) = 6\overline{u} - 3 \varepsilon^2 K^2$ of linear
traveling wave solutions $\sim e^{i (Kx - \Omega t)}$ to the KdV
equation (\ref{kdv}).  The remaining velocity in (\ref{eq:40}) is
$v_3 \to 6 \overline{u}$, where $\lambda_3 = \overline{u}$ coincides
with the mean field, 0-phase modulation \eqref{eq:86}.  The mean field
completely decouples from the linear wave modulation, which is
expressed in the spectrum (\ref{eq:27}) as a degenerate point embedded
within the semi-infinite band
\begin{equation}
  \label{eq:42}
  {\cal S}_1^{(12)} \equiv  \lim_{\lambda_2
    \to \lambda_1} {\cal 
    S}_1 = (-\infty,\lambda_{12}] \cup [\lambda_{12},\lambda_3] =
  (-\infty,\lambda_3] .
\end{equation}
The superscript $^{(12)}$ notation represents the merger of the
semi-infinite band $(-\infty,\lambda_1]$ with the finite band
$[\lambda_2,\lambda_3]$, so that ${\cal S}_1^{(12)}$ is the
\textit{merged} 1-phase spectrum.  This merging process is depicted in
Fig.~\ref{fig:merged_collapsed_1phase_spectrum}.  Modulations of the
spectrum $\mathcal{S}_1^{(12)}$ have been used to describe the
transmission and trapping of a linear wavepacket with a RW in
\cite{congy}.

The DSW spectral evolution shown in
Fig.~\ref{fig:DSW_spectrum_evolution} also exhibits spectral merger at
the DSW's harmonic edge when the finite band expands to merge with the
semi-infinite band for $x \to -12 c^2 t$.

\subsubsection{Soliton Limit:  Collapsed 1-Phase Spectrum}
\label{sec:solit-limit:-coll}

When $\lambda_2 \to \lambda_3$, the cnoidal wave \eqref{eq:17} limits
to
\begin{equation}
  \label{eq:24}
  u(\xi/\varepsilon,x,t) = \lambda_1 + 2(\lambda_{23} - \lambda_1)
  \mathrm{sech}^2 \left ( 
    \sqrt{\lambda_{23} - \lambda_1} \, \xi/\varepsilon \right ),
  \quad \xi =  x-(2 \lambda_1 + 4 \lambda_{23})t ,
\end{equation}
which corresponds to the soliton solution of the KdV equation
(\ref{kdv}).  Note that this limit is singular ($k \to 0$) so that the
phase $\Theta$ in (\ref{eq:17}) is undefined and we identify
$\xi/\varepsilon$ as the fast variable according to
\begin{equation}
  \label{eq:25}
  \lim_{\lambda_2 \to \lambda_3} \frac{K(m)}{\varepsilon \pi} \Theta =
  \frac{\sqrt{\lambda_{23} - \lambda_1} \, \xi}{\varepsilon} .
\end{equation}
The $N = 1$ Whitham velocities (\ref{eq:19}) limit to
\begin{align}
  \label{eq:16}
  \lim_{\lambda_2 \to \lambda_3} v_1 &= 6 \lambda_1, \quad
  v_{23} = 2 \lambda_1 + 4 \lambda_{23} ,
\end{align}
so that the 1-phase Whitham modulation equations \eqref{eq:13} become
\begin{subequations}
  \label{eq:51}
  \begin{align}
    \label{eq:49}
    \frac{\partial \lambda_1}{\partial t} + 6 \lambda_1 \frac{\partial
      \lambda_1}{\partial x} &= 0, \\
    \label{eq:50}
    \frac{\partial \lambda_{23}}{\partial t} + (2 \lambda_1 + 4 \lambda_{23})
    \frac{\partial \lambda_{23}}{\partial x} &= 0 .
  \end{align}
\end{subequations}
Using \eqref{eq:20} and \eqref{eq:21}, we can express the Riemann
invariants in this limit in terms of the wave mean and amplitude as
\begin{equation}
  \label{eq:23}
  \lambda_1 = \overline{u}, \quad \lambda_{23} = \frac{a}{2} +
  \overline{u} .
\end{equation}
\begin{figure}
  \centering
  \includegraphics[scale=0.7]{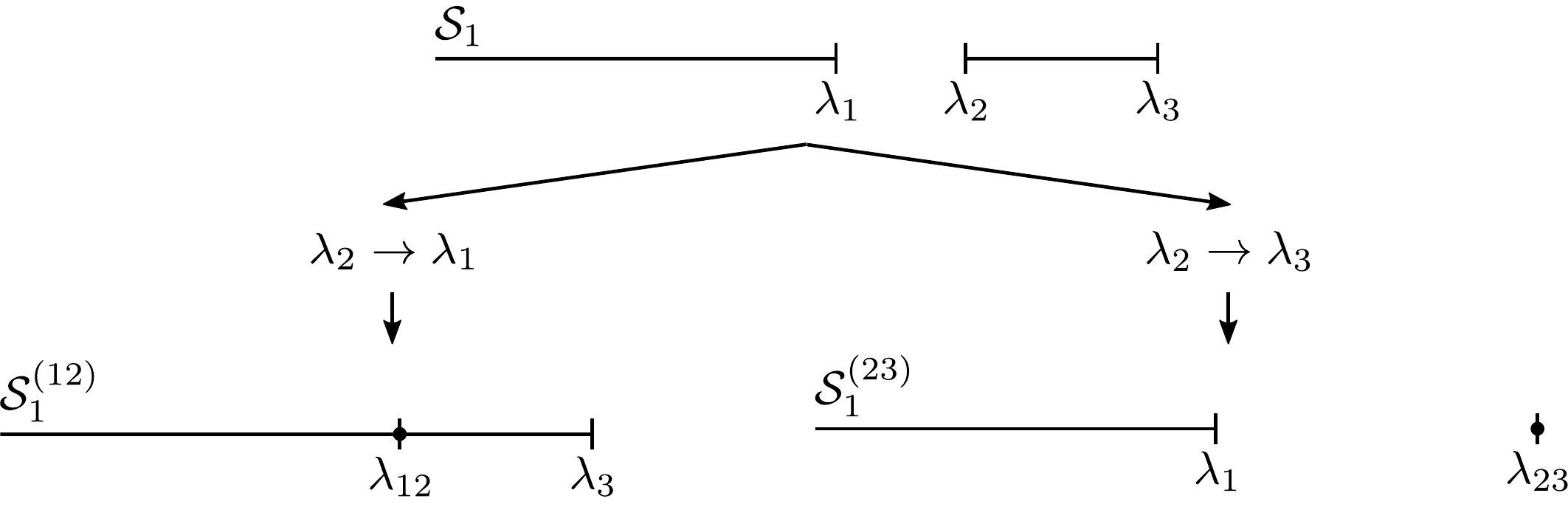}
  \caption{Degeneration pathways for the 1-phase spectrum
    $\mathcal{S}_1$.  When $\lambda_2 \to \lambda_1$, the two bands
    merge into a single, semi-infinite band, the merged spectrum
    $\mathcal{S}_1^{(12)}$.  When $\lambda_2 \to \lambda_3$, the
    finite band collapses into a single point, the collapsed spectrum
    $\mathcal{S}_1^{(23)}$.}
  \label{fig:merged_collapsed_1phase_spectrum}
\end{figure}
Consequently, we can identify the 1-phase modulation equations
\eqref{eq:51} in the limit $\lambda_2 \to \lambda_3$ with the 0-phase
mean field modulation equation \eqref{eq:49} for $\lambda_1$
(cf.~eq.~\eqref{eq:86}) and the equation \eqref{eq:50} for
$\lambda_{23}$ represents modulation of the soliton amplitude.  The
double Whitham velocity $v_{23}$ is the soliton velocity
\begin{equation}
  \label{eq:7}
  v_{23} = s(a,\overline{u}) = 6\overline{u} + 2 a,
\end{equation}
and the soliton trajectory $x_s(t)$ is the characteristic
\begin{equation}
  \label{eq:6}
  \frac{\mathrm{d}x_s}{\mathrm{d} t} = s, \quad x_s(0) = x_0 .
\end{equation}
These soliton-mean field modulations equations were also obtained
using multiple scale perturbation theory in \cite{Grimshaw1979}.

The mean field modulation fully decouples from the soliton amplitude
modulation, which manifests in the degenerate spectrum
\begin{equation}
  \label{eq:46}
  {\cal S}_1^{(23)} \equiv \lim_{\lambda_2 \to \lambda_3} {\cal
    S}_1 = (-\infty,\lambda_1] \cup \{\lambda_{23}\}
\end{equation}
consisting of the semi-infinite band of the 0-phase spectrum and the
point spectrum or eigenvalue $\lambda_{23}$.  In fact, the spectrum
${\cal S}_1^{(23)}$ corresponds to the classical 1-soliton solution on
a constant background in the framework of IST, which we will describe
in detail in Sec.~\ref{IST_section}.  Here, the superscript $^{(23)}$
corresponds to the collapse of the second band $[\lambda_2,\lambda_3]$
in the 1-phase spectrum to a point and we refer to ${\cal S}_1^{(23)}$
as the \textit{collapsed} 1-phase spectrum.  An example of this is
shown in Fig.~\ref{fig:merged_collapsed_1phase_spectrum}.

The DSW spectral evolution shown in
Fig.~\ref{fig:DSW_spectrum_evolution} also exhibits
spectral collapse when the finite band collapses to a point at $x \ge -2
c^2 t$.

\subsubsection{Soliton-RW Interaction}
\label{sec:solit-rw-inter}

We are now in a position to approximate the interaction of a soliton
and a slowly varying mean field by modulations of the collapsed 1-phase
spectrum ${\cal S}_1^{(23)}$ (\ref{eq:46}) whose band edge $\lambda_1$
and point spectrum $\lambda_{23}$ evolve according to the degenerate
1-phase Whitham equations \eqref{eq:51}.  We focus on the soliton-RW
interaction resulting from step-up initial data (\ref{general_IC})
($+$ sign) and approximate the initial soliton $v(x,0;x_0)$ according
to a spatially translated, modulated soliton (\ref{eq:24})
\begin{equation}
  \label{eq:22}
  u(x,0) = \lambda_1 + 2(\lambda_{23}-\lambda_1) \mathrm{sech}^2\left (
    \sqrt{\lambda_{23} - \lambda_1}(x - x_0)/\varepsilon \right ) .
\end{equation}
The step is represented by the initial mean field
\begin{equation}
  \label{eq:44}
  \lambda_1(x,0) =
  \begin{cases}
    0 & x < 0 \\ c^2 & x > 0
  \end{cases}
  ,
\end{equation}
which, when evolved according to \eqref{eq:49}, is the RW solution
(cf.~\eqref{eq:37})
\begin{equation}
  \label{eq:55}
  \lambda_1(x,t) =
  \begin{cases}
    0 & x < 0 \\[3mm]
    \displaystyle \frac{x}{6t} & 0 < x < 6c^2 t \\[3mm]
    c^2 & 6c^2 t < x
  \end{cases} .
\end{equation}

The sign of $x_0$ determines the relative location of the soliton with
respect to the step, situated at the origin.  The magnitude $|x_0| \gg
1$ is assumed to be large, so that the initial soliton position is
well-separated from the step and more accurately approximates the true
soliton solution, if one exists.  One of the corollaries of our
analysis in this paper is the determination of whether a genuine
soliton solution actually exists.  See Sec.~\ref{IST_section}.

It remains to prescribe initial data for $\lambda_{23}$.  While it is
natural to prescribe the initial data \eqref{eq:44} for the
semi-infinite band edge $\lambda_1$ because it directly corresponds to
the initial background, how do we prescribe the initial point spectrum
$\lambda_{23}$?  Herein lies the key observation in \cite{Hoefer2}:
\begin{quote}
  Soliton-mean field modulation is described by a \textit{simple wave}
  solution of the Whitham modulation equations.
\end{quote}
We can justify this from a well-posedness argument.  The completely
prescribed initial band edge distribution $\lambda_1(x,0)$ evolves
according to the 0-phase mean field equation \eqref{eq:49}.  The unique
solution, for \textit{smooth data} $\lambda_1(x,0)$, is the implicitly
defined simple wave
\begin{equation}
  \label{eq:47}
  x - 6 \lambda_1 t = F(\lambda_1),
\end{equation}
for $t$ less than the critical time of singularity formation and where
$F$ is the inverse function of the initial data.  This mean field
evolution is completely decoupled from soliton evolution.  There is
just one initial soliton, say of amplitude $a = a_0$ centered at the
point $x = x_0$.  So, we can use \eqref{eq:23} to associate the
initial soliton amplitude at the point $x_0$ with $\lambda_{23}$.  In
the absence of any additional information, we therefore must extend
the initial point spectrum as 
\begin{equation}
  \label{eq:48}
  \lambda_{23}(x,0) = \lambda_0 = \frac{a_0}{2} + \lambda_1(x_0,0)
  \quad  \textrm{for all} \quad x \in \mathbb{R} ,
\end{equation}
i.e., we take $\lambda_{23}$ to initially be \textit{constant}.
Otherwise, the initial value problem for the modulation equations
\eqref{eq:51} is ill-posed as stated.

According to eq.~\eqref{eq:50}, the evolution of $\lambda_{23}$ is
trivial
\begin{equation}
  \label{eq:54}
  \lambda_{23}(x,t) = \lambda_0, \quad x \in \mathbb{R}, \quad t > 0 ,
\end{equation}
i.e., the solution is a 1-wave and $\lambda_{23} \equiv \lambda_0$
represents an integral of motion.

In order to extend this argument to the discontinuous mean data
\eqref{eq:44}, we take a pointwise limit of smooth approximations to
the step and arrive at the continuous, piecewise defined RW solution
\eqref{eq:37} for the mean field.  Then the point spectrum is the
constant \eqref{eq:48}, which depends on the sign of the soliton's
initial location $x_0$:
\begin{equation}
  \label{eq:52}
  \lambda_{23}(x,t) = \lambda_0 =
  \begin{cases}
    \displaystyle \frac{a_0}{2} & \mathrm{if} ~ x_0 < 0, \\[5mm]
    \displaystyle \frac{a_0}{2} + c^2 & \mathrm{if} ~ x_0 > 0
  \end{cases} .
\end{equation}
The point spectrum $\lambda_{23} = \lambda_0$ is an \textit{adiabatic
  invariant} of soliton-mean field interaction.  In fact, for a genuine
soliton solution, the point spectrum is a \textit{global invariant} of
the full KdV dynamics.  In Sec.~\ref{IST_section}, we confirm by IST
analysis that $\lambda_0$ defined in \eqref{eq:48} for soliton-RW and
soliton-DSW interaction from modulation theory is precisely the proper
eigenvalue for a genuine soliton solution or a pseudo-embedded
eigenvalue when a genuine soliton solution does not exist.  It has
recently been shown that the spectrum for the trapped, pseudo soliton
for the soliton-RW interaction is identified by a resonant pole in the
complex plane \cite{Mucalica2022}.

\begin{figure}
  \centering
    \includegraphics[scale=0.7]{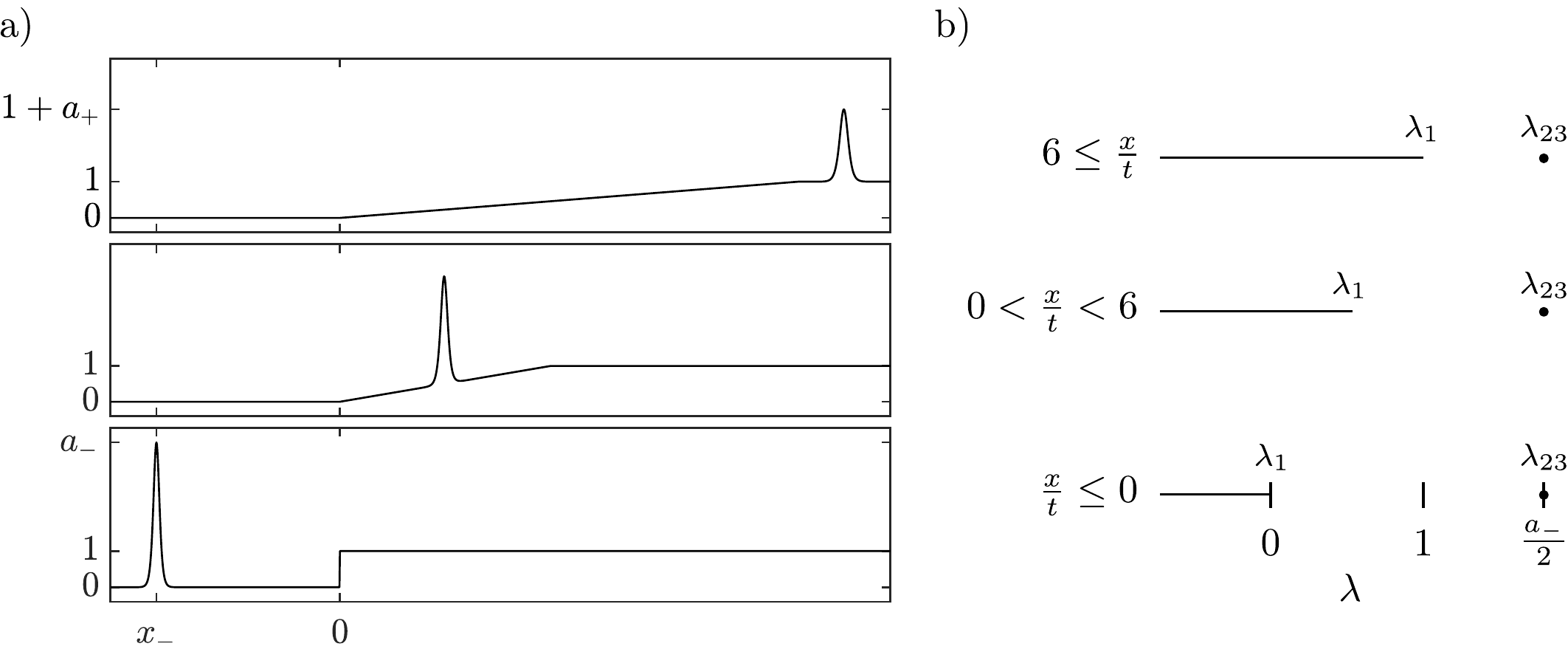}
    \caption{a) Soliton transmission through a RW when $a_- > 2 c^2$.
      b) The variation of the associated soliton-RW spectrum
      $\mathcal{S}_1^{(23)}$.  Here, $c = 1$ without loss of
      generality.}
    \label{fig:soliton_RW_tunnel_spectrum}
\end{figure}
There are a number of implications of eq.~\eqref{eq:52}, which we now
investigate.  First, we need to distinguish between where the soliton
center $x_0$ is initialized.  When $x_0 > 0$, the soliton always
outruns the mean field because the soliton characteristic \eqref{eq:6}
lies to the right of the RW fan \eqref{eq:37} due to the velocity
ordering \eqref{eq:28}.  Thus, the case $x_0 > 0$ does not give rise
to soliton-RW interaction.

When $x_0 < 0$, there are two possibilities---soliton transmission or
soliton trapping---the determination of which is based on the
spatio-temporal structure of the spectrum $\mathcal{S}_1$.  Soliton
transmission requires that the spectrum remain of the collapsed type,
i.e., of the form $\mathcal{S}_1^{(23)}$ \eqref{eq:46}.  Soliton
trapping occurs when the spectrum becomes the merged type
$\mathcal{S}_1^{(12)}$ \eqref{eq:42}.  The distinction boils down
to the ordering of the Riemann invariants.

To examine this in detail, it will be useful to identify the initial
soliton amplitude and position as $a_- = a_0$ and $x_- = x_0$,
respectively, on the mean background $\overline{u}_- = 0$, denoting
the fact that the initial soliton is located on the negative $x$-axis.

\begin{figure}
  \centering
  \includegraphics{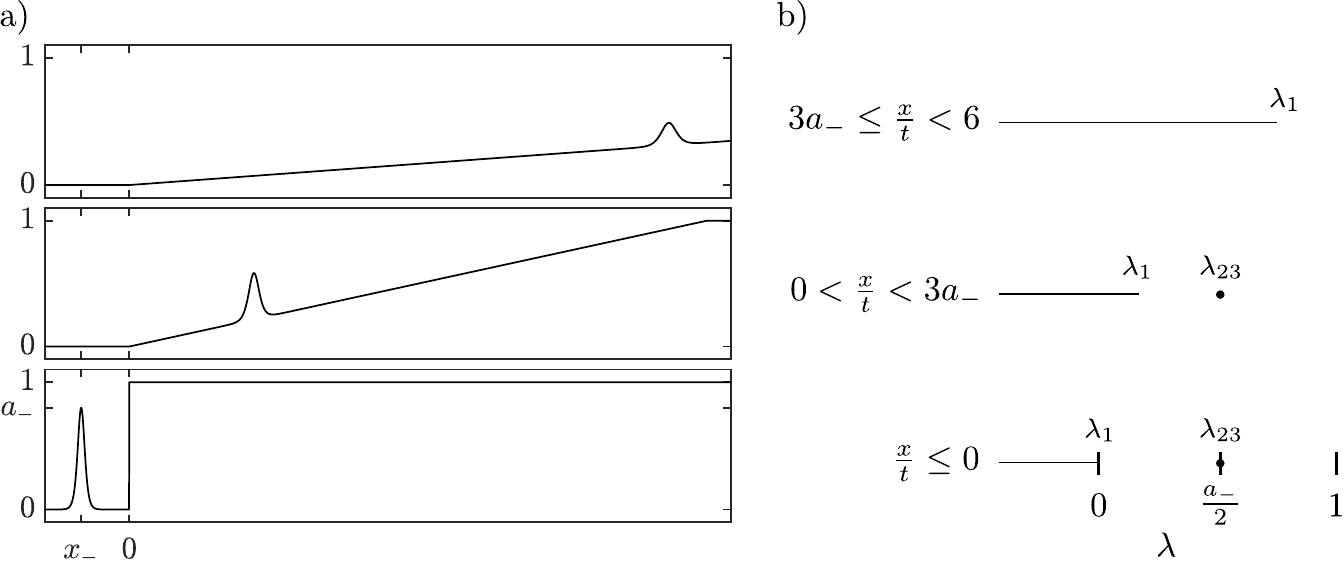}
  \caption{a) Soliton trapping by a RW when $0 < a_- \le 2c^2$.  b)
    Variation of the associated soliton-RW spectrum
    $\mathcal{S}_1^{(23)}$.  The soliton never overtakes the RW
    because for $x > 3a_- t$, the semi-infinite spectral band absorbs
    the collapsed band. Here, $c = 1$ without loss of generality.}
    \label{fig:soliton_RW_trap_spectrum}
\end{figure}

When the initial soliton amplitude is sufficiently large
($a_- > 2 c^2$), we see in the example of
Fig.~\ref{fig:soliton_RW_tunnel_spectrum} (where we set $c = 1$
without loss of generality) that the spectrum $\mathcal{S}_1^{(23)}$
consists of the semi-infinite band describing the mean field and an
eigenvalue corresponding to the soliton.  The eigenvalue never
intersects the semi-infinite band corresponding to soliton
transmission because the eigenvalue persists.  Contrast that with the
example in Fig.~\ref{fig:soliton_RW_trap_spectrum} (again with
$c = 1$) where the semi-infinite band overtakes the spectral point for
$x > 6c^2 t$.  This case corresponds to a sufficiently small initial
soliton amplitude $0 < a_- \le 2 c^2$ and is associated with soliton
trapping.

Now we describe the details of soliton-RW transmission.  Since
$\lambda_{23}(x,t) = a_-/2 > c^2$, we can identify the transmitted
soliton by its amplitude $a_+$ propagating on the mean $u_+ = c^2$.
By the constancy of $\lambda_{23}$ and the mapping in
eq.~\eqref{eq:23}, we have the relationship
\begin{equation}
  \label{eq:30}
  \frac{a_-}{2} + \overline{u}_- = \frac{a_+}{2} + \overline{u}_+ .
\end{equation}
The transmitted soliton amplitude $a_+$ is determined solely in terms
of the jump in the mean $c^2$ and the incident soliton amplitude
$a_-$.  More generally, the soliton amplitude field $a(x,t)$ is
determined by the constancy of $\lambda_{23}$ according to
\begin{equation}
  \label{eq:33}
  \begin{split}
    a(x,t) &= a_- - 2\overline{u}(x,t) 
    =
    \begin{cases}
      a_- & x \le 0, \\[3mm]
      \displaystyle a_- - \frac{x}{3t} & 0 < x < 6c^2 t, \\[3mm]
      a_+ & 6c^2 t < x
    \end{cases} .
  \end{split}
\end{equation}
Knowing $\lambda_1(x,t)$ and $\lambda_{23}$ or, equivalently, the
amplitude field and the mean field, we can determine the soliton
trajectory $x_s(t;x_-)$ by integrating the characteristic equation
\eqref{eq:6}
\begin{equation}
  \label{eq:34}
  x_s(t;x_-) =
  \begin{cases}
    x_- + 2a_- t & t \le t_* \\
    3a_- t^{1/3}\left ( t^{2/3} - t_*^{2/3} \right ) & t_* < t < t_{**} \\
    x_+ + (2a_++6c^2)t & t_{**} \le t
  \end{cases},
\end{equation}
where
\begin{equation}
  \label{eq:43}
   t_* = \frac{|x_-|}{2 a_-}, \quad t_{**} = \left (\frac{a_-}{a_+}
   \right )^{3/2} t_*, \quad x_+ = \left ( \frac{a_-}{a_+} \right
   )^{1/2} x_- .
\end{equation}
The soliton amplitude as a function of time is therefore
eq.~\eqref{eq:33} evaluated at $x = x_s(t;x_-)$.  The difference
between the post and pre interaction $x$-intercepts of the soliton
trajectory is its phase shift due to RW interaction
\begin{equation}
  \label{eq:45}
  \Delta = x_+ - x_- = \left ( \sqrt{\frac{a_-}{a_+}}
    - 1 \right ) x_- .
\end{equation}
Since $a_- > a_+$ and $x_- < 0$, the soliton phase shift is negative
$\Delta < 0$.

The results for the soliton trajectory \eqref{eq:34} and phase shift
\eqref{eq:45}, obtained by modulation theory, are the same as the
results we obtained in Sec.~(\ref{asymptotic_sec}) using soliton
perturbation theory.

There is an alternative way to determine the soliton phase shift due
to RW interaction.  For this, we utilize the additional Riemann
invariant for the 1-phase Whitham modulation equations when $0 <
\lambda_3 - \lambda_2 \ll 1$, in which the cnoidal wave exhibits the
small wavenumber (cf.~eq.~\eqref{eq:18})
\begin{equation}
  \label{eq:53}
  k \sim - \frac{2 \pi \sqrt{\lambda_3 - \lambda_1}}{\ln(\lambda_3 -
    \lambda_2)} \ll 1.
\end{equation}
This small wavenumber regime corresponds to a weakly interacting
soliton train \cite{whitham_linear_1974,Hoefer2}.  We now consider the
interaction of this soliton train with a RW modeled by the 1-wave
solution of the 1-phase Whitham modulation equations \eqref{eq:19} in
which $\lambda_2$, $\lambda_3$ are constant and $\lambda_1 =
\lambda_1(x/t)$ varies in a continuous, self-similar fashion, i.e., it
is piecewise defined satisfying either $v_1(\boldsymbol{\lambda}) =
x/t$ or $\lambda_1$ constant.  For $0 < \lambda_3 - \lambda_2 \ll 1$,
the 1-phase Whitham equations are close to the soliton limit and can
be approximated as such, i.e., $v_1 \sim 6\lambda_1$
(cf.~\eqref{eq:16}), $\lambda_1 \sim \overline{u}$ is approximately
the RW solution \eqref{eq:55}, and $\lambda_3 \sim a_-/2$ as in
\eqref{eq:52}.  The additional modulation parameter $\lambda_2$
determines the small modulation wavenumber \eqref{eq:53}, which can be
shown to approximately satisfy the conservation of waves in the form
\begin{equation}
  \label{eq:57}
  k_t + \left ( s k \right )_x = 0,
\end{equation}
where $s$ is the soliton velocity \eqref{eq:7}.  To determine this
1-wave solution, we must select a value of $\lambda_2$.  We now show
that the specific value of $\lambda_2$, other than it being close to
$\lambda_3$, is irrelevant for our purposes.

Considering the initial data \eqref{eq:44} for $\lambda_1$, constant
$\lambda_2$, $\lambda_3$ determines a relationship between the
wavenumber \eqref{eq:53} $k = k_-$ for $x < 0$ and $k = k_+$ for $x >
0$ where
\begin{equation}
  \label{eq:58}
  k_- \sim - \frac{2\pi \sqrt{\lambda_3}}{\ln(\lambda_3 - \lambda_2)},
  \quad k_+ \sim - \frac{2\pi \sqrt{\lambda_3 - c^2}}{\ln(\lambda_3 -
    \lambda_2)} .
\end{equation}
The ratio $k_-/k_+$ is independent of $\lambda_2$ so we can take the
zero wavenumber, soliton limit to obtain
\begin{equation}
  \label{eq:59}
  \lim_{\lambda_2 \to \lambda_3} \frac{k_-}{k_+} =
  \sqrt{\frac{\lambda_{23}}{\lambda_{23} - c^2}} =  \sqrt{\frac{a_-}{a_+}} .
\end{equation}
The quantity $k_-/k_+$ is an invariant of the modulation dynamics in
the soliton limit that determines the soliton phase shift $\Delta$
\eqref{eq:45} due to RW interaction.  Conservation of waves
motivate the defining relationship
\begin{equation}
  \label{eq:60}
  \lim_{\lambda_2 \to \lambda_3} \frac{k_-}{k_+} = \frac{x_+}{x_-} ,
\end{equation}
which verifies $\Delta$ in \eqref{eq:45}.  The way to interpret this
is through the concept of a weakly interacting soliton train.  What we
have shown in \eqref{eq:59} is that for any $k_- > 0$ sufficiently
small, the 1-wave modulation solution satisfies
\begin{equation}
  \label{eq:61}
  k_+ \sim \sqrt{\frac{a_+}{a_-}} k_-, \quad k_- \to 0 .
\end{equation}
Since $2\pi/k$ is the pulse spacing, we can track the propagation of
two adjacent pulses of the weakly interacting soliton train, one
located at $x = x_- < 0$, and the other at $x_- - 2\pi/k_-$.  Each
pulse's propagation traces out a characteristic, which are
well-separated by $2\pi/k_-$ for $x < 0$.  By the conservation of
$\lambda_2$ and $\lambda_3$, the two characteristics will be
well-separated by the distance $2\pi/k_+$ when $x > 6c^2 t$.
Therefore, we can measure the phase shift of the pulse initially at $x
= x_-$ by its position shift relative to the other pulse, i.e.
\begin{equation}
  \label{eq:62}
  \Delta = x_+ - x_- = \left ( \frac{2\pi/k_+}{2\pi/k_-} - 1 \right )
  x_- \sim \left ( \sqrt{\frac{a_-}{a_+}} - 1 \right )
  x_- ,
\end{equation}
by using \eqref{eq:61}.  Taking the soliton limit $k_- \to 0$, we
obtain the same result \eqref{eq:45} found by direct integration of
the soliton characteristic.  This physically inspired approach to
determining the soliton phase shift will prove to be applicable to
soliton-DSW interaction as well.

As noted earlier, when $0 < a_- < 2 c^2$, the initial soliton becomes
trapped in the interior of the RW.  The soliton's characteristic
$x_s(t;x_-)$ is the same as that given in eq.~\eqref{eq:34} except
$t_{**} \to \infty$.  The soliton amplitude in the interior of the RW
is \eqref{eq:33} evaluated at $x = x_s(t;x_-)$ for $t > t_*$
\begin{equation}
  \label{eq:63}
  a_s(t) = a_- \left ( \frac{t_*}{t} \right )^{2/3} = a_-^{1/3} \left
    ( \frac{|x_-|}{2 t} \right )^{2/3},
\end{equation}
hence decays algebraically as $t \to \infty$.  To see that the soliton
is trapped, we note that its velocity satisfies
\begin{equation}
  \label{eq:64}
  \frac{\mathrm{d} x_s}{\mathrm{d} t} = a_- \left ( 3 - \left (
      \frac{t_*}{t} \right )^{2/3} \right )  \le 3 a_- < 6 c^2,
\end{equation}
therefore, its characteristic cannot overtake the right edge of the RW
with velocity $6c^2$.

\begin{table}
  \centering
  \begin{tabular}{llll}
    conditions & result & amplitude & position \\
    \hline
    $x_0 > 0$ & no interaction & & \\
    $x_0 < 0$, $a_- > 2 c^2$ & transmission & $a_+ =
    a_- - 2 c^2$ & $\Delta/x_0 = \sqrt{a_-/a_+} - 1$ \\
    $x_0 < 0$, $a_- < 2 c^2$ & trapping & $a_s(t) =
    a_-^{1/3}\left ( \frac{|x_0|}{2 t} \right )^{2/3}$
    & $x_s(t) = 3 a_- \left ( t - \left ( t \frac{x_0^2}{4a_-^2}
      \right )^{1/3} \right )$
  \end{tabular}
  \caption{Predictions for soliton-RW
    interaction with a RW emanating from $x = 0$ at $t = 0$ and an
    initial soliton of amplitude $a_-$ centered at $x = x_0$ with
    $|x_0| \gg 1$.  $a_+$ and $\Delta$ are the transmitted soliton
    amplitude and phase shift, respectively;
    $a_s(t)$ and $x_s(t)$ are the trapped
    soliton amplitude and position, respectively.}
  \label{tab:soliton-RW}
\end{table}
We summarize our findings for soliton-RW interaction in Table
\ref{tab:soliton-RW}, all of which agree with the results obtained in
Sec.~\ref{sol_pert_th} using soliton perturbation theory.

\subsection{Soliton-Dispersive Mean Field Interaction:  2-phase
  modulations}
\label{sec:solit-oscill-mean}

\begin{figure}
  \centering
  \includegraphics[scale=0.7]{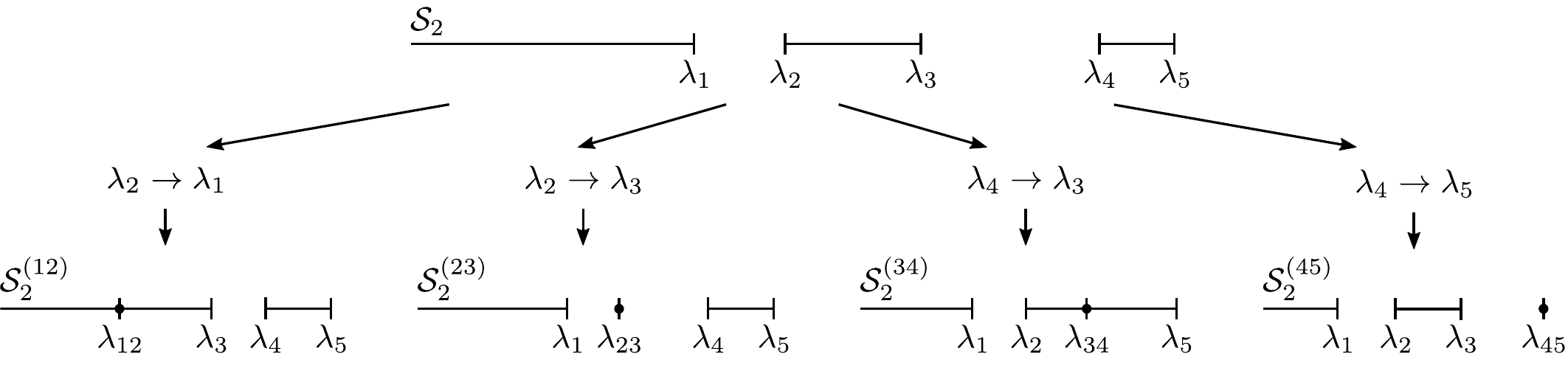}
  \caption{Degeneration pathways for 2-phase spectrum $\mathcal{S}_2$.
    For soliton-DSW interaction, the collapsing cases when $\lambda_2
    \to \lambda_3$ and $\lambda_4 \to \lambda_5$ correspond to the
    trapped and transmission scenarios, respectively.  When $\lambda_2
    \to \lambda_1$ or $\lambda_4 \to \lambda_3$, one band merges with
    another.  These cases correspond to transmission and trapped
    linear wavepacket-DSW interaction, respectively (see
    Sec.~\ref{sec:line-wavep-dsw}).}
  \label{fig:2-phase-spectrum}
\end{figure}

Recall that a DSW resulting from the Riemann problem is described by a
special solution \eqref{eq:36} of the 1-phase Whitham equations.  A
soliton propagating on a slowly varying mean field is described by the
degenerate 1-phase Whitham equations \eqref{eq:51} in which the mean
field evolves independent of the soliton.  In order to describe the
interaction of a soliton and a DSW, the modulations must
simultaneously represent both the DSW and the soliton.  As we now
demonstrate, this is achieved by consideration of the degenerate
2-phase Whitham equations (\ref{eq:13}) with a collapsed spectral
profile, either $\mathcal{S}_2^{(23)}$ or $\mathcal{S}_2^{(45)}$
depicted in Fig.~\ref{fig:2-phase-spectrum}.  We relate the modulation
of the spectrum $\mathcal{S}_2^{(45)}$ to the problem of soliton-DSW
transmission and modulations of $\mathcal{S}_2^{(23)}$ to soliton-DSW
trapping.  We provide a direct proof that the DSW modulation evolves
independent of the soliton in the case of soliton-DSW transmission and
we determine the soliton's characteristic velocity in both the
transmission and trapping cases.

First, we introduce convenient notation.  Recalling the characteristic
velocities~\eqref{eq:29} for the $N = 2$ case, the entries of the 2x2
matrices $\mathrm{Q}$ and $\mathrm{P}$ are terms of the form
\begin{equation}
  \label{eq:56}
  I_r^n(\lambda_k) = \int_{\lambda_{2r-1}}^{\lambda_{2r}}
  \frac{(\lambda_k - \mu) \mu^n\, \mathrm{d}\mu}{R(\mu)} , \quad r \in
  \{1,2\}, \quad n \in \{0,1,2\} ,
\end{equation}
where $R(\mu) = \sqrt{(\mu-\lambda_1)(\mu-\lambda_2) (\mu-\lambda_3)
  (\mu-\lambda_4)(\mu-\lambda_5)}$.  The hyperelliptic integrals
\eqref{eq:56} will be simplified to elliptic integrals upon following
the degeneration pathways noted in
Fig.~\ref{fig:2-phase-spectrum}. For the various degenerate cases we
consider, it is helpful to define quantities associated with the
decoupled 1-phase DSW modulation by the parameters $\lambda_i \le
\lambda_j \le \lambda_k$
\begin{equation}
  \label{eq:10}
  \begin{split}
    R_{ijk}(\mu) &\equiv \sqrt{(\mu-\lambda_i)(\mu - \lambda_j)(\mu -
      \lambda_k)}, \\ \\
    m_{ijk} &\equiv \frac{\lambda_j - \lambda_i}{\lambda_k -
      \lambda_i} , \quad V_{ijk} \equiv 2(\lambda_i + \lambda_j +
    \lambda_k) .
  \end{split}
\end{equation}

\subsubsection{Soliton-DSW Transmission}
\label{sec:solit-dsw-transm}

We consider the soliton-DSW interaction resulting from step down
initial data \eqref{general_IC} ($-$ sign) and approximate the initial
soliton $v(x,0;x_0)$ according to the spatially translated, modulated
soliton
\begin{subequations}
  \label{eq:90}
  \begin{equation}
    \label{eq:88}
    v(x,0;x_0 = x_-) = \lambda_2 + 2(\lambda_{45} - \lambda_2)
    \mathrm{sech}^2 \left 
      ( \sqrt{\lambda_{45} - \lambda_2}(x-x_-)/\varepsilon \right ),
  \end{equation}
  for $x_- < 0$ where $\lambda_{45}$ is constant and $\lambda_2$ is
  \begin{equation}
    \label{eq:89}
    \lambda_2(x,0) =
    \begin{cases}
      -1, & x < 0, \\
      0, & x > 0 .
    \end{cases}
  \end{equation}
\end{subequations}
The remaining spectral parameters are constant $\lambda_1 = -1$,
$\lambda_3 = 0$.  We have scaled the step down to unit amplitude $c^2
= 1$ without loss of generality.

We remark that selecting piecewise constant initial data that leads to
a global modulation solution for the Whitham equations is known as
initial data regularization and was originally introduced in
\cite{Bloch} (see also \cite{hoefer_dispersive_2006-1}).

For the case of soliton-DSW transmission, we consider the limit
$\lambda_4 \to \lambda_5$, resulting in the collapsed spectrum
$\mathcal{S}_2^{(45)}$ depicted in Fig.~\ref{fig:2-phase-spectrum}.
The asymptotics of the integrals \eqref{eq:56} for $r = 1$ when
$\lambda_4 \to \lambda_5$ are
\begin{equation}
  \label{eq:65}
  I_1^n(\lambda_k) \to \int_{\lambda_1}^{\lambda_2} \frac{(\lambda_k -
    \mu) \mu^n\,\mathrm{d}\mu}{(\lambda_{45} - \mu) R_{123}(\mu)}, \quad
  \lambda_4 \to \lambda_5 .
\end{equation}
These integrals can be expressed in terms of complete elliptic
integrals.  When $r = 2$, the result depends upon $k$
\begin{equation}
  \label{eq:67}
  I_2^n(\lambda_k) \sim
  \begin{cases}
    \displaystyle \lambda_{5}^n \frac{(\lambda_{5} - \lambda_k)
      \ln(\lambda_{5} -
      \lambda_4)}{R_{123}(\lambda_{5})}, & k \in \{1,2,3\}, \\
    \displaystyle \int_{\lambda_3}^{\lambda_{45}} \frac{\mu^n \,
      \mathrm{d}\mu}{R_{123}(\mu)}, & k \in \{4,5\} ,
  \end{cases} \quad \lambda_4 \to \lambda_{5} .
\end{equation}
The integrals \eqref{eq:67} for $k \in \{4,5\}$ can be expressed in
terms of incomplete elliptic integrals.

We are now in a position to calculate the Whitham modulation
velocities \eqref{eq:29}.  When $k \in \{1,2,3\}$,
\begin{equation}
  \label{eq:68}
  \begin{split}
    \lim_{\lambda_4 \to \lambda_5} v_k &= -6(\lambda_1 + \lambda_2 +
    \lambda_3 + 2 \lambda_{45}) + 12 \lambda_k + \lim_{\lambda_4 \to \lambda_5} 12
    \frac{I_2^2(\lambda_k) I_1^0(\lambda_k)  - 
      I_2^0(\lambda_k) I_1^2(\lambda_k)}{I_2^1(\lambda_k)I_1^0(\lambda_k)  -
                                         I_2^0(\lambda_k) I_1^1(\lambda_k) } \\
    &= -6(\lambda_1 + \lambda_2 + \lambda_3 + 2 \lambda_{45}) + 12 \lambda_k +
    12 \frac{ \lambda_{45}^2 I_1^0(\lambda_k)  -
      I_1^2(\lambda_k) } {\lambda_{45} I_1^0(\lambda_k) 
      - I_1^1(\lambda_k)} \\
    &= -6(\lambda_1 + \lambda_2 + \lambda_3 + 2 \lambda_{45}) + 12 \lambda_k +
    12 \frac{ \displaystyle \int_{\lambda_1}^{\lambda_2} \frac{(\lambda_{45} +
      \mu)(\lambda_k - \mu)  \, \mathrm{d}\mu}{R_1(\mu)}}
  {\displaystyle \int_{\lambda_1}^{\lambda_2} \frac{(\lambda_k - \mu)
      \, \mathrm{d}\mu}{R_1(\mu)}}  \\
  &= -6(\lambda_1 + \lambda_2 + \lambda_3) + 12 \lambda_k +
    12 \frac{ \displaystyle \int_{\lambda_1}^{\lambda_2}
      \frac{(\lambda_k - \mu) \mu \, \mathrm{d}\mu}{R_1(\mu)}}
    {\displaystyle \int_{\lambda_1}^{\lambda_2} \frac{(\lambda_k - \mu)
      \, \mathrm{d}\mu}{R_1(\mu)}} .
  \end{split}
\end{equation}
The last line is precisely the definition \eqref{eq:29} of the 1-phase
Whitham velocities \eqref{eq:19} for the variables $\lambda_1$,
$\lambda_2$, $\lambda_3$.

\begin{figure}
  \centering
  \includegraphics[scale=0.7]{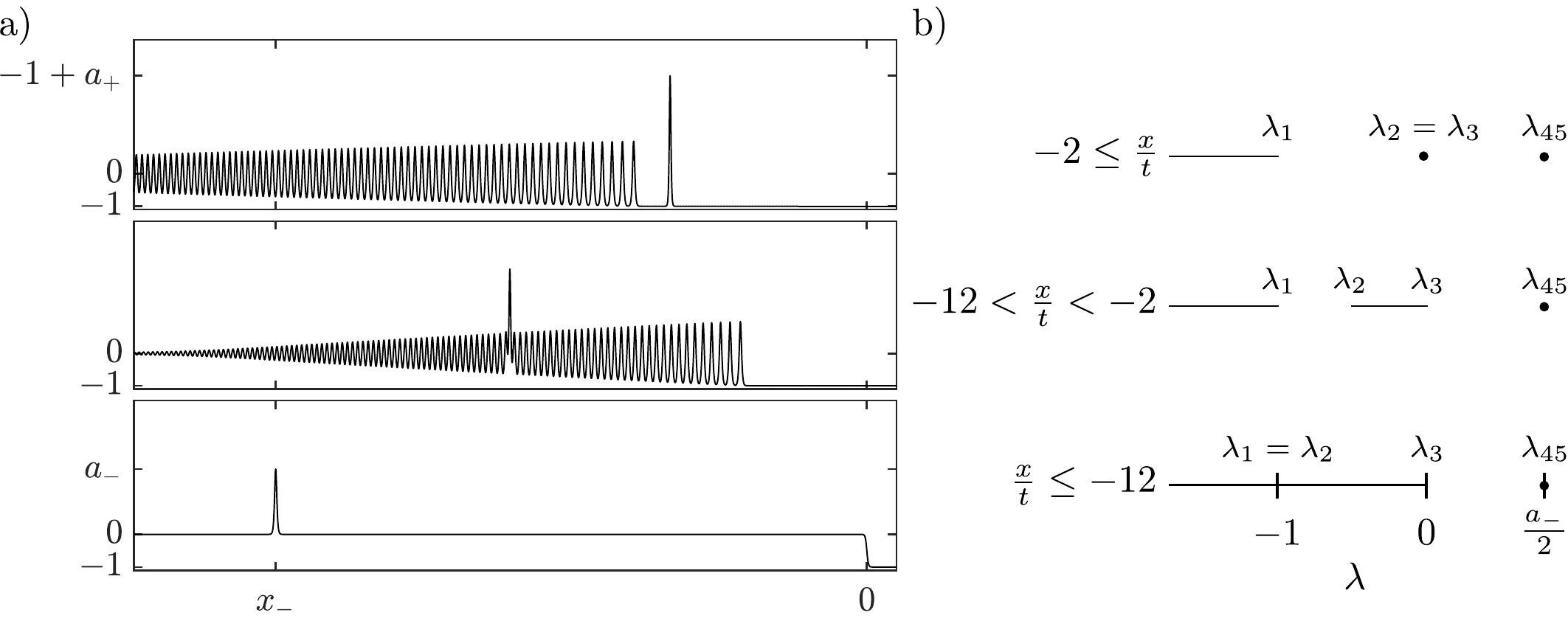}
  \caption{Soliton transmission through a DSW.  a) Evolution of the
    solution $u(x,t)$.  b) Evolution of the associated soliton-DSW
    spectrum $\mathcal{S}_2^{(45)}$.}
  \label{fig:soliton-dsw-transmission}
\end{figure}
For the cases $k \in \{4,5\}$, the integrals in \eqref{eq:65} and
\eqref{eq:67} are independent of $\lambda_4$.  Consequently, the
velocities coalesce in the limit $v_{45} \equiv \lim_{\lambda_4 \to
  \lambda_5} v_4 = \lim_{\lambda_4 \to \lambda_5} v_5$. After a
calculation (the elliptic integral reference \cite{Byrd} is helpful),
we obtain
\begin{equation}
  \label{eq:70}
  \begin{split}
    v_{45} &= V_{123} + \frac{4(\lambda_{45} - \lambda_2)}{1 -
      \sqrt{\frac{(\lambda_{45} - \lambda_2)(\lambda_3 -
          \lambda_1)}{(\lambda_{45} - \lambda_3)(\lambda_{45} -
          \lambda_1)}} Z(\varphi,m_{123})} , \quad \sin{\varphi} =
    \sqrt{\frac{\lambda_{45} - \lambda_3}{\lambda_{45} - \lambda_2}} .
  \end{split}
\end{equation}
The function
\begin{equation}
  \label{eq:71}
  Z(\varphi,m) \equiv E(\varphi,m) - \frac{E(m)}{K(m)} F(\varphi,m)
\end{equation}
is the Jacobian zeta function and involves the complete ($K(m)$,
$E(m)$) and incomplete ($F(\varphi,m)$, $E(\varphi,m)$) elliptic
integrals of the first and second kinds \cite{Byrd}.  

In the DSW's harmonic limit $\lambda_2 \to \lambda_1$, we have
$m_{123} \to 0$ and $Z(\varphi,m_{123}) \to 0$ so that the further
degeneration (merger) of $\mathcal{S}_2^{(45)}$ results in the
limiting characteristic velocity
\begin{equation}
  \label{eq:78}
  \lim_{\lambda_2 \to \lambda_1} v_{45} = 2 \lambda_3 + 4 \lambda_{45},
\end{equation}
which is the speed of a soliton associated with the eigenvalue
$\lambda_{45}$ propagating on the mean $\lambda_3$
(cf.~eq.~\eqref{eq:24}).  In the DSW's soliton limit
$\lambda_2 \to \lambda_3$, we have
$\varphi = \pi/2 + \mathcal{O}((\lambda_3-\lambda_2)^{1/2})$ and
$m_{123} = 1 + \mathcal{O}(\lambda_3 - \lambda_2)$ so that
$Z(\varphi,m_{123}) \to 0$ and the spectrum $\mathcal{S}_2^{(45)}$
further collapses.  The resulting modulation velocity under this limit
becomes
\begin{equation}
  \label{eq:79}
  \lim_{\lambda_2 \to \lambda_3} v_{45} = 2 \lambda_1 + 4 \lambda_{45},
\end{equation}
which is the speed of a soliton associated with $\lambda_{45}$
propagating on the mean $\lambda_1$.

Figure \ref{fig:soliton-dsw-transmission}(a) details the case of a
soliton with amplitude $a_- > 0$ that is initially located at
$x = x_- = x_0 < 0$ on the mean value $\overline{u}_- = 0$.  The
negative step leads to the generation of a DSW with which the soliton
interacts for some finite time.  Post interaction, the soliton with
amplitude $a_+ > a_-$ on the background $\overline{u}_+ = -c^2 = -1$
(recall that we have scaled $c^{2}=1$) emerges to the right of the
DSW.  Its trajectory is $x_s(t) = (6 \overline{u}_+ + 2 a_+)t + x_+$
and its phase shift due to DSW interaction is $\Delta = x_+ - x_-$.
We now interpret and analyze Fig.~\ref{fig:soliton-dsw-transmission}a)
in terms of modulation theory.

\begin{figure}
  \centering
  \includegraphics[scale=0.333333]{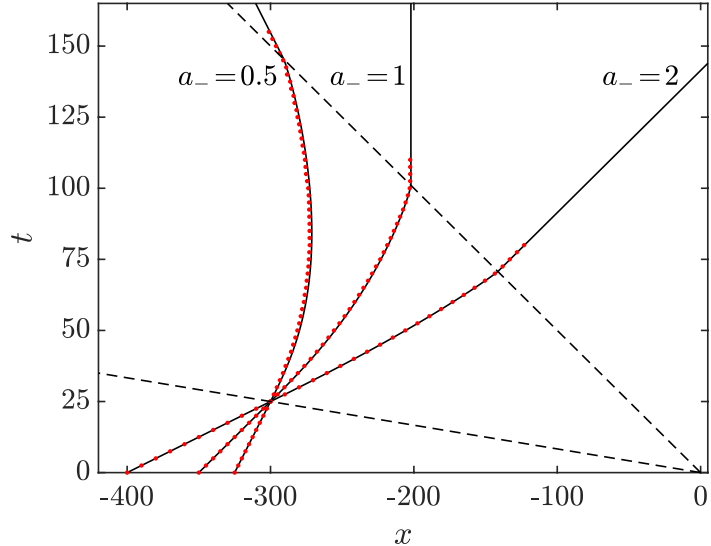}
  \caption{Soliton trajectories for the case of soliton-DSW
    transmission with varying incident soliton amplitudes $a_- \in
    \{0.5,1,2\}$.  The solid curves are the characteristics
    \eqref{eq:12} and the dots are the positions of the soliton
    extracted from numerical simulations of KdV \eqref{kdv}
    ($\varepsilon = 1$) subject to \eqref{eq:90}.  The dashed lines
    are the DSW's space-time boundaries.}
  \label{fig:soliton_dsw_tunneling_trajectories}
\end{figure}
Figure \ref{fig:soliton-dsw-transmission}(b) depicts the evolution of
the spectrum $\mathcal{S}_2^{(45)}$ according to the GP DSW modulation
solution \eqref{eq:36}.  The modulation solution is a 2-wave in which
only $\lambda_2$ varies with characteristic velocity $v_2$ while the
other modulation variables are constant.  The soliton's trajectory
$x_s(t)$ is completely determined by the characteristic
\begin{equation}
  \label{eq:12}
  \frac{\mathrm{d} x_s}{\mathrm{d} t} =
  v_{45}(-1,\lambda_2(x_s/t),0,\lambda_{45},\lambda_{45}), \quad
  x_s(0) = x_0 .
\end{equation}
Since $v_{45} > v_2$ by strict hyperbolicity, the soliton trajectory
$x_s(t)$ passes through the DSW if and only if the soliton is
initialized to the left of the step: $x_0 = x_- < 0$.  Prior to
soliton-DSW interaction, the spectrum is doubly degenerate so that we
identify (cf.~eq.~\eqref{eq:23})
\begin{subequations}
  \label{eq:82}
  \begin{equation}
    \label{eq:66}
    \lambda_2 \to \lambda_1: \quad \lambda_3 = \overline{u}_- = 0, \quad
    \lambda_{45} = \frac{a_-}{2} + \overline{u}_-
  \end{equation}
  and the soliton velocity $v_{45}$ is \eqref{eq:78}.  Post soliton-DSW
  interaction, the spectrum degenerates again so that we find the
  relations
  \begin{equation}
    \label{eq:80}
    \lambda_2 \to \lambda_3: \quad \lambda_1 = \overline{u}_+ = -c^2,
    \quad \lambda_{45} = \frac{a_+}{2} + \overline{u}_+,
  \end{equation}
\end{subequations}
and the soliton velocity $v_{45}$ is now \eqref{eq:79}.  The four
relations in \eqref{eq:82} determine the constant modulation variables
$\lambda_1$, $\lambda_3$, and $\lambda_{45}$ that, along with the GP
modulation solution \eqref{eq:36}, are the soliton-DSW 2-wave
modulation.  But $\lambda_{45}$ is overdetermined, yielding a
constraint on the soliton's amplitude post DSW interaction
\begin{equation}
  \label{eq:84}
  \frac{a_+}{2} + \overline{u}_+ = \frac{a_-}{2} + \overline{u}_- .
\end{equation}
This equation relating the soliton amplitude and mean field pre and
post DSW interaction is the same as the relation \eqref{eq:30} for
soliton-RW interaction.

\begin{figure}
  \centering
  \includegraphics[scale=0.4]{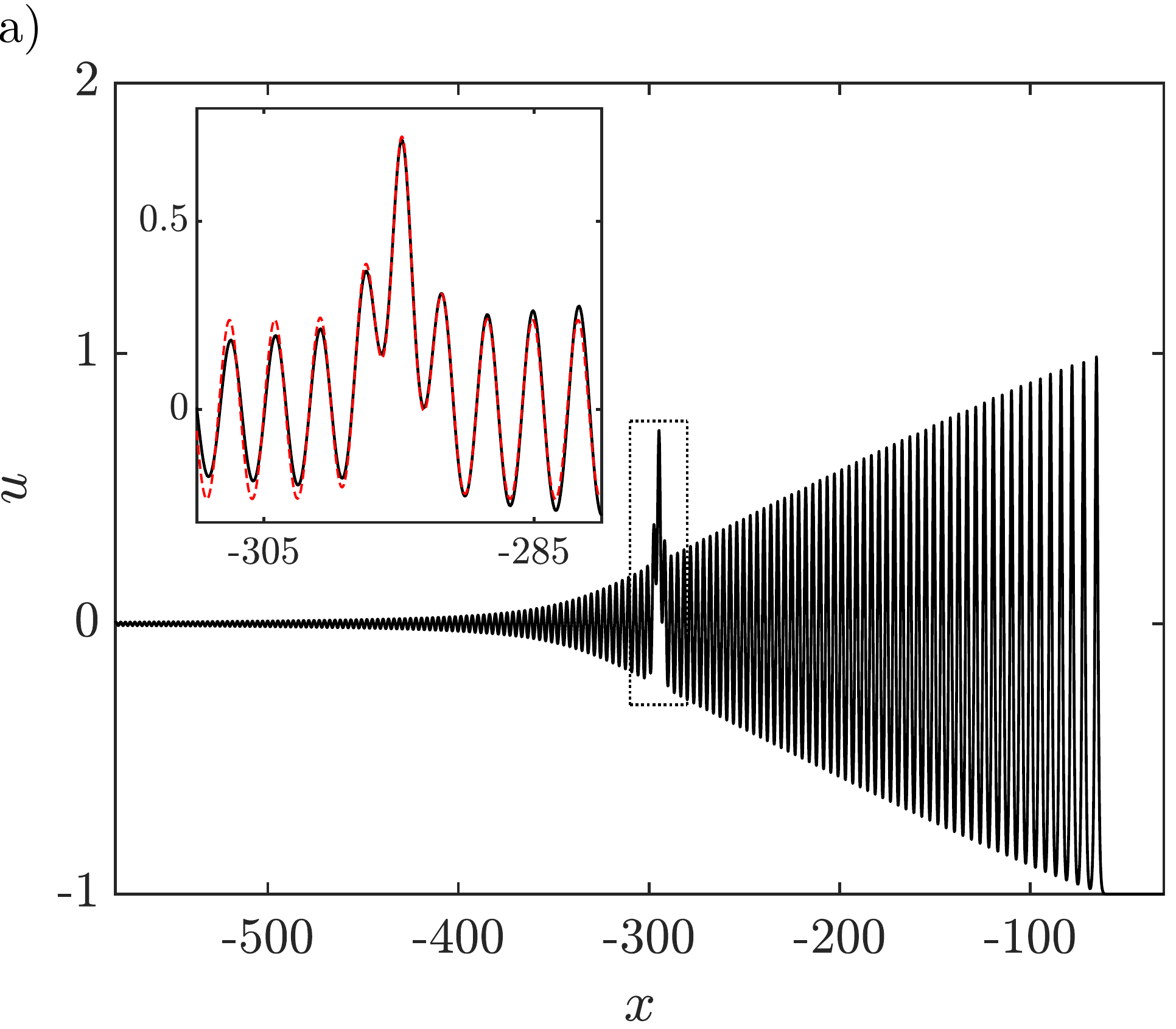}
  \includegraphics[scale=0.4]{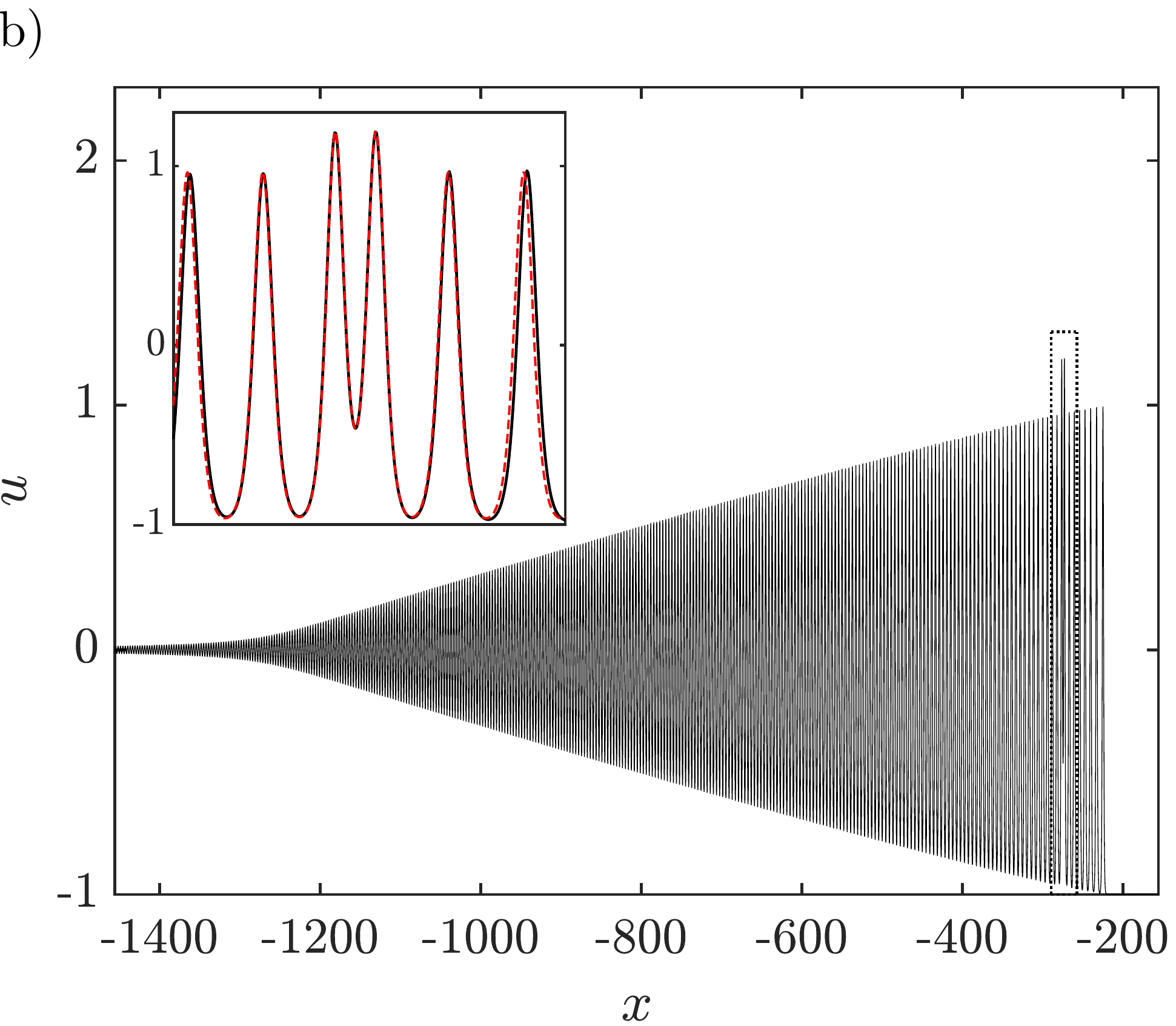}
  \caption{Numerical simulation of soliton-DSW transmission for $a_- =
    0.5$, $x_- = -325$ ($\lambda_{45} = 0.25$).  Insets: Local
    description of the simulation (solid, black curve) by the 2-phase
    solution with spectrum $\mathcal{S}_2^{(45)}(t) = (-\infty,-1]
    \cup [\lambda_2(x_s(t)/t),0] \cup \{\lambda_{45}\}$ (dashed, red
    curve).  a) Snapshot at $t = 30$. b) Snapshot at $t = 109.3$.}
  \label{fig:soliton_dsw_tunneling_2phase}
\end{figure}
By again introducing a non-interacting soliton train in which $0 <
\lambda_5 - \lambda_4 \ll \lambda_{45}$, the same argument as
described in Sec.~\ref{sec:solit-rw-inter} for soliton-RW interaction
also holds for soliton-DSW interaction, resulting in the same soliton
phase shift as in eq.~\eqref{eq:62}
\begin{equation}
  \label{eq:87}
  \Delta = x_+ - x_- = \left ( \sqrt{\frac{a_-}{a_+}} - 1 \right ) x_-
    .
\end{equation}
This time, $a_- < a_+$ so that the soliton phase shift due to DSW
interaction is positive, $\Delta > 0$. We will demonstrate that this
is the same phase shift as obtained through (\ref{DSW_phase_asym}) in
the small dispersion IST analysis. Consistency requires that the
direct integration of the soliton-DSW characteristic \eqref{eq:12}
yields the same phase shift as in eq.~\eqref{eq:87}.  We have
numerically verified this to be the case, to the precision of the
numerical method, by numerical integration of the characteristic
equation \eqref{eq:12} for a range of initial soliton amplitudes
$a_-$.

Formulas \eqref{eq:84} and \eqref{eq:87} relate the soliton post DSW
interaction to the soliton pre DSW interaction.  We now investigate
the predictions from modulation theory for the soliton during DSW
interaction.  

The soliton trajectories \eqref{eq:12} in soliton-DSW transmission for
different choices of $a_-$ are shown in
Fig.~\ref{fig:soliton_dsw_tunneling_trajectories}. The 2-phase
characteristics obtained by integrating eq.~\eqref{eq:12} (solid
curves) compared with the soliton trajectories extracted from
numerical simulations of the KdV equation (dots) are visually
indistinguishable.  Modulation theory can also be used to reconstruct
an approximation to the full solution by inserting the modulation
solution into the degenerate 2-phase solution (\eqref{eq:8} with $N =
2$ in the limit $\lambda_4 \to \lambda_5$)
\begin{equation}
  \label{eq:85}
  u(x,t) = \lambda_1 + \lambda_2 + \lambda_3 - 2\mu_1(x,t) -
  2(\mu_2(x,t) - \lambda_{45}) .
\end{equation}
The functions $\mu_1$, $\mu_2$ satisfy the coupled nonlinear system
\eqref{eq:9}, which we solve numerically.  In fact, explicit
representations of this solution have been obtained
\cite{Kuznetsov1975,Hu,Lou,Bertola} and we identify them as KdV
breather solutions because they exhibit two time scales: one
associated with their propagation and the other associated with their
internal oscillations.  In \cite{Bertola}, $N+1$-phase solutions
\eqref{eq:14} are analyzed in the case that $N$ finite bands collapse.
This scenario describes $N$ breathers propagating on a cnoidal wave
background.  We use elevation, bright breathers to investigate the
local structure of the solution within the vicinity of the soliton
trajectory at certain times.  Figure
\ref{fig:soliton_dsw_tunneling_2phase} displays a numerical simulation
of soliton-DSW interaction for the case $a_- = 0.5$, $x_- = -325$,
corresponding to the trajectory shown in
Fig.~\ref{fig:soliton_dsw_tunneling_trajectories}.  At two different
times, we plot the numerical solution.  We also evaluate the DSW
modulation solution $\lambda_2(x/t)$ \eqref{eq:36} at $x = x_s(t)$ to
obtain the local spectral profile
$\mathcal{S}_2^{(45)} = (-\infty,\lambda_1] \cup [\lambda_2,\lambda_3]
\cup \{\lambda_{45}\}$ at the soliton characteristic $x_s(t)$ at time
$t$.  The remaining parameters $\lambda_1 = -1$ and $\lambda_3 = 0$
are obtained from the initial step down and $\lambda_{45} = 0.25$
corresponds to the initial soliton.  The initial phases $\mu_j(0,0)$,
$j = 1,2$ for the 2-phase solution are chosen so that the 2-phase
solution best matches the numerical simulation.  The insets in
Fig.~\ref{fig:soliton_dsw_tunneling_2phase} display the 2-phase
solution for the spectrum $\mathcal{S}_2^{(45)}$ (dashed) overlaid on
top of the numerical simulation (solid), showing excellent agreement
in the vicinity of the soliton trajectory.  The deviation is due to
the modulation of the 2-phase wave which we have not incorporated into
our approximate solution.  At the early time depicted in
Fig.~\ref{fig:soliton_dsw_tunneling_2phase}(a), the soliton is
interacting with the DSW harmonic edge and is approximately a linear
superposition of a soliton and a cosine traveling wave.  At the later
time in Fig.~\ref{fig:soliton_dsw_tunneling_2phase}(b), the soliton
interacts with the DSW's soliton edge.  Here, the solution is
approximately a soliton interacting with a soliton train.  We have
demonstrated that the KdV soliton-DSW interaction is well-described by
a modulated bright breather solution of the KdV equation.

\subsubsection{Soliton-DSW Trapping}
\label{sec:solit-dsw-trapp}

\begin{figure}
  \centering
  \includegraphics{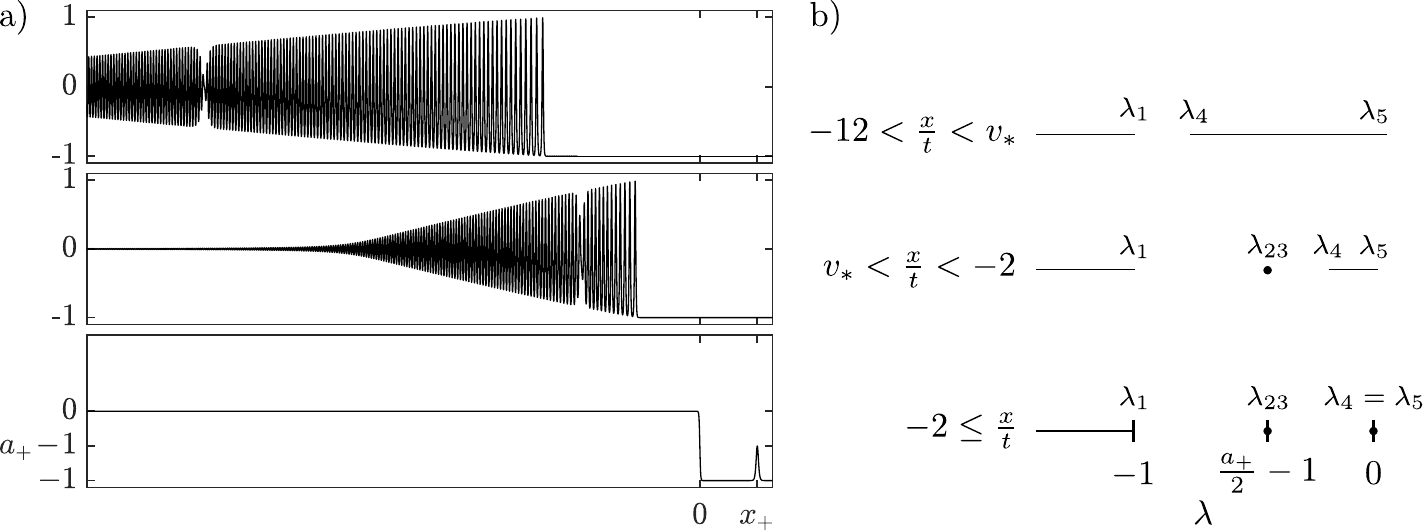}
  \caption{Soliton trapping by a DSW.  a) Evolution of the solution
    $u(x,t)$.  b) Evolution of the associated soliton-DSW spectrum
    $\mathcal{S}_2^{(23)}$.}
  \label{fig:soliton_dsw_trapping_spectrum}
\end{figure}
We now consider the soliton-DSW interaction from step down initial
data \eqref{general_IC} ($-$ sign) and approximate the initial soliton
$v(x,0;x_0)$ according to the spatially translated, modulated soliton
\begin{subequations}
  \label{eq:97}
  \begin{equation}
    \label{eq:91}
    v(x,0;x_0 = x_+) = \lambda_4 + 2(\lambda_{23} - \lambda_4) \mathrm{sech}^2
    \left ( \sqrt{\lambda_{23} - \lambda_4}(x-x_+)/\varepsilon \right
    ) ,
  \end{equation}
  where $x_+ > 0$, $\lambda_{23}$ is constant and there is an initial
  jump in $\lambda_4$
  \begin{equation}
    \label{eq:95}
    \lambda_4(x,0) =
    \begin{cases}
      -1 & x < 0, \\
      0 & x > 0 ,
    \end{cases}
  \end{equation}
\end{subequations}
where we have scaled $c^2 = 1$, without loss of generality.  The
remaining spectral parameters are constant with the values
$\lambda_1 = -1$, $\lambda_5 = 0$ because the step has been scaled to
unit amplitude.  The reason for requiring the soliton to be located at
$x = x_+ > 0$ is because the case where $x = x_- < 0$ necessarily
leads to soliton transmission as shown in the previous subsection.

An example numerical evolution of $u(x,t)$ and the modulation
parameters is shown in
Fig.~\ref{fig:soliton_dsw_trapping_spectrum}. In contrast to the case
of soliton-DSW transmission, the soliton eigenvalue $\lambda_{23}$
here eventually coincides with the DSW modulation solution
$\lambda_4(x/t)$.  This is the reason that the soliton is trapped.

For the soliton-DSW trapping problem, we require the asymptotics of
the hyperelliptic integrals $I_r^n(\lambda_k)$ \eqref{eq:56} in the
limit $\lambda_2 \to \lambda_3$.  For the characteristic velocity
$v_{23} \equiv \lim_{\lambda_2 \to \lambda_3} v_2 = \lim_{\lambda_2
  \to \lambda_3} v_3$, note that
\begin{equation}
  \label{eq:92}
  I_1^n(\lambda_k) \sim \int_{\lambda_1}^{\lambda_{23}}
  \frac{\mu^n}{R_{145}(\mu)}
  \mathrm{d}\mu, \quad I_2^n(\lambda_k) \sim \int_{\lambda_{23}}^{\lambda_4}
  \frac{\mu^n}{R_{145}(\mu)}
  \mathrm{d}\mu, \quad \lambda_2 \to \lambda_3
\end{equation}
when $k \in \{2,3\}$.  These are incomplete elliptic integrals that,
after inserting into equation \eqref{eq:29} and simplifying, result
in the soliton's characteristic velocity
\begin{equation}
  \label{eq:74}
  \begin{split}
    v_{23} &= V_{145} - 4(\lambda_5 - \lambda_{23})
    \frac{\sqrt{\frac{(\lambda_4 - \lambda_{23})(\lambda_{23} -
          \lambda_1)}{(\lambda_5 - \lambda_{23})(\lambda_5 -
          \lambda_1)}}}{Z(\varphi,m_{145})}, \quad \sin{\varphi} =
    \sqrt{\frac{\lambda_{23} - \lambda_1}{\lambda_4 - \lambda_1}} ,
  \end{split}
\end{equation}
where, again, $Z$ is the Jacobian zeta function \eqref{eq:71}.

\begin{figure}
  \centering
  \includegraphics[scale=0.4]{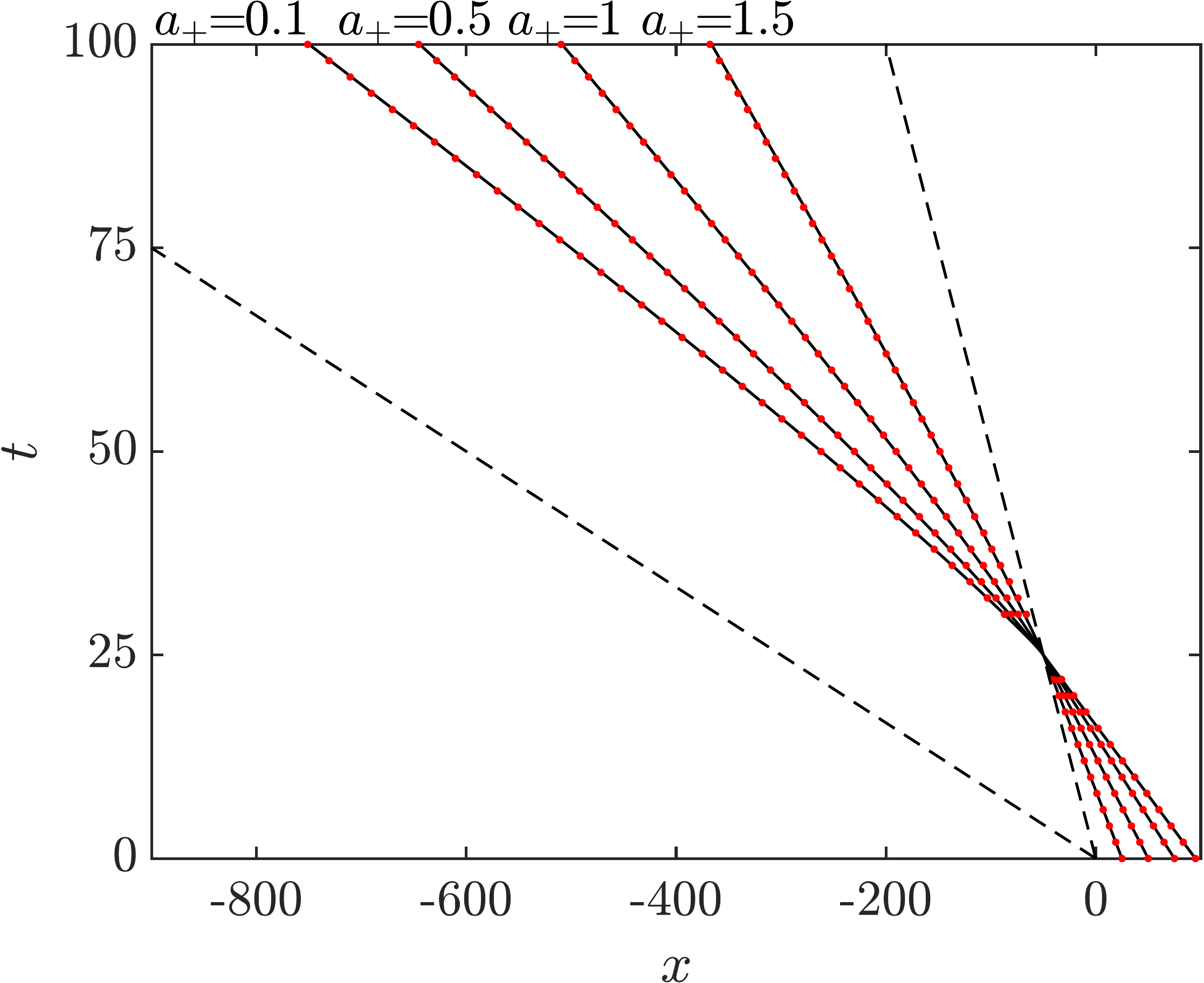}
  \caption{Soliton trajectories for the case of soliton-DSW trapping
    with varying incident soliton amplitudes $a_+ \in
    \{0.1,0.5,1,1.5\}$.  The solid curves are the characteristics
    \eqref{eq:94} and the dots are the positions of the soliton
    extracted from numerical simulations of KdV \eqref{kdv}
    ($\varepsilon = 1$) subject to \eqref{eq:97}.  The dashed lines
    are the DSW's space-time boundaries.}
  \label{fig:dsw_soliton_trapping_trajectories}
\end{figure}
In the DSW's soliton limit,
\begin{equation}
  \label{eq:77}
  \lim_{\lambda_4 \to \lambda_5} v_{23} = 2\lambda_1 + 4 \lambda_{23},
\end{equation}
which is the speed of the soliton associated with $\lambda_{23}$ on
the mean $\lambda_1$.  When $\lambda_4 \to \lambda_{23}$, we obtain
\begin{equation}
  \label{eq:76}
  v_* \equiv \lim_{\lambda_4 \to \lambda_{23}} v_{23} = 2(\lambda_1 + \lambda_{23} +
  \lambda_5) - 4 (\lambda_{23} - \lambda_1)
  \frac{(1-m_*)K(m_*)}{E(m_*)-(1-m_*)K(m_*)} , \quad m_* =
  \frac{\lambda_{23} - \lambda_1}{\lambda_5 - \lambda_1} .
\end{equation}
This is precisely the DSW modulation velocity
$v_4(-1,\lambda_{23},0) = v_*$ (cf.~$v_2$ in eq.~\eqref{eq:19}).
Since $v_4(-1,\lambda_4,0) < v_*$ for all $\lambda_4 < \lambda_{23}$,
the soliton cannot exit the DSW.  It is trapped within the interior of
the DSW and propagates no slower than $v_*$.  In fact, this is not a
true soliton solution; rather, it corresponds to the scenario of a
pseudo soliton that is described in Sec.~\ref{IST_section}.

Figure \ref{fig:soliton_dsw_trapping_spectrum}(a) shows the evolution
of a soliton with initial amplitude $0< a_+ < 2$ in front of the
negative step $-c^2 = -1$.  The step generates a DSW that, upon
interaction with the soliton, exhibits a defect.  The defect manifests
as a localized depression in the DSW's envelope that resembles a dark
envelope solitary wave.  These are dark breather solutions of the KdV
equation \cite{Kuznetsov1975}.  Although the dark breather migrates
closer to the DSW harmonic, trailing edge, it remains trapped within
the DSW.

Figure \ref{fig:soliton_dsw_trapping_spectrum}(b) depicts the
evolution of the spectrum $\mathcal{S}_{2}^{(23)}$ according to the GP
DSW modulation for $\lambda_4$ while the remaining $\lambda$s are
constant.  When $\lambda_4(x/t) < \lambda_{23}$, the modulation
spectrum exhibits a merger into the 1-phase spectrum $(-\infty,-1]
\cup [\lambda_4(x/t),0]$.  This merger distinguishes soliton-DSW
trapping from transmission.

\begin{figure}
  \centering
  \includegraphics[scale=0.4]{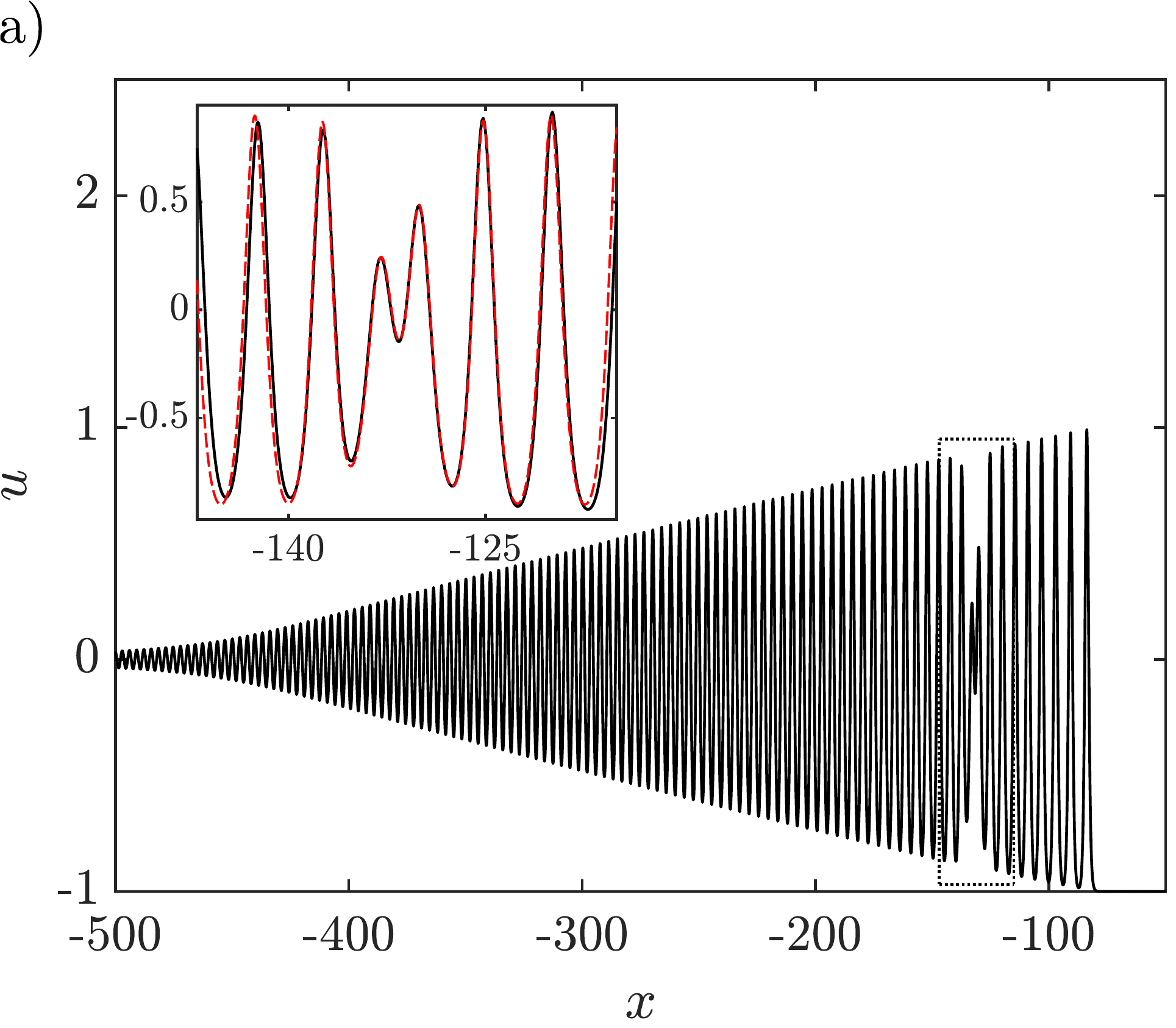}
  \includegraphics[scale=0.4]{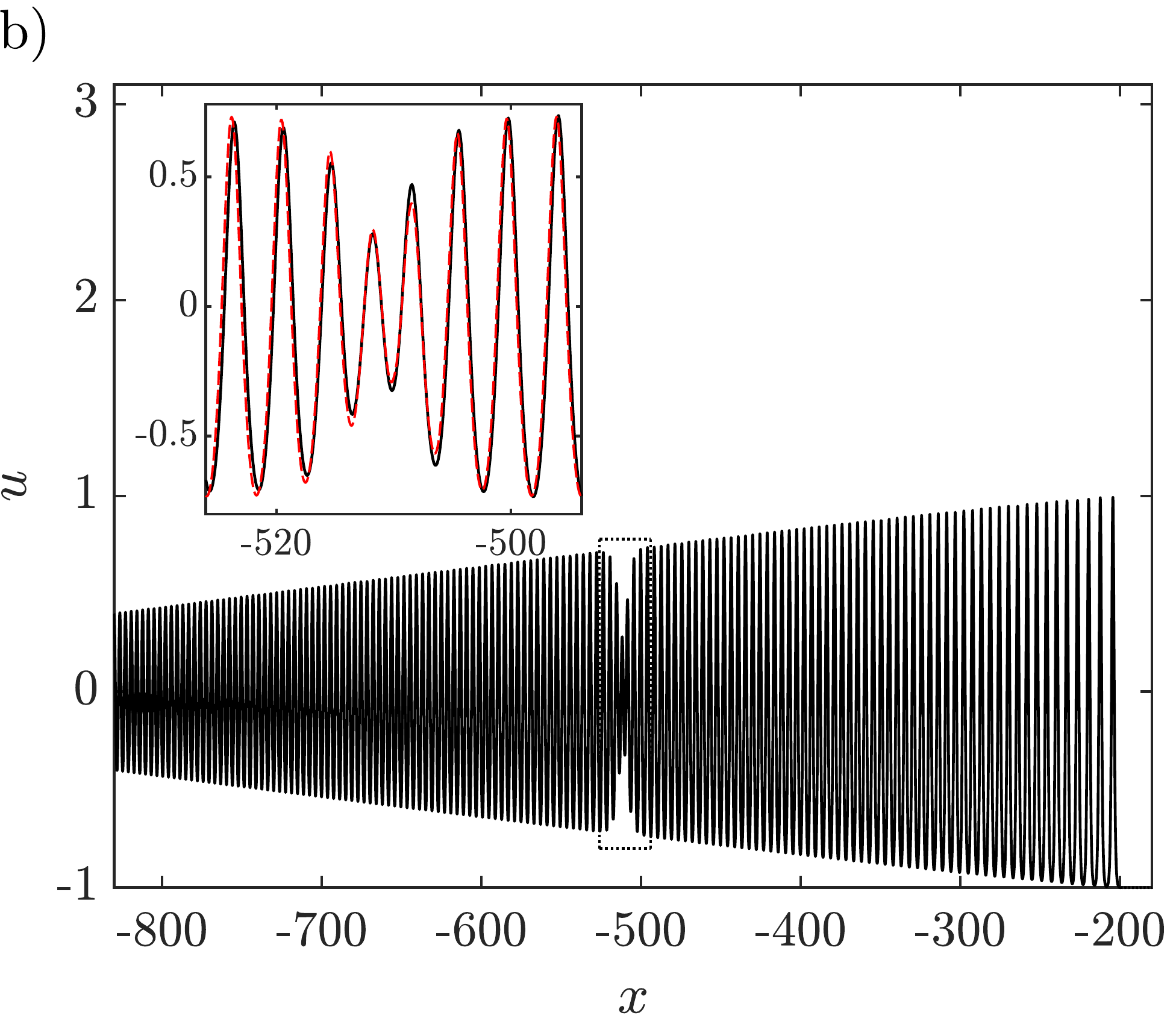}
  \caption{Numerical simulation of soliton-DSW trapping for $a_+ = 1$
    ($\lambda_{23} = -0.5$) with 2-phase local description.  Insets:
    Local description of the simulation (solid, black curve) by the
    2-phase solution with spectrum $\mathcal{S}_2^{(23)}(t) =
    (-\infty,-1]\cup \{\lambda_{23}\} \cup [\lambda_4(x_s(t)/t),0]$
    (dashed, red curve). a) Snapshot at $t = 40$. b) Snapshot at $t =
    100$.}
  \label{fig:soliton_dsw_trapping_2phase}
\end{figure}
The soliton's trajectory is completely determined by the
characteristic
\begin{equation}
  \label{eq:94}
  \frac{\mathrm{d} x_s}{\mathrm{d} t} =
  v_{23}(-1,\lambda_{23},\lambda_{23},\lambda_4(x_s/t),0), \quad
  x_s(0) = x_+.
\end{equation}
So long as $\lambda_{23} < \lambda_4 \le 0$, $v_{23} < v_4$ so that a
soliton initially located at $x = x_+ > 0$, will necessarily interact
with the DSW forming behind it.  Prior to soliton-DSW interaction, the
spectrum is doubly degenerate so that we identify
(cf.~eq.~\eqref{eq:23})
\begin{equation}
  \label{eq:98}
  \lambda_4 \to \lambda_5: \quad \lambda_1 = \overline{u}_+ = -1,
  \quad \lambda_{23} = \frac{a_+}{2} + \overline{u}_+ 
\end{equation}
then the soliton velocity $v_{23}$ is \eqref{eq:77}.  The second
eigenvalue $\lambda_{45}$ corresponds to the DSW's soliton leading
edge, which, from \eqref{eq:36}, is $\lambda_{45} = \overline{u}_-$.
Therefore, the requirement on the initial soliton amplitude $a_+$ for
soliton-DSW trapping is
\begin{equation}
  \label{eq:93}
  \lambda_{23} < \lambda_{45} \quad \iff \quad a_+ < 2(\overline{u}_-
  - \overline{u}_+) .
\end{equation}

Trapped soliton trajectories, extracted from numerical simulations of
KdV, are favorably compared in
Fig.~\ref{fig:dsw_soliton_trapping_trajectories} with the
characteristics \eqref{eq:94}.  The local, 2-phase description of a
trapped soliton is shown in
Fig.~\ref{fig:soliton_dsw_trapping_2phase}.  As in soliton-DSW
transmission, the solution is locally described by a 2-phase solution
whose spectrum ${\cal S}_{2}^{(23)}$ is determined by the DSW
modulation evaluated on the soliton's trajectory
$\lambda_4(x_s(t)/t)$.  Deviation between the 2-phase solution and the
numerical simulation are due to the DSW modulation, which is not
accounted for here.

During soliton-DSW interaction, the trapped soliton's velocity
$v_{23}$ decreases as $\lambda_4$ approaches $\lambda_{23}$.  For
$\lambda_4$ sufficiently close to $\lambda_{23}$, we can estimate the
trapped soliton's propagation.  The modulation velocities admit the
asymptotic expansions
\begin{equation}
  \label{eq:96}
  \begin{split}
    v_{23} &\sim v_* + \frac{1}{3}
    G(\lambda_1,\lambda_{23},\lambda_5)(\lambda_4-\lambda_{23}) +
    \cdots \\
    v_4 &\sim v_* +
    G(\lambda_1,\lambda_{23},\lambda_5)(\lambda_4-\lambda_{23}) +
    \cdots,
  \end{split}
\end{equation}
where
\begin{equation}
  \label{eq:100}
  G(\lambda_1,\lambda_{23},\lambda_5) = 2 K(m_*)
  \frac{(1-m_*)(\lambda_1 - 3 \lambda_{23} + 2
    \lambda_5) K(m_*) - 2(\lambda_1 - 2
    \lambda_{23}+\lambda_5)E(m_*)}{(\lambda_5 - \lambda_1)((1-m_*)K(m_*) -
    E(m_*))^2} .
\end{equation}
Combining these asymptotic expansions with the modulation solution
\eqref{eq:36} evaluated at $x = x_s(t)$, we express the small
parameter as
\begin{equation}
  \label{eq:102}
  \lambda_4 - \lambda_{23} \sim \frac{x_s(t)/t - v_*}{G} .
\end{equation}
Then, expanding the characteristic ODE \eqref{eq:94}, we obtain
\begin{equation}
  \label{eq:101}
  \frac{\mathrm{d} x_s}{\mathrm{d} t} \sim \frac{2}{3} v_* + \frac{x_s(t)}{3t} ,
\end{equation}
which admits the general solution
\begin{equation}
  \label{eq:103}
  x_s(t) \sim v_* t + C t^{1/3}, \quad C \in \mathbb{R} .
\end{equation}
The trapped soliton's velocity, as $\lambda_4$ approaches
$\lambda_{23}$, asymptotes to the interior DSW modulation velocity
$v_*$ as $t \to \infty$ with correction proportional to $t^{-2/3}$.
 

\subsection{Linear Wavepacket-DSW interaction:  2-phase modulations}
\label{sec:line-wavep-dsw}

In the case of soliton-DSW interaction, we considered the collapsed
spectra $\mathcal{S}_2^{(2\to 3)}$ and $\mathcal{S}_2^{(4 \to 5)}$.
It turns out that the consideration of the merged spectra
$\mathcal{S}_2^{(1\to 2)}$ and $\mathcal{S}_2^{(3 \to 4)}$, depicted
in Fig.~\ref{fig:2-phase-spectrum}, can be used to describe the
interaction of a linear wavepacket and a DSW.  For completeness, we
briefly report the degenerate 2-phase modulation velocities and refer
the reader to \cite{congy} for more information on the application to
wavepacket-DSW interaction.

First, we compute the limit $\lambda_1 \to \lambda_2 = \lambda_{12}$:
\begin{equation}
  \label{eq:69}
  \begin{split}
    v_{12} &\equiv \lim_{\lambda_1 \to \lambda_{2}} v_1 =
    \lim_{\lambda_1 \to
      \lambda_{2}} v_2 \\
    &= V_{345} - 4 K(m_{345}) \frac{3 \lambda_{12}^2 - \lambda_{12} V_{345}
      + \lambda_3 \lambda_4 + \lambda_3 \lambda_5 + \lambda_4
      \lambda_5}{(\lambda_5 - \lambda_{12}) K(m_{345}) - (\lambda_5 -
      \lambda_3) E(m_{345})} .
  \end{split}
\end{equation}
The DSW modulation corresponds to the variation of $\lambda_3$,
$\lambda_4$, and $\lambda_5$.  In the DSW harmonic and soliton limits,
we find $\lim_{\lambda_4 \to \lambda_3} v_{12} = 12 \lambda_{12} - 6
\lambda_5$ and $\lim_{\lambda_4 \to \lambda_5} v_{12} = 12
\lambda_{12} - 6 \lambda_3$, respectively, which corresponds to the
linear group velocity $\omega_k = 6 \overline{u} - 3 \varepsilon^2 k^2$
where $(k,\overline{u}) = (2\sqrt{\lambda_5 -
  \lambda_{12}},\lambda_5)$ and $(k,\overline{u}) = (2\sqrt{\lambda_3
  - \lambda_{12}},\lambda_3)$, respectively.  Because both of these
limits exist, it corresponds to the case of linear wavepacket-DSW
transmission.

Next, we compute the limit $\lambda_3 \to \lambda_4 = \lambda_{34}$:
\begin{equation}
  \label{eq:81}
  \begin{split}
    v_{34} &\equiv \lim_{\lambda_3 \to \lambda_4} v_3 = \lim_{\lambda_3
      \to \lambda_4} v_4 \\
    &= V_{125} - 4 K(m_{125}) \frac{3 \lambda_{34}^2 - \lambda_{34} V_{125} +
      \lambda_1 \lambda_2 + \lambda_1 \lambda_5 + \lambda_2
      \lambda_5}{(\lambda_5 - \lambda_{34})K(m_{125}) - (\lambda_5 -
      \lambda_1)E(m_{125})} .
  \end{split}
\end{equation}
In this case, only the DSW harmonic limit corresponds to the linear
group velocity $\lim_{\lambda_2 \to \lambda_1} v_{34} = 12
\lambda_{34} - 6 \lambda_5$ where $(k,\overline{u}) =
(2\sqrt{\lambda_5 - \lambda_{34}},\lambda_5)$.  The DSW soliton limit
$\lambda_2 \to \lambda_5$ cannot be reached.  Instead,
$\lim_{\lambda_2 \to \lambda_{34}} v_{34}$ is the 1-phase modulation
velocity corresponding to $v_2$ in \eqref{eq:19}.  This corresponds to
linear wavepacket-DSW trapping.

\subsection{Discussion}
\label{sec:discussion}

The fact that the soliton's trapping or transmission, amplitude
\eqref{eq:84}, and phase shift \eqref{eq:87} post dispersive
hydrodynamic interaction only depend upon the boundary conditions and
not on the details of the intermediate hydrodynamic state itself is a
reflection of the isospectrality of the KdV equation.  It is a
generalization of soliton-soliton interaction and has been shown to
approximately hold for a large class of nonlinear dispersive wave
equations \cite{Hoefer2}.  This property, termed \textit{hydrodynamic
  reciprocity}, does not rely on the particular details of 2-phase
modulation theory we have used here.  Rather, all that is needed is a
degenerate 1-phase modulation theory, which yields the existence of
two adiabatic invariants that provide the needed relations.  But this
theory only describes the soliton pre and post hydrodynamic
interaction.

Two phase modulation theory provides a detailed description of the
solution structure and soliton trajectory during DSW interaction.
When one of the bands in the 2-phase spectrum degenerately collapses,
the new spectrum consists of a 1-phase spectrum, whose modulation
corresponds to the DSW, and a single point corresponding to the
soliton.  This degenerate 2-phase solution is a KdV breather.  The
modulation velocity corresponding to the degenerate point spectrum
determines the trajectory of the breather as it moves through the DSW.
The soliton's initial position $x_0$ and the location of the
degenerate point in the spectrum determine the type of soliton-DSW
interaction (transmission or trapping) and polarity of the breather
(elevation/bright or depression/dark).  The local structure of the
soliton-DSW interaction is well-approximated by a breather solution.

While we used the degenerate Whitham velocities \eqref{eq:70} and
\eqref{eq:74} to determine the trajectories of a soliton interacting
with a DSW that results from step initial data, they can also be used
to describe soliton propagation in other modulated 1-phase fields.  For
example, a localized disturbance \cite{El2002}, a DSW interacting with
a RW \cite{El2002,Baldwin2013}, or a DSW resulting from a cubic
wavebreaking profile \cite{Gurevich}.  These modulation velocities
could also be used to describe the boundaries of modulated 2-phase
DSW-DSW interactions \cite{Grava2002,Baldwin2013}.

\section{Inverse Scattering Transform}
\label{IST_section}

In this section, the IST is applied to (\ref{kdv}), which is an
integrable equation. Unlike the other methods considered in this work,
this approach is {\it exact}. The overview of the method is to
construct eigenfunctions and scattering data for the time-independent
Schr\"odinger scattering problem, i.e., the first half of the Lax pair
(\ref{E:lax 1}).  The potential of the Schr\"odinger operator is
reconstructed through solving an associated integral equation. The
time dependence of the scattering data is derived from the second half
of the Lax pair (\ref{E:time evolution}), which yields the time
evolution of the potential, i.e., the solution of the KdV equation. In
this work, the potentials are not reflectionless, yet we still find
discrete eigenvalues.  Taking the far-field limit, when the soliton is
well to the left or right of either a RW or DSW, we obtain
exponentially accurate formulas that are found to agree with
those obtained in the previous two sections by asymptotic expansions.

The IST for the KdV equation with rapidly decaying data relies on
developing the inverse scattering associated with the time-independent
Schr\"odinger equation (\ref{E:lax 1}),
cf.~\cite{Faddeev67,Deift79,GGKM,Marchenko2011}. For step-like
potentials without solitons, the inverse scattering for the
time-independent Schr\"odinger equation was first analyzed in
\cite{BF62}. Over the years, the inverse scattering has been
rigorously investigated and used to analyze the KdV equation for
step-like initial data,
cf.~\cite{Cohen84,CohenKaep85,Atko99,Tesch2015}.

\subsection{General Scattering Theory Formulation}

The linear Lax pair associated with (\ref{kdv}) is 
\begin{equation}
\label{E:lax 1}
v_{xx}+\left(\frac{u(x,t)}{\varepsilon^{2}}+\frac{k^{2}}{\varepsilon^{2}}\right)v=0 ~ ,
\end{equation}
\begin{equation}
\label{E:time evolution}
v_{t}=\left(u_{x}+\gamma\right)v+\left(4k^{2}-2u \right)v_{x} ~ ,
\end{equation}
where $k$ is a time-independent spectral parameter and $\gamma$ a
constant. In order to satisfy the compatibility condition
$(v_{xx})_t = (v_t)_{xx}$, the potential $u(x,t)$ must solve
(\ref{kdv}). Hence, the goal of this section is to find eigenvalues
and eigenfunctions associated with the Schr\"odinger operator
corresponding to step up/down BCs and from them reconstruct the
potential.  The eigenfunctions of (\ref{E:lax 1}) are characterized by
their asymptotic behavior, cf.~\cite{Ablowitz1},
\begin{align}
\label{E:asymptotic 1}
& \phi(x,k)\sim e^{-\frac{i k x}{\varepsilon}}, \ \ \ \overline{\phi}(x,k)\sim e^{\frac{ i k x}{\varepsilon}} ~ , ~~~~ \text{as}~x\rightarrow -\infty ~, \\
\label{E:asymptotic 2}
& \psi(x,\lambda)\sim e^{\frac{i \lambda x}{\varepsilon}}, \ \ \ \overline{\psi}(x,\lambda)\sim e^{-\frac{i \lambda x}{\varepsilon}} ~ , ~~~~ \text{as}~x\rightarrow \infty ~ ,
\end{align}
where $\lambda(k) \equiv \sqrt{k^{2} \pm c^{2}}$, depending on the
boundary condition as $x \rightarrow \infty$. Specifically, for the
step up BCs ($+c^2$), $\lambda = \sqrt{k^{2} + c^{2}}$ and we take the
branch cut $k\in [-ic, ic]$ so that
$\text{sgn}( \Im k) = \text{sgn} (\Im \lambda)$. The values of
$\lambda$ along the branch cut are located in the interval $ [-c,
c]$. If, on the other hand, we take step down BCs ($-c^2$), then
$\lambda=\sqrt{k^{2} - c^{2}}$. In this case, we take the branch cut
of $\lambda(k)$ to be $k\in [-c, c]$, which in the $\lambda$-plane
corresponds to the imaginary interval $ [-ic, ic]$ and again
$\text{sgn}( \Im k) = \text{sgn} (\Im \lambda) .$

The left (linearly independent) eigenfunctions can be expressed as a
linear combination of the right eigenfunctions in the form
\begin{align}
\label{E:linear combination 1}
& \phi(x,k)=b(k)\psi(x,\lambda(k))+a(k)\overline{\psi}(x,\lambda(k)) ~ ,  \\
\label{E:linear combination 2}
& \overline{\phi}(x,k)=\overline{a}(k)\psi(x,\lambda(k))+\overline{b}(k)\overline{\psi}(x,\lambda(k)) ~ ,
\end{align}
where $k\in\mathbb{R}$. The coefficients are called the left
scattering data and they are computed via
\begin{equation}
\label{E: scattering data}
a(k)=\frac{\varepsilon}{2i\lambda}W(\phi, \psi) ~ , \  \ b(k)=\frac{\varepsilon}{2i\lambda}W(\overline{\psi},\phi) ~ , \ \ \overline{a}(k)=\frac{\varepsilon}{2i\lambda}W(\overline{\psi} , \overline{\phi}) ~ , \ \  \overline{b}(k)=\frac{\varepsilon}{2i\lambda}W(\overline{\phi},\psi) ~ ,
\end{equation}
where $W(f,g) \equiv f g_x - f_x g $. The scattering data can be
expressed in terms of either $k$ or $\lambda$, which are related by
$\lambda^2 = k^2 \pm c^2,$ the $\pm$ sign depending on the BCs. On the
other hand, the right eigenfunctions can be expressed in terms of the
left eigenfunctions as
\begin{align}
\label{E:linear combination 3}
& \psi(x,\lambda)=\alpha(k)\overline{\phi}(x,k(\lambda))+\beta(k)\phi(x,k(\lambda)) ~ , \\
 \label{E:linear combination 4}
 & \overline{\psi}(x,\lambda)=\overline{\alpha}(k)\phi(x,k(\lambda))+\overline{\beta}(k)\overline{\phi}(x,k(\lambda)) ~ ,
\end{align}
where $\alpha(k)$, $\overline{\alpha}(k)$, $\beta(k)$ and
$\overline{\beta}(k)$ are the right scattering data. When we analyze
Eqn.~(\ref{E:linear combination 1})--(\ref{E:linear combination 2}) we
say that we are solving the left scattering problem, and
(\ref{E:linear combination 3})--(\ref{E:linear combination 4}) is the
right scattering problem.  The left and right scattering data are
related by
\begin{equation}
\overline{\alpha}(k)=\frac{\lambda}{k}\overline{a}(k) ~ , \ \ \ \overline{\beta}(k)=-\frac{\lambda}{k}b(k) ~ , \ \ \
\beta(k)=-\frac{\lambda}{k}\overline{b}(k) ~ , \ \ \ \alpha(k)=\frac{\lambda}{k} a(k) ~ .
\end{equation}
For values of $k, \lambda \in \mathbb{R} $, it can be shown that the
eigenfunctions possess the following symmetries
\begin{equation}
\label{symmetries}
\psi(x,-\lambda)=\overline{\psi}(x,\lambda) =\psi^{*}(x,\lambda) ~ , \ \ \ \phi(x,-k)=\overline{\phi}(x,k) =  \phi^{*}(x,k) ~ .
\end{equation}

Below we derive a set of integral equations from the left
(\ref{E:linear combination 1}) and right (\ref{E:linear combination
  3}) scattering problems and with them compute solutions, where
possible. A complete description of the soliton when it is inside
either a RW or DSW is technical. Instead, we focus on soliton modes
either well before or well after the interaction with either the RW or
DSW. From this we are able to derive {asymptotic} 
soliton formulas with
phase shift as well as give the 
transmission condition in terms
of spectral parameters.

Both the step up and step down cases involve a similar calculation,
hence we present them simultaneously with differences noted.  The main
difference between the two problems is the way in which the branch cut
is taken for the different spectral parameter conditions. As a result,
an additional term can appear in the integral equation that describes
the system. We only consider phase shifts for the 1-soliton case
because we are focused on the interaction of a single soliton with a
RW or DSW mean field.  Multi-soliton phase shifts can be calculated
with a little more effort.

\subsubsection{Right Scattering Problem}

In this section, we consider the right scattering problem
(\ref{E:linear combination 3}), where the right eigenfunctions are
expressed as a linear combination of the left eigenfunctions. Doing so
gives us insight into soliton modes that are initially placed {well}
to left of the jump. To consider both boundary conditions 
simultaneously we use the notation $\pm$, where the top sign
corresponds to the BC $u \rightarrow +c^2$ as $x \rightarrow \infty$
and the bottom sign to the BC $u \rightarrow - c^2 $ as
$x \rightarrow \infty$.

Define the right reflection coefficient $\widetilde{\rho}(k; t) \equiv -\frac{\overline{b}(k; t)}{a(k; t)}$ in terms of the scattering data given in (\ref{E: scattering data}). Then (\ref{E:linear combination 3}) is rewritten as
\begin{equation}
\label{E:RH'}
\frac{\psi(x,\lambda(k))}{a(k)}=\frac{\lambda}{k}\left[ ~\overline{\phi}(x,k)+\widetilde{\rho}(k)   \phi(x,k)\right] ~ ,
\end{equation}
where we have suppressed the time-dependence.
We assume that $\phi$ and $\overline{\phi}$ have the following triangular forms
\begin{equation}
\label{E:integral equation 3}
\phi(x,k)=e^{-\frac{i k x}{\varepsilon}}+\int_{-\infty}^{x}\widetilde{G}(x,s,t)e^{-\frac{i k s }{\varepsilon}}ds ~ ,
\end{equation}
\begin{equation}
\label{E:integral equation 4}
\overline{\phi}(x,k)=e^{\frac{i k x}{\varepsilon}}+\int_{-\infty}^{x}\widetilde{G}(x,s,t)e^{\frac{i k s }{\varepsilon}}ds ~ ,
\end{equation}
with $\widetilde{G}(x,s,t)\equiv 0$ for $x>s$. 
From these integral forms one can show through a Green's function that $\psi(x;k)$ and $\phi(x,k)$ are analytic in the upper half plane of both $k$ and $\lambda$. Consequently, the scattering data $a(k)$ defined in (\ref{E: scattering data}) is also analytic in the upper half plane. Next we substitute (\ref{E:integral equation 3}) and (\ref{E:integral equation 4}) into (\ref{E:RH'}), and apply the Fourier transform $\frac{1}{2\pi \varepsilon}\int_{-\infty}^{\infty}e^{- \frac{i \lambda y}{\varepsilon}}d\lambda$ for $y < x$ and exploit the analyticity of $\phi(x,k), \psi(x,\lambda(k))$ and $a(k)$ in the upper half plane. The resulting Gel'fand-Levitan-Marchenko (GLM) equation obtained is 
\begin{equation}
\label{GLM4}
\widetilde{G}(x,y,t)+\widetilde{\Omega}(x+y,t)+\int_{-\infty}^{x}\widetilde{\Omega}(s+y,t)\widetilde{G}(x,s,t)ds=0 ~ ,
\end{equation}
where
\begin{equation}
\label{kernelC}
\widetilde{\Omega}(z,t)= \frac{1}{2\pi \varepsilon}\int_{-\infty}^{\infty} \widetilde{\rho}(k;t)e^{-\frac{ i k z}{\varepsilon}}dk
- \widetilde{C}(t) e^{-\frac{ \kappa_0 z}{\varepsilon}} + \overline{\Omega}_{\pm}(z,t) ~ ,
\end{equation}
with normalization constant
{\begin{equation}
\widetilde{C}(t) =-\frac{i \overline{b}({k}; t)}{\varepsilon \frac{ \partial a }{\partial k}({k}; t) } \bigg|_{k =  i k_0} ~ , ~~~~~~ k_0 > 0 
\end{equation}}
and we remind the reader that $\pm$ represents the step up(+)/down(-) cases. 
The {eigenvalue $ k = i k_0$} is a simple zero of $a(k)$. A zero exists
when $\kappa_0 > c$ in the step up case and any $\kappa_0 > 0$ in the
step down scenario. {The eigenvalues satisfy $k_0 \ge \kappa_0$ with equality at $\frac{\kappa_0 x_0}{\varepsilon} = -\infty$.}  Note that $ \overline{\Omega}_{-}(z,t) \equiv 0 $
in (\ref{kernelC}) and
\begin{equation}
\label{omega_plus}
\overline{\Omega}_{+}(z,t) = \frac{1}{2\pi \varepsilon}\int_{0}^{c}\frac{\kappa  e^{\frac{\kappa z}{\varepsilon}} }{\sqrt{c^2 - \kappa^2} |a(i\kappa;t)|^{2}}d\kappa ~ .
\end{equation}
This additional term is due to the branch cut $[-ic , ic]$ in the upper half $k$-plane for the step up boundary condition; the step down boundary condition has a branch cut lying on the real axis.
The solution to the GLM equation in (\ref{GLM4}) is related to the potential  by
\begin{equation}
\label{kdv4}
u(x,t)=-2\varepsilon^{2}\frac{d}{dx}\widetilde{G}(x,x,t) ~ .
\end{equation}

From the  temporal equation of the Lax pair (\ref{E:time evolution}) we obtain the time-evolution of the scattering data where $\gamma = \frac{2 i \lambda}{\varepsilon} (- 2 k^2 \pm c^2)$ and 
\begin{align}
& a_{\pm}(k;t)=a_{\pm}(k;0)e^{i (4k^{3}-4k^{2}\lambda_{\pm}  \pm  2c^{2}\lambda_{\pm} )t/\varepsilon} ~ , \\ \nonumber
&  \overline{b}_{\pm}(k;t) = \overline{b}_{\pm}(k;0)e^{ - i (4k^{3}+4k^{2}\lambda_{\pm}  \pm  2c^{2}\lambda_{\pm})t/\varepsilon} ~ ,
\end{align}
such that $\lambda_{\pm}(k) = \sqrt{k^2 \pm c^2}$ corresponds to step up and step down, respectively. From this we obtain the time-evolution of the reflection coefficient
\begin{equation}
\widetilde{\rho}_{\pm}(k;t) = - \frac{\overline{b}_{\pm}( k; t)}{a_{\pm}(k;t)}   = \widetilde{\rho}_{\pm}(k;0)e^{-8ik^{3}t/\varepsilon} ~ ,
\end{equation}
and normalization constant
{\begin{equation}
\widetilde{C}_{\pm}(t)= - \frac{i \overline{b}_{\pm}(k; t)}{\varepsilon \frac{\partial a_{\pm}(k; t)}{\partial k} } \bigg|_{k = i k_0} = \widetilde{C}_{\pm}(0)e^{-8 k_0^3t/\varepsilon} ~ . 
\end{equation}}

In order to obtain a solution to (\ref{GLM4}) we fix $t= 0$ and then
let $z \rightarrow - \infty$ so that the continuous spectrum in
(\ref{kernelC}) goes to zero due to the Riemann-Lebesgue
Lemma. Additionally, if we place a soliton initially far to the left
of the origin 
($x_0 \ll -\varepsilon/\kappa_0$), then the branch cut term
(\ref{omega_plus}) also decays to zero exponentially fast leaving only
the discrete spectrum term. Solving the resulting system for
$ \kappa_0 > 0 $ we find
\begin{equation}
  \label{KdvSol2BB}
  u(x,t)\sim 2\kappa_0^{2}~ \text{sech}^{2}
  \left[\frac{\kappa_0}{\varepsilon}
    (x-4\kappa_0^{2}t-x_{0}^{\pm})\right] ~ , ~~~ x_0^{\pm} =
  \frac{\varepsilon}{2 \kappa_0} \log\left( -\frac{ 2
      \kappa_0}{\varepsilon \widetilde{C}_{\pm}(0)} \right)  ~ .
\end{equation}
This is the form of a soliton when it is far to the left of the initial jump {($k_0 \approx \kappa_0$)} and any subsequent radiation it emits.

\subsubsection{Left Scattering Problem}
\label{sec_IST_left_scatter}

Let us next consider the left scattering problem (\ref{E:linear combination 1}), where the left eigenfunctions are expressed as a linear combination of the right eigenfunctions. Here we will obtain a formula for the soliton when it is well to the right of the jump,  {where the radiation from the continuous spectrum is exponentially small.}
These formula describe a soliton which transmits left-to-right through either a RW or DSW. Again, when we write $\pm$, the top (bottom) sign corresponds to the step up (down) boundary condition at $x \rightarrow \infty$.

Begin by defining the reflection coefficient $\rho(\lambda; t) \equiv \frac{b(\lambda; t)}{a(\lambda; t)}$ in terms of the scattering data (\ref{E: scattering data}). Then (\ref{E:linear combination 1}) can be rewritten as
\begin{equation}
\label{E:RH}
\frac{\phi(x,k(\lambda))}{a(\lambda)}=\rho(\lambda)\psi(x; \lambda)+\overline{\psi}(x; \lambda) ~ ,
\end{equation}
where the time dependence has been suppressed.
Next we assume that $\psi$ and $\overline{\psi}$ are {expressed in the Volterra integral equations form}
\begin{equation}
\label{E:integral equation 1}
\psi(x,t; \lambda)=e^{\frac{i \lambda x}{\varepsilon}}+\int_{x}^{\infty}G(x,s,t)e^{\frac{i \lambda s}{\varepsilon}}ds ~ ,
\end{equation}
\begin{equation}
\label{E:integral equation 2}
\overline{\psi}(x,t; \lambda)=e^{-\frac{i \lambda x}{\varepsilon}}+\int_{x}^{\infty}G(x,s,t)e^{-\frac{i\lambda s}{\varepsilon}}ds ~ ,
\end{equation}
with $G(x,s,t)\equiv 0$ when $s<x$. 
 Again, $\phi(x,k(\lambda)), \psi(x,\lambda)$ and $a(\lambda)$ are analytic functions in the upper half plane of $\lambda$. Taking into account this analyticity we substitute (\ref{E:integral equation 1}) and (\ref{E:integral equation 2}) into (\ref{E:RH}) and then operate on the resulting equation by $\frac{1}{2\pi \varepsilon}\int_{-\infty}^{\infty}e^{\frac{i \lambda y }{\varepsilon}}d\lambda$ for $y > x$ to obtain the GLM integral equation
\begin{equation}
\label{E:GLM+}
G(x,y,t)+\Omega(x+y,t)+\int_{x}^{\infty}\Omega(y+s,t)G(x,s,t)ds=0 ~, ~~y > x ~ ,
\end{equation}
with kernel 
\begin{equation}
\label{KernelGLM+}
\Omega(z,t)= \frac{1}{2\pi \varepsilon}\int_{-\infty}^{\infty}\rho(\lambda;t)e^{\frac{i\lambda z}{\varepsilon}}d\lambda
-C(t) e^{- \frac{ \eta_0 z }{\varepsilon}} + \Omega'_{\pm}(z,t) ~ ,
\end{equation}
and normalization constant
{\begin{equation}
\label{norm_const1}
C(t)=\frac{ i b(\lambda; t)}{\varepsilon  \frac{ \partial a}{\partial \lambda}(\lambda; t)  } \bigg|_{\lambda = i \ell_0} ~ ,
 \end{equation}}
where {$\lambda = i \ell_0, \ell_0 >0$} is a simple zero of $a(\lambda)$. In the step up case, $\Omega'_+(z,t) \equiv 0$  and for step down there is an additional branch cut contribution of 
\begin{equation}
\label{omega_neg}
\Omega'_-(z,t) = \frac{1}{2\pi \varepsilon}\int_{0}^{c}\frac{\eta e^{- \frac{\eta z}{\varepsilon}} }{\sqrt{c^2 - \eta^2 } |\alpha(i \eta;t)|^2}   d\eta  ~ .
\end{equation}
The potential can be related to the solution of the GLM equation in (\ref{E:GLM+}) by
\begin{equation}
\label{soln1}
u_{\pm}(x,t)= \pm c^{2}+2\varepsilon^{2}\frac{d}{dx}G(x,x,t) ~ ,
\end{equation}
where the positive (negative) sign is taken in the $+c^2$ ($-c^2$) boundary condition case.

Next we find the time evolution of the relevant scattering data defined in (\ref{E: scattering data}) from (\ref{E:time evolution}). Here $\gamma = \frac{4 i k^3}{\varepsilon}$ and so
\begin{align}
& a_{\pm}(\lambda;t)=a_{\pm}(\lambda;0)e^{i(4k_{\pm}^{3}-4k_{\pm}^{2}\lambda \pm 2c^{2}\lambda)t/\varepsilon} ~ , \\
& b_{\pm}(\lambda;t)=b_{\pm}(\lambda;0)e^{i(4k_{\pm}^{3}+4k_{\pm}^{2}\lambda \mp 2c^{2}\lambda)t/\varepsilon} ~ ,
\end{align}
where $k_{\pm} = \sqrt{\lambda^2 \mp c^2}.$
The reflection coefficient and normalization constant that form the kernel in (\ref{KernelGLM+}) are given, respectively, by
\begin{equation}
\rho_{\pm}(\lambda; t)= \frac{b_{\pm}(\lambda; t)}{a_{\pm}(\lambda; t)} =\rho_{\pm}(\lambda; 0)e^{i(8k_{\pm}^{2} \lambda \mp 4c^{2}\lambda)t/\varepsilon} ~ ,
\end{equation}
and
{\begin{equation}
C_{\pm}(t)=C_{\pm}(0)e^{ \ell_0 ( 8 k_0^2 \pm 4c^2)t/\varepsilon} ~ , ~~~ C_{\pm}(0) = \frac{i b_{\pm}(\lambda; 0)}{ \varepsilon \frac{ \partial a_{\pm}}{\partial \lambda}(\lambda; 0) } \bigg|_{\lambda = i \ell_0}   ~ ,
\end{equation}}
{where $k_0^{\pm} = \sqrt{ \ell_0^2  \pm c^2}.$ CHECK THIS ONE}
Let us consider the case in which {$\ell_0$} is significantly greater than $c$. Then as $t \rightarrow \infty$ the extra term in Eq.~(\ref{omega_neg}) decays in time and the only significant contribution to the kernel is the normalization constant term. From the reconstruction formula (\ref{soln1}) we compute the 1-soliton solution in the $t \rightarrow \infty, x \gg 1$ asymptotic limit
\begin{equation}
\label{KdvSol2A}
u_{\pm}(x,t)\sim {\pm} c^{2}+2(\eta_{0}^{\pm})^{2} ~ \text{sech}^{2} \left[\frac{\eta_{0}^{\pm}}{\varepsilon}\left(x-(4(\eta_{0}^{\pm})^{2} {\pm} 6c^{2})t -x_{s}^{\pm}\right)\right] ~, ~~~ x_{s}^{\pm}=\frac{\varepsilon}{2 \eta_0^{\pm}} \log \left( -\frac{\varepsilon C_{\pm}(0)}{2 \eta_0^{\pm}} \right)  ~ ,
\end{equation}
with $ \eta_0^{\pm} = \sqrt{\kappa_0^2 {\mp} c^2} > 0 $ corresponding
to step up and step down, respectively. {Here, we have taken the scale separation limit so that $\ell_0^{\pm} \rightarrow \eta_0^{\pm} $.} At this point, the continuous
spectrum contribution from the reflection coefficient in
(\ref{KernelGLM+}) is exponentially small. This is what we refer to as
a proper soliton with a phase shift of
$\Delta \equiv x_s^{\pm} - x_0.$

\subsection{Discussion of IST Results}
\label{discuss_IST_sec}

Let us discuss the significance of Eq.~(\ref{KdvSol2A}). Consider an
initial soliton of the form (\ref{KdvSol2BB}) with 
$x_0 \ll -\varepsilon/\kappa_0$ and step up BCs. {Hence, the eigenvalue is $k_0$ ($\ell_0$) is exponentially close to $\kappa_0$ ($\eta_0$), the amplitude parameter.} If $\kappa_0 > c$,
then the soliton will transmit through the resulting rarefaction wave
and it will be of the form given by the positive root in
(\ref{KdvSol2A}). The transmitted soliton will be of smaller amplitude
$( \eta_0^+ < \kappa_0)$, but larger velocity in comparison to the
incident soliton. In terms of the scattering problem (\ref{E:lax 1}),
this mode is a proper soliton and corresponds to a proper discrete
eigenvalue \cite{ALC}. The phase shift for this case is calculated
below.

If on the other hand $\kappa_0 < c$, then the soliton will become
trapped within the rarefaction ramp, never reaching the top. In this
case, formula (\ref{KdvSol2A}) does not apply since there are no
proper eigenvalues of the direct scattering problem; that is,
eigenvalues with corresponding eigenfunctions that decay rapidly to
zero as $|x| \rightarrow \infty$. Instead, the inverse scattering
yields modes that resemble solitons until they reach the rarefaction
ramp \cite{ALC}, hence we call them quasi or pseudo solitons. The
terms of the inverse scattering that yield pseudo solitons were called
pseudo-embedded eigenvalues, however they were not associated with
zeros of scattering data like normal eigenvalues, rather they
corresponded to the dominant contribution of the branch cut integral
(\ref{omega_plus}) for $-\infty < x_0 \ll -\varepsilon/\kappa_0$.
Indeed, a more recent work \cite{Mucalica2022} has identified this
case as corresponding to resonant poles that leave the imaginary
axis. The associated eigenfunctions are not bound states.  To be
clear, proper solitons with true eigenvalues {\it always} transmit
through a rarefaction wave. Finding explicit formula that describe the
dynamics of the soliton while it is on the ramp from an IST
point of view is an open problem.

We point out that if an initial soliton of the form in
(\ref{KdvSol2A}) with fixed $x_0^{+} = x_0 >0$ and $\eta_0^+ > 0$ is
taken, then the soliton will never become trapped since the right edge
of the ramp moves to the right slower than the soliton.

Now let us examine (\ref{KdvSol2A}) in the case of step down
BCs. First, consider an initial soliton of the form (\ref{KdvSol2BB})
with $x_0 \ll -\varepsilon/\kappa_0$. {Again, we replace $\ell_0$ ($k_0$) with $\eta_0$ ($\kappa_0$) in the formulas.} Regardless of the size of
$\kappa_0$, this soliton will tunnel through the resulting DSW and
when it reaches the other side, it will have a form like the negative
root in (\ref{KdvSol2A}). In this case
$\eta_0^- = \sqrt{\kappa_0^2 + c^2} > c$, so this is a proper soliton
associated with a true eigenvalue and
$L^{2}(\mathbb{R})$-eigenfunctions of (\ref{E:lax 1}). The transmitted
soliton will have a larger amplitude $(\eta_0^- > \kappa_0)$, but
smaller velocity in comparison to the initial state. The phase shift
is computed below.

The final case to consider is when a soliton is initially placed to
the right of a step down. The initial condition will look like the
negative root of (\ref{KdvSol2A}) with fixed
$x_0 = x_s^{-} >0, \eta_0 = \eta_0^{-} > 0$ at $ t = 0$.  The soliton
can have positive, negative or zero velocity, depending on the
amplitude relative to the step height. When $4 \eta_0^2 > 6c^2$
($4 \eta_0^2 < 6c^2$) the soliton velocity is positive (negative); if
$4 \eta_0^2 = 6c^2$, the soliton velocity is zero. The DSW that
emanates from the jump will have negative velocity. By the Whitham
approach, the right (solitonic) edge of the DSW region moves with
velocity $- 2 c^2$. A proper soliton corresponds to $\eta_0 > c$ and
has velocity greater than $- 2c^2$; hence a proper soliton never
reaches the right edge of the DSW. This is true even if the soliton
has negative velocity since it is not negative enough to reach the
nonlinear, soliton edge of the DSW. When $\eta_0 < c$ the pseudo
soliton velocity is less than $- 2 c^2$ and there is no true
eigenvalue of (\ref{E:lax 1}) and thus no proper solitons. Here, the
velocity of the {pseudo} soliton is less than $- 2 c^2$ and the
{pseudo} soliton will eventually catch up to the DSW and embed
itself inside it; i.e. the {pseudo} soliton will eventually become
trapped. This is again an instance of a pseudo-embedded eigenvalue
(resonant poles).

\subsection{Spectral Data for a Soliton Mode plus Heaviside Initial Condition}
\label{phase_shift_section}

Consider an initial condition of the form
\begin{equation}
\label{general_IC_soliton}
u(x,t) = 2\kappa_0^{2}~ \text{sech}^{2}\left[\frac{\kappa_0}{\varepsilon}(x - x_{0})\right] \pm c^2 H(x) ~ , ~~~ x_0 < 0 ~ ,
\end{equation}
where $H(x)$ is the Heaviside function \ref{Heaviside}. The positive
(negative) sign corresponds to step up (down) BCs and results in a RW
(DSW). The associated half space initial eigenfunctions to (\ref{E:lax
  1}) for the pure 1-soliton solution are
\begin{align}
\label{eig_func_neg_x}
& \phi(x,0) =  e^{- \frac{i k x}{\varepsilon}} \left[1 - \frac{2 i \kappa_0}{(k + i \kappa_0) \left( 1 + e^{-\frac{2 \kappa_0 (x-x_0)}{\varepsilon}} \right) } \right] ~ , ~~~~  x < 0 ~ , \\ \label{eig_func_pos_x}
& \psi(x,0) =  e^{ \frac{i \lambda x}{\varepsilon}} \left[1 - \frac{2 i \kappa_0}{(\lambda + i \kappa_0) \left( 1 + e^{\frac{2 \kappa_0 (x-x_0)}{\varepsilon}} \right) } \right] ~ , ~~~~  x > 0  ~ .
\end{align}
The function in Eq.~(\ref{eig_func_neg_x}) can be obtained from the
Rieman-Hilbert formulation of the inverse problem with one eigenvalue
and zero reflection coefficient cf.~\cite{Ablowitz1}.  The
eigenfunctions $\overline{\phi}(x,0)$ and $\overline{\psi}(x,0)$ are
obtained from the symmetry relations in (\ref{symmetries}).  The
scattering data is then computed from (\ref{E: scattering data}) and
evaluated at $x = 0$ to yield
\begin{align}
\label{RW_scat_data}
& a_{\pm}(\lambda; 0)  =   \frac{(k_{\pm}  + \lambda) \left[ \kappa_0^2 + k_{\pm}  \lambda - i \kappa_0 (k_{\pm}  - \lambda) \text{tanh}\left( \frac{\kappa_0 x_0}{\varepsilon} \right) \right] }{2 \lambda (k_{\pm}  + i \kappa_0)(\lambda + i \kappa_0)} ~ , \\ \nonumber
& b_{\pm}(\lambda; 0)  =  \frac{(k_{\pm}  - \lambda) \left[ \kappa_0^2 - k_{\pm}  \lambda - i \kappa_0 (k_{\pm}  + \lambda) \text{tanh}\left( \frac{\kappa_0 x_0}{\varepsilon} \right) \right] }{2 \lambda (k_{\pm}  + i \kappa_0)(\lambda - i \kappa_0)} ~ ,
\end{align}
such that $k_{\pm} \equiv \sqrt{\lambda^2 \mp c^2}$ for the step up
and step down cases, respectively.  As a result, the reflection
coefficient is
\begin{equation}
\rho_{\pm}(\lambda; 0) = {\frac{b_{\pm}(\lambda; 0)}{a_{\pm}(\lambda; 0)}}= \frac{(k_{\pm} - \lambda)(\lambda + i \kappa_0) \left[ \kappa_0^2 - k_{\pm} \lambda - i \kappa_0 (k_{\pm} + \lambda)~ \text{tanh}\left( \frac{\kappa_0 x_0}{\varepsilon} \right) \right]}{(k_{\pm} + \lambda)(\lambda -  i \kappa_0) \left[ \kappa_0^2 + k_{\pm} \lambda - i \kappa_0 (k_{\pm} - \lambda) ~\text{tanh}\left( \frac{\kappa_0 x_0}{\varepsilon} \right) \right]} ~ . 
\end{equation}
This reflection coefficient is interesting from an integrability point
of view since it is {\it not} reflectionless and it {\it is} exact. {Let us now consider a soliton initially well-separated from the step, that is $x_0 \ll - \varepsilon/\kappa_0$. In this case, the discrete eigenvalues are exponentially close to $\lambda_{\pm} = i\eta_0^{\pm}  = i \sqrt{\kappa_0^2 \mp c^2}$, where $\kappa_0$ is the amplitude parameter in (\ref{general_IC_soliton}).} The
normalization constant for $\kappa_0 > c$ in the RW case and
$\kappa_0 > 0$ in the DSW case is given by
\begin{align}
\label{norm_constant_exact}
&\varepsilon C_{\pm}(0) = \frac{i b_{\pm}(\lambda; 0)}{ \frac{ \partial a_{\pm}}{\partial \lambda}(\lambda; 0) } \bigg|_{\lambda = i \eta_0}    = \\ \nonumber
& \frac{2 \kappa_0^2 \eta_0^{\pm} ( \kappa_0 + \eta_0^{\pm})^3 \left[ 1 + \tanh\left( \frac{\kappa_0 x_0}{ \varepsilon} \right) \right]}{ - 8 \kappa_0^4 ( \kappa_0 + \eta_0^{\pm} ) -c^4 (5 \kappa_0 + \eta_0^{\pm} ) \pm c^2 \kappa_0^2 (11 \kappa_0 + 5 \eta_0^{\pm} ) + c^2 \left[ c^2 (\kappa_0 + \eta_0^{\pm} ) \mp \kappa_0^2 (3 \kappa_0 + 5 \eta_0^{\pm} ) \right] \tanh\left( \frac{\kappa_0 x_0}{ \varepsilon} \right)} ,
\end{align}
where $ \eta_0^{\pm} = \sqrt{\kappa_0^2 {\mp} c^2} $ for step up and
step down, respectively.  From this 
{formula} 
we are
able to calculate the IST phase shift below in both the step up and
step down BC cases. 

\subsubsection{Scale Separation 
Asymptotics}
\label{small_eps_sec}

There is a direct relationship between the soliton phase in
(\ref{KdvSol2A}) and the asymptotic phases in (\ref{rare_phase_shift})
and (\ref{eq:45}). Namely, in the {limit $\frac{\kappa_0 x_0 }{\varepsilon} \ll -1$} 
the
IST phase shift approaches that of the asymptotics. To see
this, we compute the IST phase $x_s^+$ using the formulae given in
(\ref{general_IC_soliton})-(\ref{norm_constant_exact}) for $x_0 < 0$
with step up BCs. Within the normalization constant given in
Eq.~(\ref{norm_constant_exact}) the only place for {scale separation effects} 
is in the
$\tanh\left( \frac{\kappa_0 x_0}{\varepsilon} \right)$ terms. Taking
the limit {$\frac{\kappa_0 x_0 }{\varepsilon} \rightarrow - \infty$} 
and expanding
$\tanh\left( \frac{\kappa_0 x_0}{\varepsilon} \right) \approx -1 + 2
e^{\frac{2\kappa_0 x_0}{\varepsilon}} $ yields, to leading order, a
normalization constant of
\begin{equation}
\label{norm_RW_smalleps}
C_+(0) \rightarrow - \frac{2 \kappa_0^2 e^{\frac{2 \kappa_0 x_0}{\varepsilon}}}{\varepsilon \eta_0^+} ~ , 
\end{equation}
where $\eta_0^+ = \sqrt{\kappa_0^2 - c^2}$.
Substituting this into the  phase formula of Eq.~(\ref{KdvSol2A}) shows
\begin{equation}
\label{RW_phase_shift_asym}
x_s^+ \rightarrow \frac{\varepsilon}{ 2 \eta_0^+} \log\left( \frac{\kappa_0^2 e^{\frac{2 \kappa_0 x_0}{\varepsilon}}}{(\eta_0^+)^2} \right) = \frac{\varepsilon}{ 2 \eta_0^+} \log\left( \frac{\kappa_0^2 }{(\eta_0^+)^2} \right)  +  \frac{\kappa_0 x_0}{\eta_0^+} \approx \frac{\kappa_0 x_0}{\eta_0^+}  = \frac{x_0}{\sqrt{1 - \left(\frac{c}{\kappa_0} \right)^2}}  ,
\end{equation}
when $\varepsilon $ approaches zero; which is exactly the result in
(\ref{rare_phase_shift}) and (\ref{eq:45}). Hence, both the soliton
perturbation theory and Whitham modulation theory, in the small
dispersion limit, reproduce the exact limiting phase shift of
$\Delta \approx x_0 ( \kappa_0 / \eta_0^+ - 1 ).$

Now consider step down BCs with a soliton placed to the left
($x_0 < 0$) of the jump. To obtain an asymptotic approximation we
again take a small $\varepsilon$ limit in the normalization constant
(\ref{norm_constant_exact}), similar to what was done in the RW case.
Doing so yields
\begin{equation}
  \label{norm_DSW_smalleps}
  C_-(0) \rightarrow - \frac{2 \kappa_0^2 e^{\frac{2 \kappa_0
        x_0}{\varepsilon}}}{\varepsilon \eta_0^-} ~ ,  
\end{equation}
as $\varepsilon \rightarrow 0$ for
$\eta_0^- = \sqrt{\kappa_0^2 + c^2}$. Substituting this into the phase
given in Eq.~(\ref{KdvSol2A}) yields
\begin{equation}
\label{DSW_phase_asym}
x_s^- \rightarrow \frac{\varepsilon}{ 2 \eta_0^-} \log\left( \frac{\kappa_0^2 e^{\frac{2 \kappa_0 x_0}{\varepsilon}}}{(\eta_0^-)^2} \right) \approx \frac{\kappa_0 x_0}{\eta_0^-}  = \frac{x_0}{\sqrt{1 + \left(\frac{c}{\kappa_0} \right)^2}} ~ .
\end{equation}
The resulting phase shift for a soliton through a DSW in the small
dispersion limit is $\Delta \approx x_0 (\kappa_0 / \eta_0^- - 1).$
Notice that this is exactly the phase shift produced by the Whitham
approach in (\ref{eq:87}).  In comparison to the RW asymptotic phase
shift in Eq.~(\ref{RW_phase_shift_asym}), the only difference is the
sign under the square root.

Using the asymptotic phase shift in (\ref{DSW_phase_asym}) for the
transmitted soliton ($x_0 < 0$) along with the leading (rightmost)
edge of the DSW region ($x = -2c^2 t$) obtained from Whitham theory,
one can predict the time and position that the soliton will exit the
DSW region. Namely, the soliton peak intersects the leading
(solitonic) edge of the DSW region when
$z(T_2) = -2c^2 T_2 = 4(\eta_0^-)^2 t + x_s^-$, or at time and
position
\begin{equation}
  \label{exit_time}
  T_2 = - \frac{x_0}{4 \kappa_0^2 \sqrt{1 + \frac{c^2}{\kappa_0^2}}} ,
  ~~~ z(T_2) = \frac{x_0 c^2}{2 \kappa_0^2 \sqrt{1 +
      \frac{c^2}{\kappa_0^2}} } , 
\end{equation}
respectively. These formulas were found to agree well with numerics.

Furthermore, the amount of time a tunneling soliton spends in the DSW
region is approximately
\begin{equation}
T_2-T_1 =   - x_0 \left(  \frac{1}{4 \kappa_0 \sqrt{\kappa_0^2 + c^2}} - \frac{1}{4 \kappa_0^2 + 12 c^2 } \right) ,
\end{equation}
where the incident time is given in (\ref{time_T1}). Asymptotically,
this quantity goes to infinity as $x_0 \rightarrow -\infty$ or
$\kappa_0 \rightarrow 0$. On the other hand, the time interval goes to
zero as $x_0 \rightarrow 0^-$ or $c \rightarrow \infty$ or
$c \rightarrow 0$ or $\kappa_0 \rightarrow \infty$.

\subsection{IST Results}

In this section, the results obtained from the IST are highlighted. It
should be emphasized that these results are  {exponentially accurate} 
for solitons
well-clear of the RW or DSW. Where relevant, comparison with the
asymptotic results obtained in Secs.~\ref{sol_pert_th} and
\ref{sec:solit-disp-hydr} are shown.

\subsubsection{Soliton-RW Results}

The phase shift, $\Delta = x_s^+ - x_0$, of a soliton that has
tunneled through a RW is analytically calculated in two ways. First,
in Eq.~(\ref{KdvSol2A}) the IST phase shift is found through the
normalization constant in (\ref{norm_constant_exact}). The other means
of approximating the phase shift is through the asymptotics and given
in Eqs.~(\ref{rare_phase_shift}) and (\ref{eq:45}).

\begin{figure} [h]
  \centering
  \includegraphics[scale=0.3]{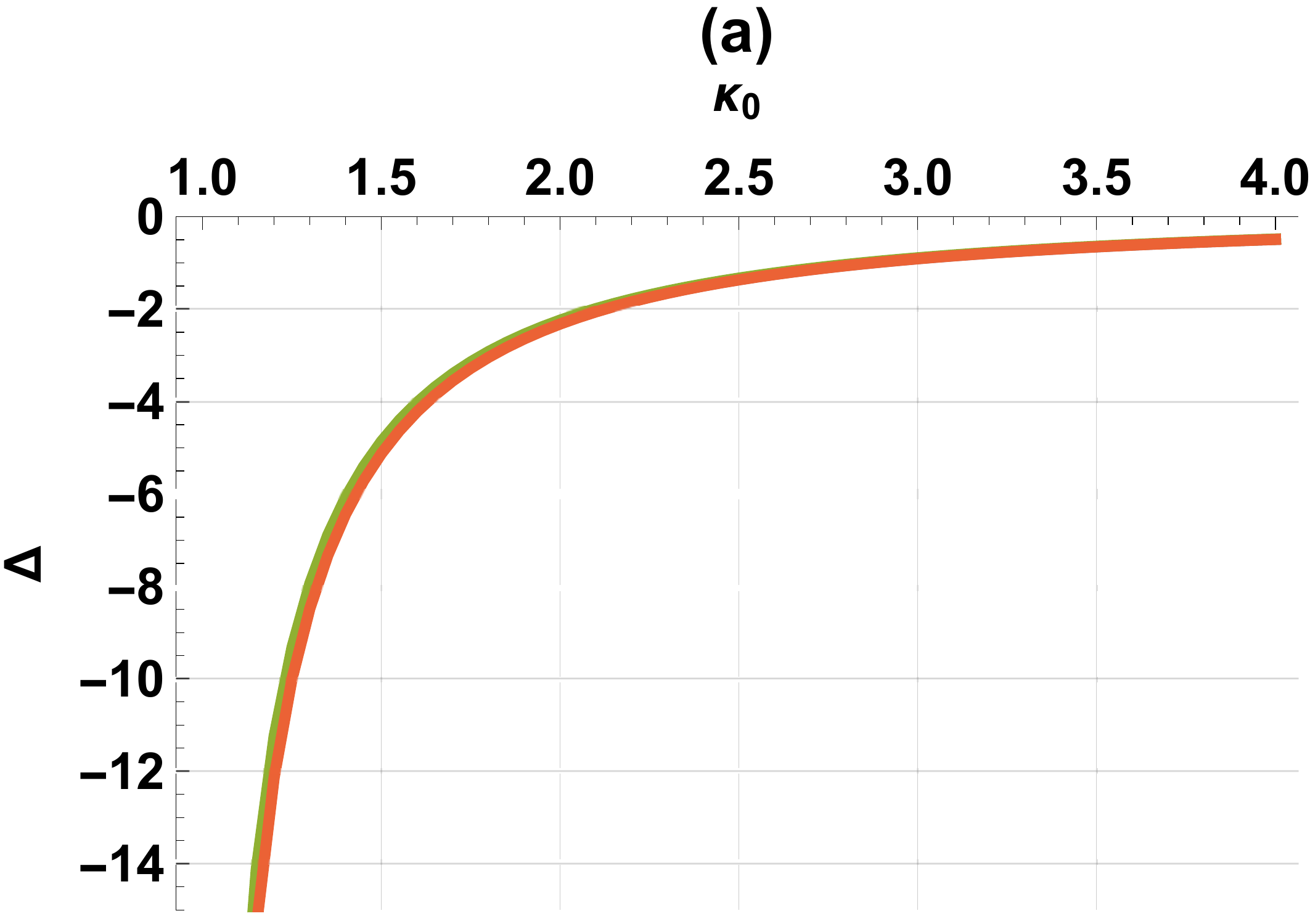}
  \includegraphics[scale=0.4]{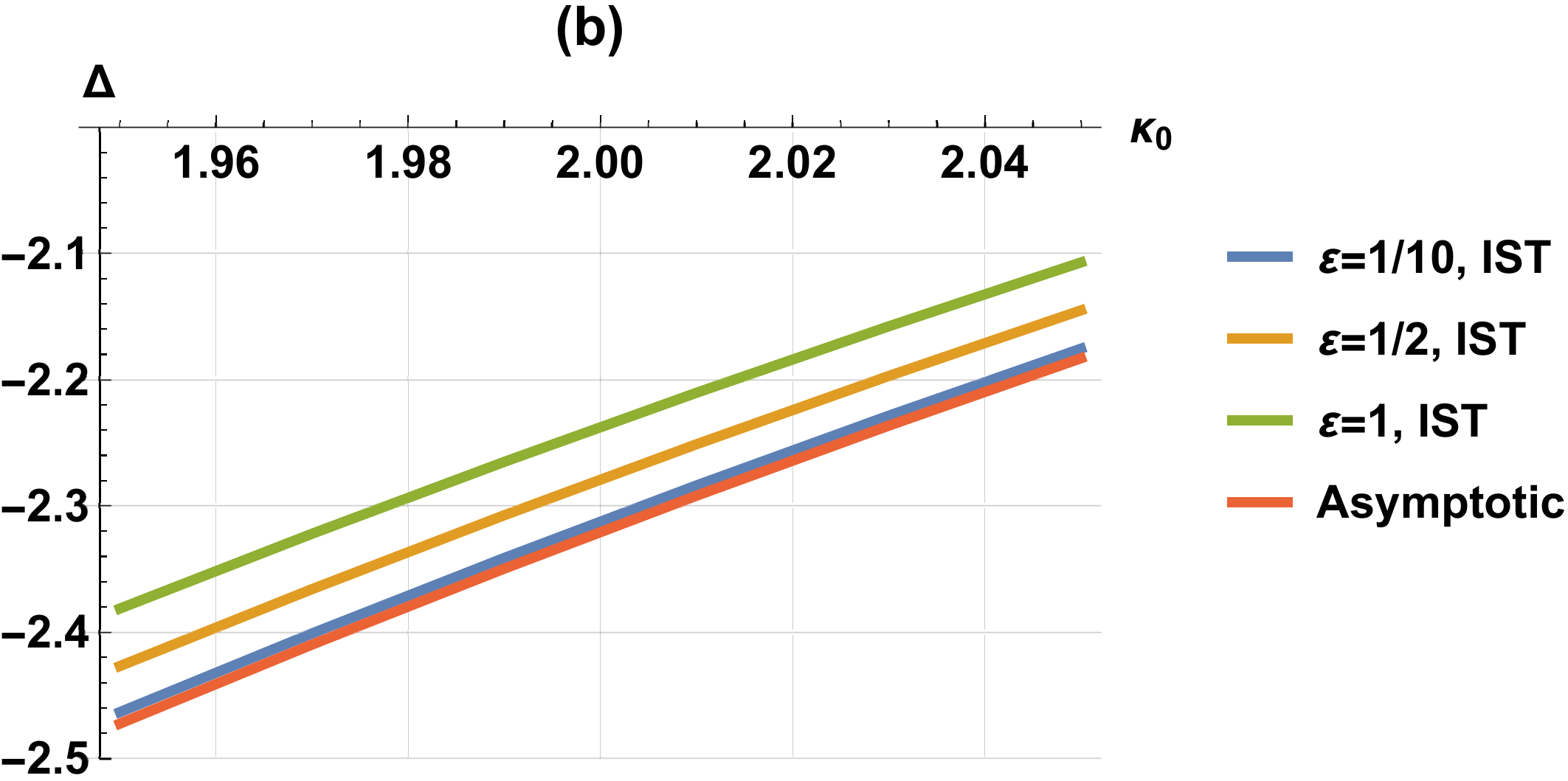}
  \caption{Phase shift of a soliton through a RW as a function of
    initial amplitude parameter $\kappa_0$. Other parameters:
    $x_0 = -15, c = 1$.}
  \label{phase_shift_varykap_diffeps}
\end{figure}

The IST phase shifts found through Eq.~(\ref{KdvSol2A}), for fixed $c$
and $x_0$, are plotted in Fig.~\ref{phase_shift_varykap_diffeps} as a
function of $\kappa_0$. The transmitted phase is less than the initial
phase, indicating a negative phase shift. The magnitude of the shift
is found to decrease as the incoming soliton amplitude relative to the
step height increases. A larger amplitude means the step is small in
comparison to the soliton and does not significantly affect it. In the
limit $\kappa_0 \rightarrow c^+$, the soliton phase shift approaches
negative infinity which corresponds to trapping. The phase shift for
different values of $\varepsilon$ are quite close, so it is difficult
to distinguish between the cases. A close-up view of the curves in
Fig.~\ref{phase_shift_varykap_diffeps}(b) shows {clearly} that as
$\varepsilon$ decreases the IST result (exact) approaches the
asymptotic result.  This observation was already noted in
Fig.~\ref{soliton_transmit_profiles} and was analytically established
Sec.~\ref{small_eps_sec}. Again, the asymptotic approaches considered
are small dispersion limits of the exact IST results.
 
\begin{figure} [h]
\centering
\includegraphics[scale=0.5]{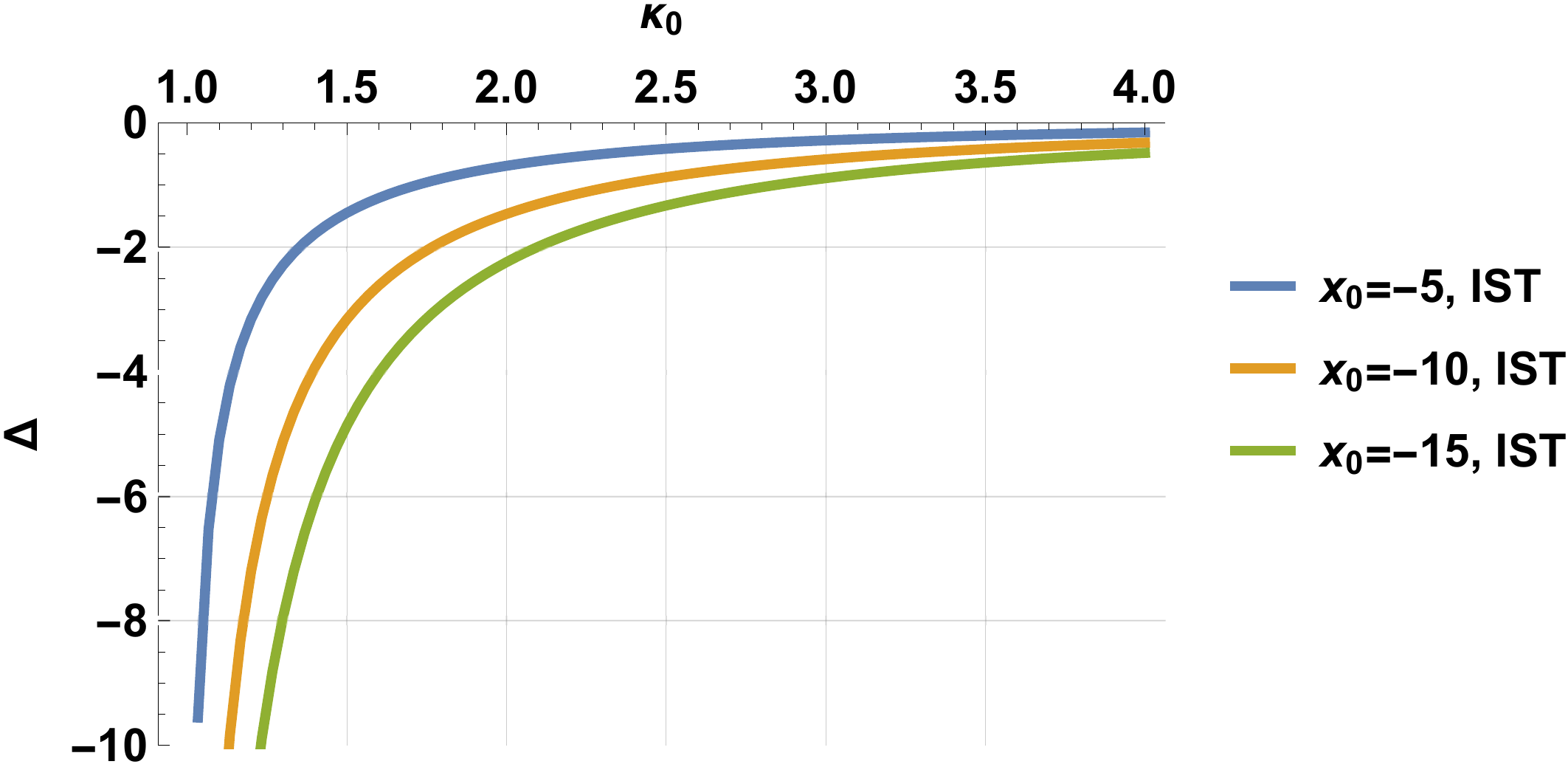}
\caption{Phase shift of a soliton through a RW as a function of
  $\kappa_0$ for different values of the initial position $x_0$. Other
  parameters: $\varepsilon = 1, c = 1$.}
\label{phase_shift_varykap_diffx0}
\end{figure}

Next, the phase shift for different initial positions is displayed in
Fig.~\ref{phase_shift_varykap_diffx0}. The difference in the curves
arises from the fact that the length of the ramp grows with
time. Hence a soliton which starts further away from the ramp (more
negative value of $x_0$) experiences a larger negative phase
shift. Here we find the initial position and the magnitude of the
phase shift are directly proportional to one another.

\begin{figure} [h]
  \centering
  \includegraphics[scale=0.35]{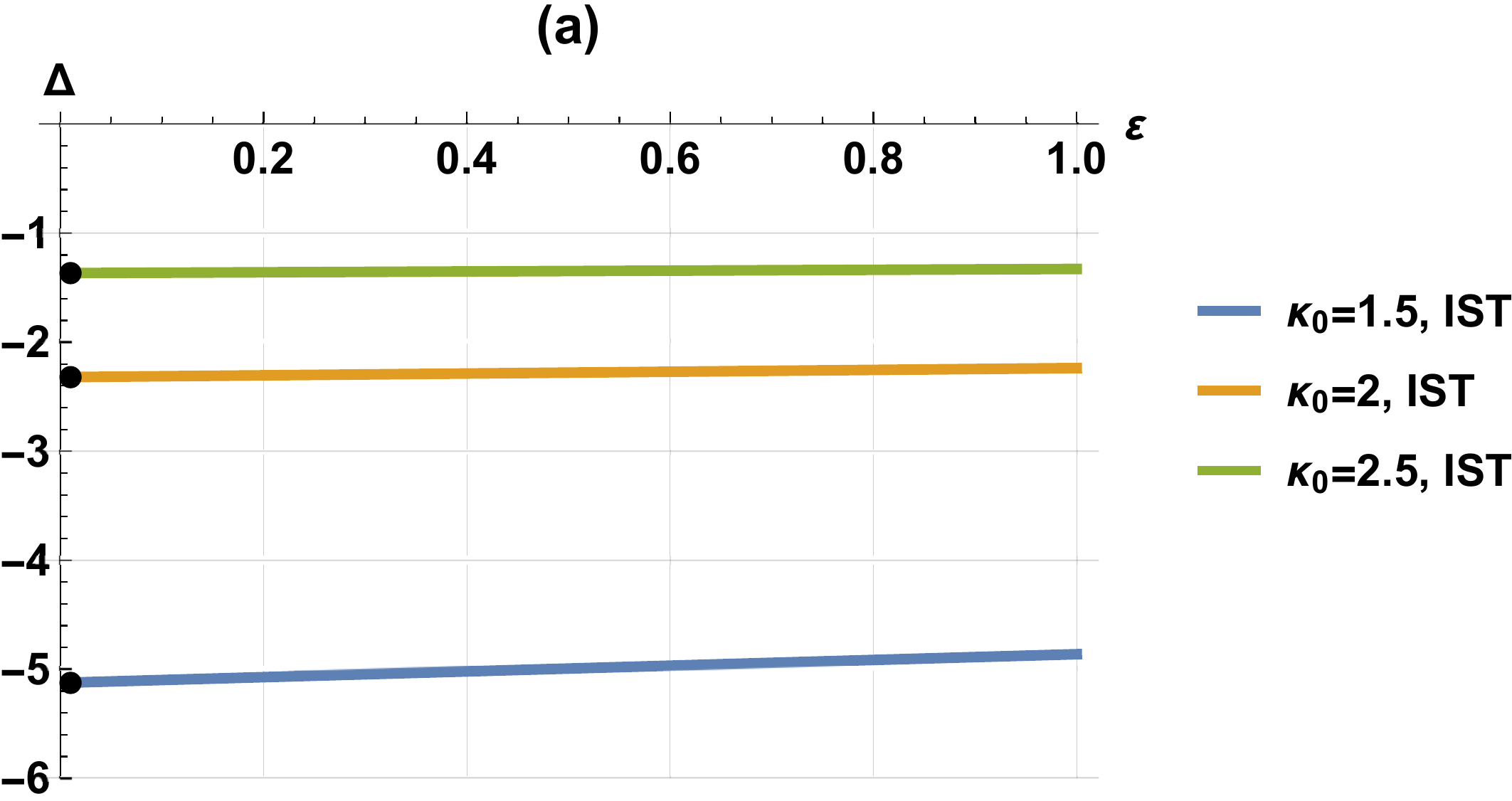}
  \includegraphics[scale=0.35]{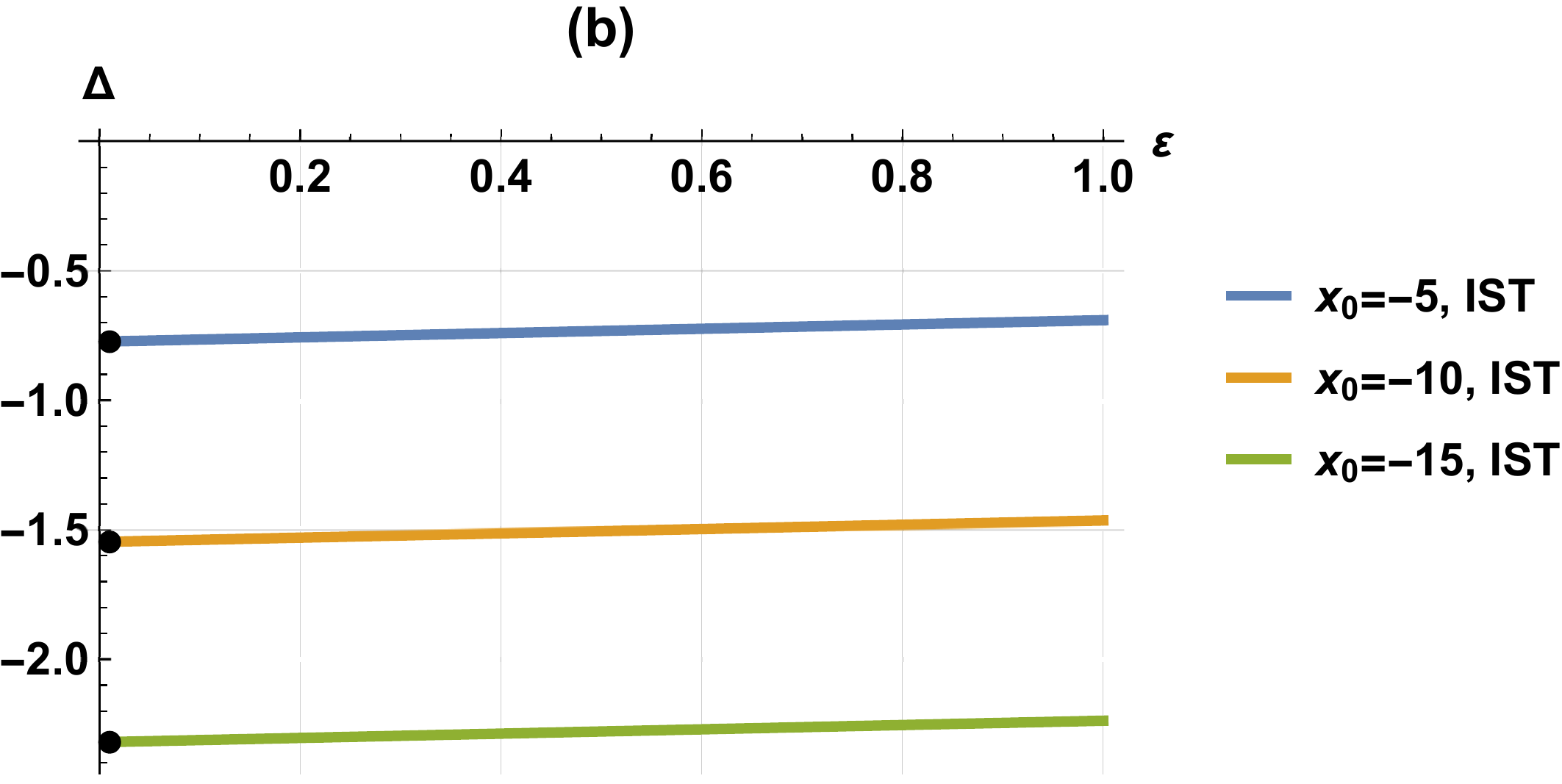}
  \caption{Phase shift of a soliton through RW as a function of
    dispersion parameter $\varepsilon$.  Other parameters: (a)
    $x_0 = -15, c = 1$ and (b) $\kappa_0 = 2, c = 1$. Black dots near
    $\varepsilon = 0$ denote the corresponding asymptotic predictions
    found through Eqs.~(\ref{rare_phase_shift}) and
        (\ref{eq:45}).}
      \label{phase_shift_varyeps}
\end{figure}

The next case to consider is when the dispersion parameter
$\varepsilon$ is varied. The plots in Fig.~\ref{phase_shift_varyeps}
show that the phase shift does not vary much when $\varepsilon$ is
relatively small ($0 < \varepsilon \le 1$). This is especially true
when $\kappa_0$ is large, as Fig.~\ref{phase_shift_varyeps}(a)
demonstrates. This suggests that for large $\kappa_0$ the asymptotic
approximation will continue to provide reasonable approximation, even
when $\varepsilon$ is not too small, i.e. provided the overall value
of $\kappa_0 / \varepsilon$ remains relatively large. The phase shift
{is} found to be directly proportional to the {(negative)} parameter
$x_0$ in Fig.~\ref{phase_shift_varyeps}(b); similar to what was
observed in Fig.~\ref{phase_shift_varykap_diffx0}. Finally, both plots
in Fig.~\ref{phase_shift_varyeps} highlight that as
$\varepsilon \rightarrow 0$ the IST formulae approach the asymptotic
approximation. 

\subsubsection*{Soliton Trapping}

In this section, we consider the scenario of a soliton becoming
trapped on a rarefaction ramp. In this case $\kappa_0 < c$ (or
$a_- < 2 c^2$) and there are no proper solitons in the IST. Instead
there are so-called pseudo embedded eigenvalues (resonant poles) that
yield pseudo solitons, which for $x_0 \ll 1$ resemble solitons prior to
soliton-RW interaction \cite{ALC}. The evolution of a trapped soliton
is displayed in Figs.~\ref{fig:interaction_scenarios}(b) and
\ref{KdV_RW_tunnel_trap}(b).  No matter how long the system is
integrated forward in time, the solitary wave will not reach the top
of the ramp.  The absence of any true eigenvalues renders the soliton
result in (\ref{KdvSol2A}) inapplicable since it assumes that the
soliton transmits through the RW.

\subsubsection{Soliton-DSW Results}

A step down boundary condition that generates a DSW is considered in
this section. The primary case that our IST results can shed insight
upon is when $x_0 < 0$ and the soliton is initially to the left of the
jump. Here, the soliton {\it always} tunnels through the DSW. The
phase shift of a transmitted soliton is calculated via asymptotic
approximation and IST. When a soliton starts to the right of a step
down ($x_0 > 0$), there are two outcomes: (a) soliton trapping (small
amplitude) (b) no interaction with DSW (large amplitude). Again, the
trapping case does not correspond to proper eigenvalues and is not
discussed in detail here. A large amplitude initial condition does
correspond to a genuine soliton (proper eigenvalue), but is not so
interesting.

\subsubsection*{Soliton Initially Left of the Step, $x_0 < 0$}

\begin{figure} [h]
  \centering
  \includegraphics[scale=0.6]{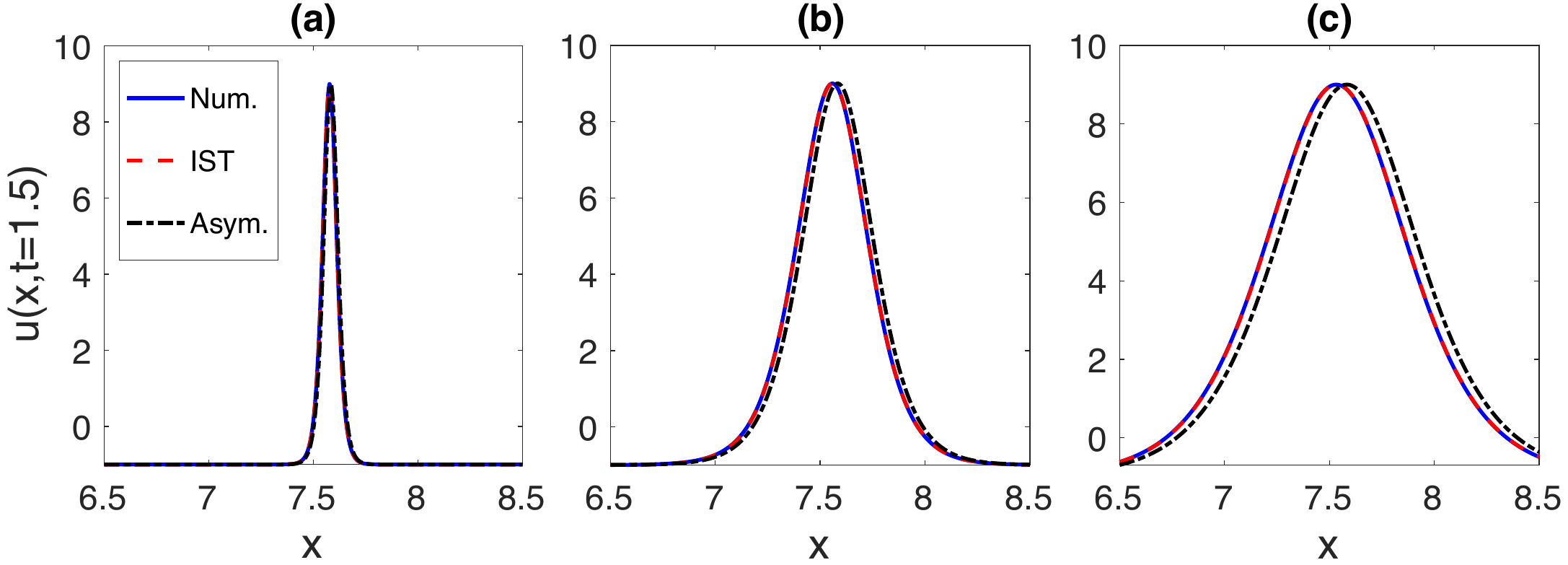}
  \caption{Soliton (post transmission) portion of the soliton-DSW
    solution for (a) $\varepsilon = 1/10$, (b) $\varepsilon = 1/2$,
    and (c) $\varepsilon = 1$. Shown are numerical, IST, and
    asymptotic results. The initial condition used is
    (\ref{general_IC_soliton}) with parameters:
    $ x_0 = -15, \kappa_0 = 2, c = 1$.}
  \label{soliton_transmit_profiles_DSW}
\end{figure}

The typical evolution of a soliton initially placed to the left of the
step down at the origin is shown in Fig.~\ref{KdV_DSW_tunnel_trap}(a).
Profiles corresponding to transmitted soliton results are presented in
Fig.~\ref{soliton_transmit_profiles_DSW} for different values of
$\varepsilon$. Three different approaches are compared: numerics, IST,
and asymptotics. The IST approximation of the phase for a transmitted
soliton is given in Eq.~(\ref{KdvSol2A}). The asymptotic approximation
of the phase shifts are obtained through the Whitham (\ref{eq:87}) or
the IST small dispersion limit (\ref{DSW_phase_asym}). Indeed,
Fig.~\ref{soliton_transmit_profiles_DSW} indicates overall good
agreement between the different approaches, with the the asymptotic
approximation improving as $\varepsilon \rightarrow 0$.

\begin{figure} [h]
  \centering
  \includegraphics[scale=0.3]{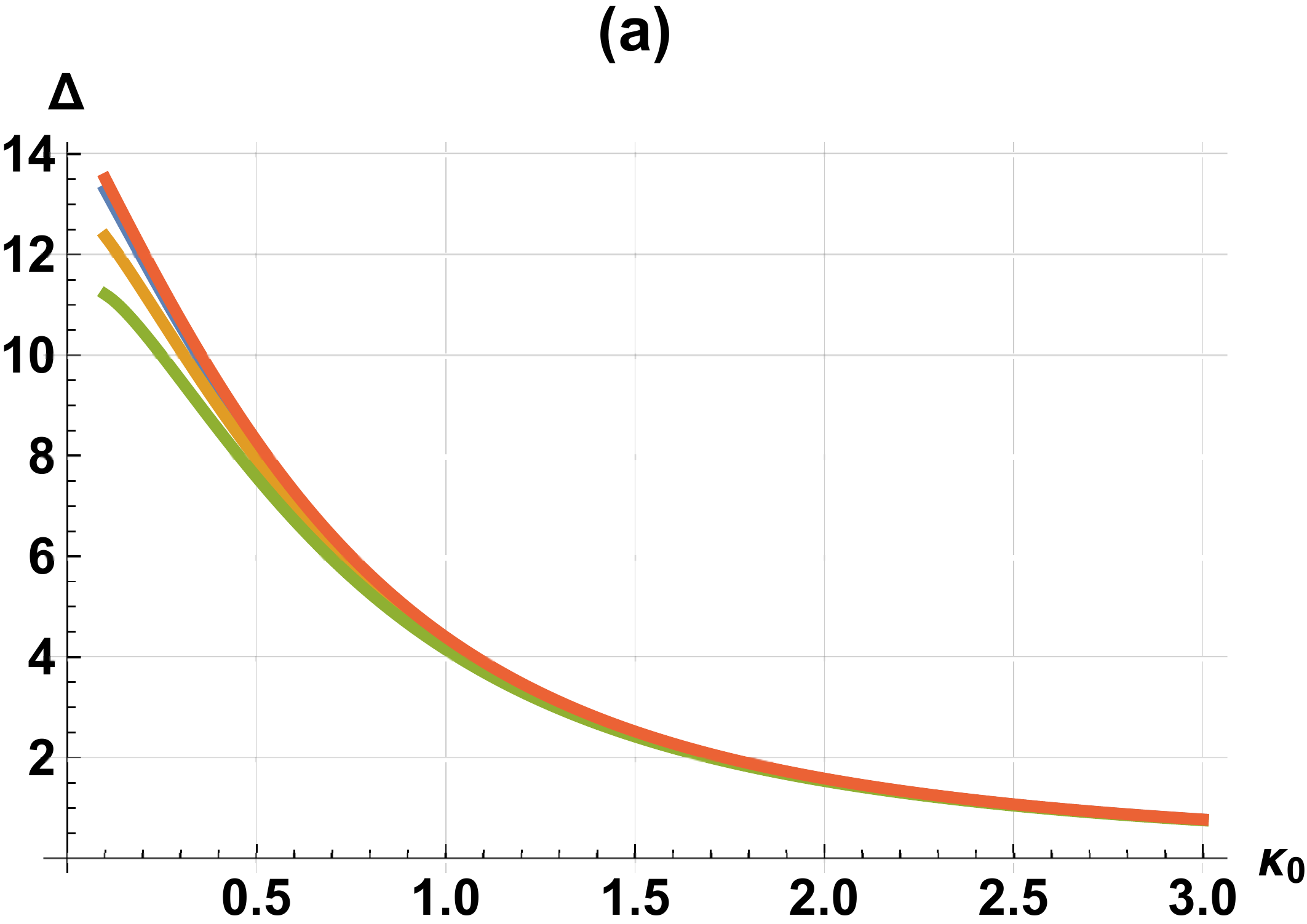}
  \includegraphics[scale=0.4]{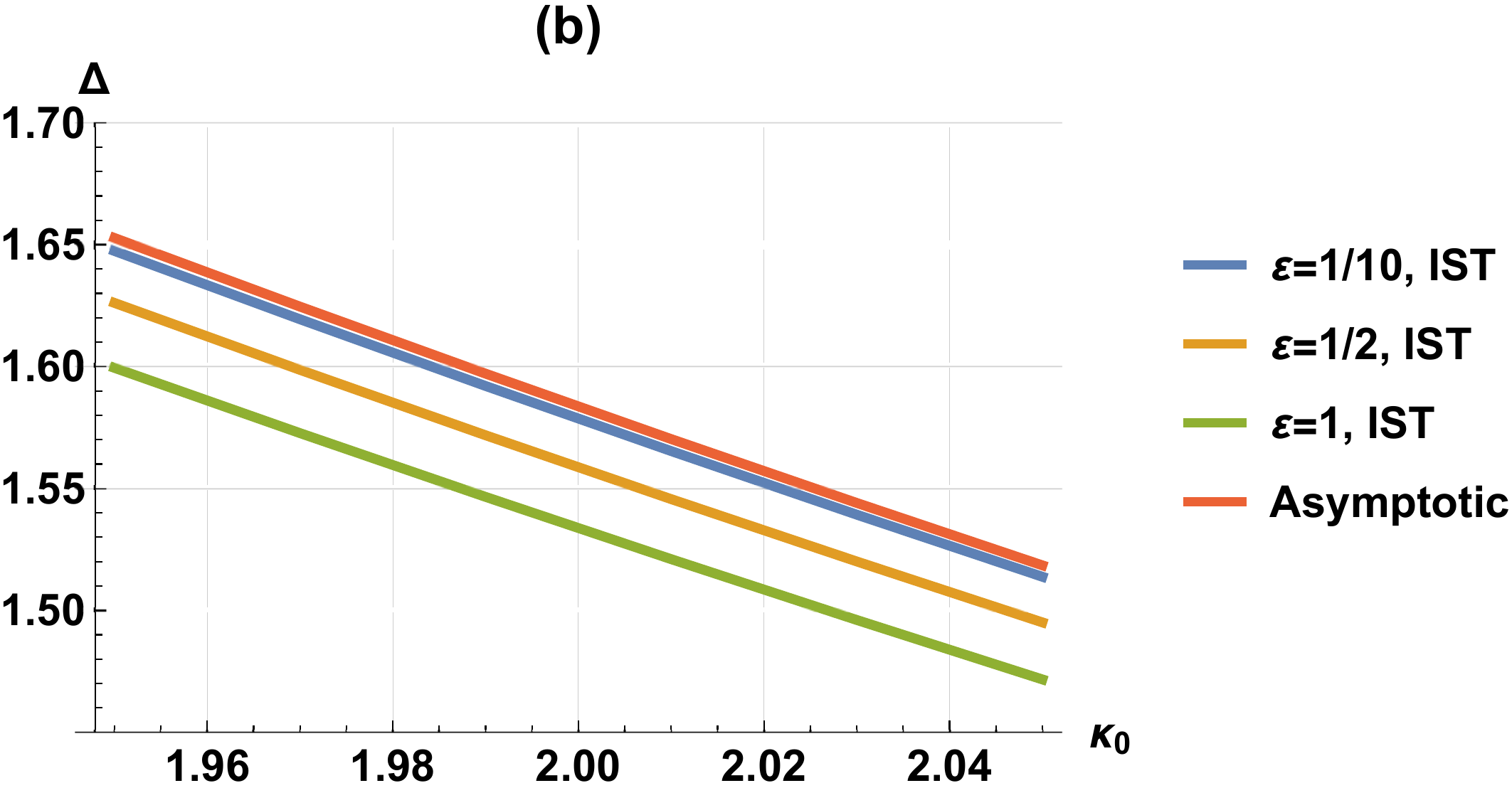}
  \caption{Phase shift of soliton through a DSW as a function of
    initial amplitude parameter $\kappa_0$ {with varying
      $\varepsilon$}. Other parameters: $x_0 = -15, c = 1$.}
  \label{phase_shift_varykap_diffeps_DSW}
\end{figure}

The phase shift for transmitted solitons is considered in more detail
next. Recall the IST phase shift is defined by $\Delta = x_s^- - x_0$
for the soliton phase given in (\ref{KdvSol2A}). The phase shift for
different values of $\kappa_0$ {is} shown in
Fig.~\ref{phase_shift_varykap_diffeps_DSW}. The figures indicate that
the soliton always experiences a positive phase shift forward
(opposite the RW case). Moreover, the phase shift approaches
{zero} 
in the large amplitude limit{,} when $\kappa_0 \rightarrow
\infty$. What is also observed from
Fig.~\ref{phase_shift_varykap_diffeps_DSW}(b) is that the phase shift
curves approach the asymptotically derived formula derived in
(\ref{DSW_phase_asym}) and (\ref{eq:87}) as
$\varepsilon \rightarrow 0$.

\begin{figure} [h]
  \centering
  \includegraphics[scale=0.5]{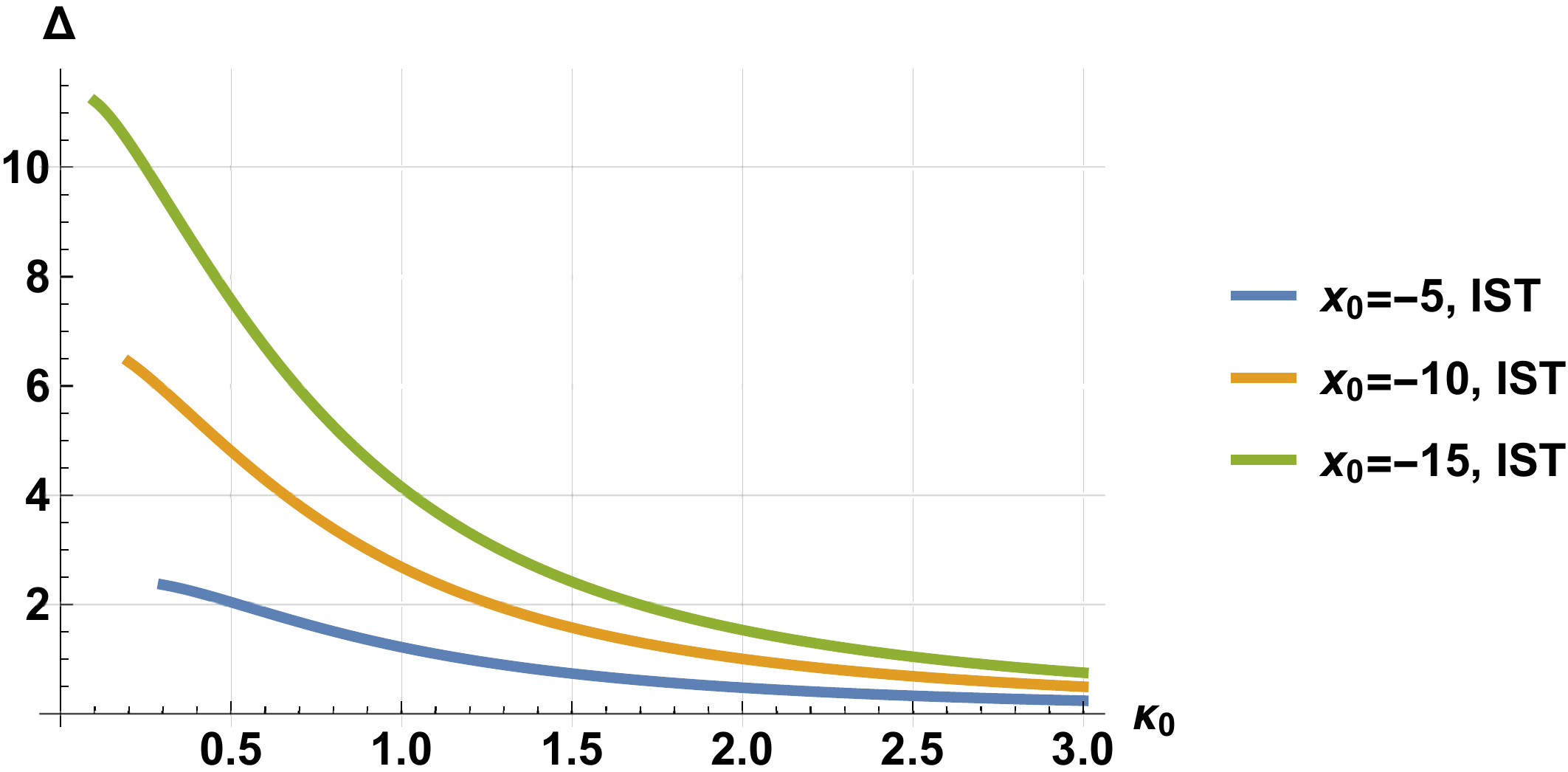}
  \caption{Phase shift of soliton through a DSW as a function of
    amplitude parameter $\kappa_0$ {with varying $x_0$}. Other
    parameters: $\varepsilon = 1, c = 1$.}
  \label{phase_shift_varykap_diffx0_DSW}
\end{figure}

Next, consider when the initial position of the soliton is different,
but still negative. The results in
Fig.~\ref{phase_shift_varykap_diffx0_DSW} show that the farther a
soliton starts from the jump, the larger the phase shift it
experiences. Intuitively, this makes since the length of the DSW
region, $10 c^2 t$, grows with time, hence a soliton that encounters a
DSW later in time experiences its dispersive effects for a longer
duration.  At fixed amplitude, the phase shift is inversely
proportional to the initial position.

\begin{figure} [h]
  \centering
  \includegraphics[scale=0.35]{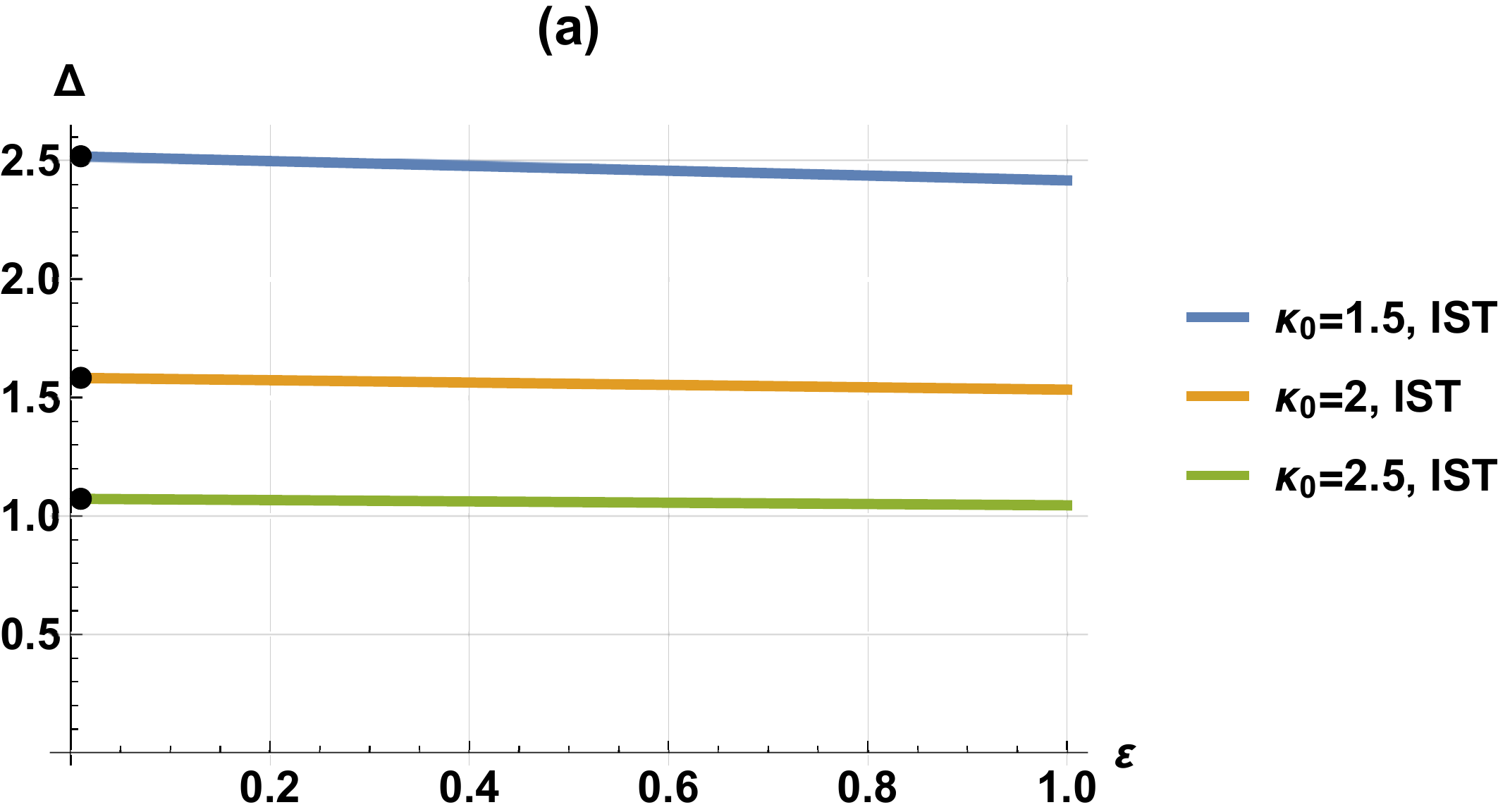}
  \includegraphics[scale=0.35]{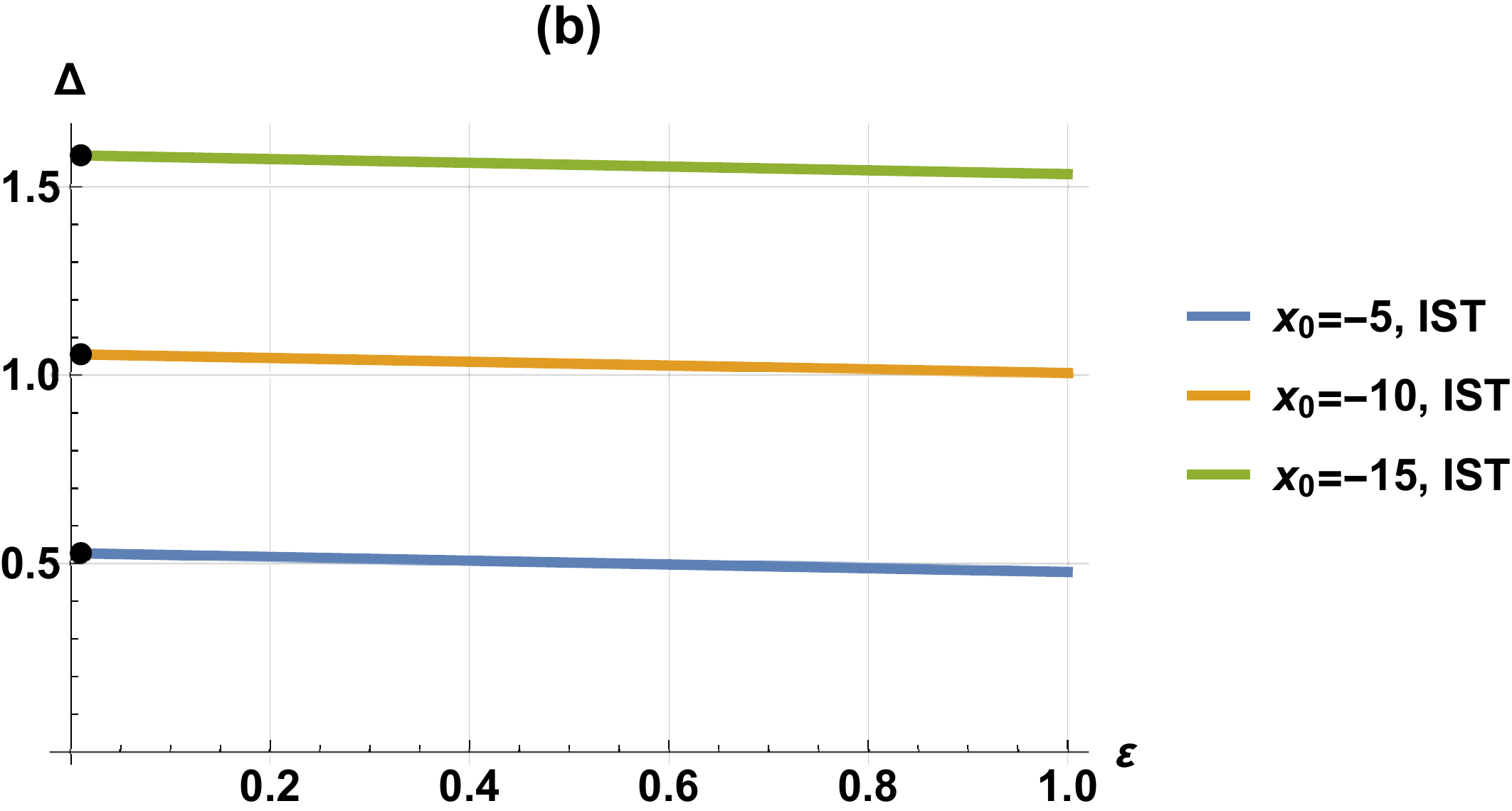}
  \caption{Phase shift of soliton through a DSW as a function of
    dispersion parameter $\varepsilon$.  Other parameters: (a)
    $x_0 = -15, c = 1$ and (b) $\kappa_0 = 2, c = 1$. Black dots near
    $\varepsilon = 0$ denote the corresponding asymptotic predictions
    found through Eqs.~(\ref{DSW_phase_asym}) and (\ref{eq:87}).}
  \label{phase_shift_varyeps_DSW}
\end{figure}

The final set of phase shift figures to consider are those when
$\varepsilon$ is varied. Two separate cases are considered in
Fig.~\ref{phase_shift_varyeps_DSW}: different values of $\kappa_0$
[see Fig.~\ref{phase_shift_varyeps_DSW}(a)] and different values of
$x_0$ [see Fig.~\ref{phase_shift_varyeps_DSW}(b)]. In both cases, the
phase shift is positive and there is little variation in the curves
for the range of $\varepsilon$ considered here, which are relatively
small in magnitude. From Fig.~\ref{phase_shift_varyeps_DSW}(a) we note
that the phase shift is inversely proportional to the amplitude. These
figures show that the initial amplitude and position most
significantly affect the soliton phase shift. As expected, the phase
shift approaches the IST asymptotics and modulation theory prediction
in Eqs.~(\ref{DSW_phase_asym}) and (\ref{eq:87}) as
$\varepsilon \rightarrow 0$.

We note recent, related work in which a mKdV soliton interacts
propagates into an oscillatory wavetrain that asymptotes to a cnoidal
wave solution as $x \to \infty$ \cite{Girotti2022}.  The initial value
problem is solved using Riemann-Hilbert methods.  An amplitude
threshold, similar to the requirement $\kappa_0 > c$ for KdV soliton
tunneling, is obtained in which the soliton completely penetrates the
expanding, modulated cnoidal wavetrain.

\subsubsection*{Soliton Initially Right of the Step, $x_0 > 0$}

A soliton that begins to the right of the ramp can either become
trapped inside the DSW or not enter the DSW region at all. In the
latter case, the solution resembles (\ref{DSW_asym_soln_x0_pos}) and
(\ref{KdvSol2A}) for all $t > 0$.  Moreover, depending on the initial
amplitude, the soliton will propagate with negative
($\eta_0 < c \sqrt{3/2}$), positive ($\eta_0 > c \sqrt{3/2}$), or zero
($\eta_0 = c \sqrt{3/2}$) velocity.  If $\eta_0 < c \sqrt{3/2}$, but
$\eta_0 > c $, then the soliton moves with negative velocity, but {\it
  never} catches the solitonic edge of the DSW. For $\eta_0 > c$, the
soliton corresponds to a proper eigenvalue of the direct scattering
problem and a proper soliton.

When $0 < \eta_ 0 < c$, there are no proper eigenvalues associated
with the Schr\"odinger operator and this mode corresponds to a pseudo
soliton \cite{ALC}.  The evolution of a soliton becoming trapped
inside a DSW is shown in Fig.~\ref{KdV_DSW_tunnel_trap}(b). The
soliton reaches the solitonic edge of the DSW region at approximately
the trapping time (\ref{trapping_time_DSW}). Once inside the DSW
region, the soliton takes a dark soliton breather form on the DSW
background (see Fig.~\ref{fig:soliton_dsw_trapping_spectrum}).

\section{Conclusion}
\label{conclude}

Separately, the evolution of solitons and jump-induced waves in the
Korteweg-de Vries equation have been thoroughly studied. This work
considered solutions composed of {\it both} a soliton and either a
rarefaction wave or a dispersive shock wave, and their mutual
interaction. Various analytical approaches, such as a soliton
perturbation theory, Whitham modulation theory, and the Inverse
Scattering Transform, are able to describe the evolution of these
interactions and agree well with direct numerics. In particular, these
methods are able to elucidate the phenomena of soliton tunneling and
trapping and its dependence on the initial data of the soliton
relative to the step. Furthermore, the step-up and step-down problems
are found to be duals of each other and related via a principle known
as reciprocity. From the point-of-view of the scattering data, this
reciprocity condition is a relationship between the discrete
eigenvalues and the jump height, corresponding to exact integrals of
motion.  Remarkably, the adiabatic invariants obtained via the
asymptotic methods presented here predict precisely the same
reciprocity condition.

The first method of analysis, soliton perturbation theory, benefits
from being a general approach to soliton-mean interaction, so long as
an approximate description of the mean field is available.  This
approach is particularly effective for soliton-RW interaction and can
readily be extended to other evolutionary equations that admit
solitary wave solutions.  The second method of analysis, Whitham
modulation theory, can similarly be extended to a broad class of
evolutionary equations \cite{Hoefer2}.  The primary limitation for
both of these methods when applying them to non-integrable equations
is obtaining a tractable, accurate, analytical description of
soliton-DSW interaction.  We overcome this limitation for the
completely integrable KdV equation using multiphase Whitham modulation
theory.  The soliton-mean interaction is approximated with a 2-phase
solution of the KdV equation whose associated Schr\"odinger operator's
$L^\infty(\mathbb{R})$ finite gap spectrum is allowed to vary slowly
in space and time.  Soliton-DSW interaction is a modulated bright or
dark breather solution of the KdV equation, depending on whether the
soliton is transmitted or trapped, respectively.  The third method of
analysis using the Inverse Scattering Transform results in the exact
solution whose small dispersion limit is analyzed.  In contrast to
multiphase Whitham modulation theory, the Schr\"odinger operator's
$L^2(\mathbb{R})$ spectrum is independent of time.  The existence of a
discrete eigenvalue implies the transmission of a soliton through a RW
or a DSW and exactly determines the transmitted soliton's amplitude.
The spectrum in the trapped case does not correspond to discrete
eigenvalues of decaying eigenfunctions, hence we refer to them pseudo
solitons.  All three methods of analysis asymptotically agree in the
limit of a soliton initialized sufficiently far away from the step or,
equivalently, in the small dispersion regime.

\section*{Acknowledgements}
\label{Acknowledge}
The authors would like to thank the Isaac Newton Institute for
Mathematical Sciences for support and hospitality during the program
Dispersive Hydrodynamics when work on this paper was undertaken.  This
work was supported by EPSRC Grant Number EP/R014604/1.  MJA was
partially supported by NSF under Grant DMS-2005343.  XL was partially
supported by NSFC under Grant 12101590.  MAH was partially supported
by NSF under Grant DMS-1816934.  GAE was partially supported by EPSRC
under Grant EP/W032759/1.


\end{document}